\documentclass{aa} 

\usepackage{graphicx}
\usepackage{txfonts}
\usepackage{natbib}
\usepackage{amsmath}
\usepackage{color}
\usepackage{array}
\usepackage{multirow}
\usepackage{subfig}
\usepackage{xspace}
\usepackage{xcolor}
\usepackage{url}
\usepackage{rotating}
\usepackage{pdflscape}

\newcommand{\equ}[1]{Eq.~\ref{eq:#1}}
\newcommand{\equs}[2]{Eqs.~\ref{eq:#1}-\ref{eq:#2}}
\newcommand{\fig}[1]{Fig.~\ref{fig:#1}}
\newcommand{\figs}[2]{Figs.~\ref{fig:#1}-\ref{fig:#2}}
\newcommand{\tab}[1]{Table~\ref{tab:#1}}

\newcommand{\sect}[1]{Sect.~\ref{sec:#1}}

\newcommand{\lcdm}[0]{$\Lambda$CDM\xspace}
\newcommand{\biso}[0]{\beta_\mathrm{iso}}
\newcommand{\blit}[0]{\beta_\mathrm{lit}}
\newcommand{\bneg}[0]{\beta_\mathrm{neg}}

\newcommand{\tabsfit}[0]{Tables~\ref{tab:tablcdm} and~\ref{tab:tabmond}\xspace}
\newcommand{\allgals}[0]{Figs.~\ref{fig:g0}~--~\ref{fig:g19}\xspace}
\newcommand{\re}[0]{$R_\mathrm{e}$\xspace}

\begin{document}

\title{Study of gravitational fields and globular cluster systems of early-type galaxies}

\titlerunning{{Gravitational fields of early-type galaxies}}

\author{M. B\'{i}lek\inst{1}\fnmsep\inst{2}
\and
S. Samurovi\'c\inst{3}
\and
F. Renaud\inst{4}
}
\institute{Astronomical Institute, Czech Academy of Sciences, Bo\v{c}n\'{i} II 1401/1a, CZ-141\,00 Prague, Czech Republic\\
\email{bilek@asu.cas.cz}
\and Faculty of Mathematics and Physics, Charles University in Prague, Ke~Karlovu 3, CZ-121 16 Prague, Czech Republic
\and
Astronomical Observatory, Volgina 7, 11060 Belgrade, Serbia
\and
Lund Observatory, Sölvegatan 27, Box 43, SE-221 00 Lund, Sweden
}

\date{Received ...; accepted ...}

\abstract{Gravitational fields at the outskirts of early-type galaxies (ETGs) are difficult to constrain observationally. It {thus} remains poorly explored how well the \lcdm and MOND hypotheses agree with ETGs. }
{ The dearth of studies on this topic motivated us to gather a large sample of ETGs and examine homogeneously which dark matter halos they occupy, whether the halos follow the theoretically predicted stellar-to-halo mass relation (SHMR) and the halo mass-concentration relation (HMCR), whether ETGs obey MOND and the radial acceleration relation (RAR) observed for late-type galaxies (LTGs), and finally whether \lcdm or MOND {perform} better in ETGs.}
{We employed Jeans analysis of radial velocities of  globular clusters (GCs). We analysed nearly all ETGs having more than about 100 archival GC radial velocity measurements available. The GC systems of our 17 ETGs extend mostly over ten effective radii. A \lcdm simulation of GC formation helped us to interpret the results.}
{Successful \lcdm fits are found for all galaxies, but compared to  the theoretical HMCR and SHMR, the best-fit halos usually have concentrations that are too low and stellar masses that are too high for their masses. This might be because of tidal stripping of the halos or because ETGs and LTGs occupy  different halos. Most galaxies can be fitted by the MOND models successfully as well, but for some of the galaxies, especially those in centers of galaxy clusters, the observed GCs velocity dispersions are too high. This might be a manifestation of the additional dark matter that MOND requires in galaxy  clusters. Additionally, we find many signs that the GC systems were perturbed by galaxy interactions. Formal statistical criteria prefer the best-fit \lcdm  models over the MOND models, but this might be due to the higher flexibility of the \lcdm models. The MOND approach can predict the GC velocity dispersion profiles better.}
{}


\maketitle

\section{Introduction} \label{sec:intro}
The missing mass problem has not been solved decisively yet. The two most discussed solutions are the standard cosmological \lcdm model (e.g., \citealp{mo}) and the MOND paradigm of modified dynamics  \citep{milg83a, famaey12, milgcjp}. In this paper we aim to test their predictions in early-type galaxies (ETGs).

Assuming the \lcdm paradigm, galaxies are surrounded by dark matter halos whose density can be described well by the Navarro-Frenk-White (NFW) profile \citep{nfw} according to dark-matter-only cosmological simulations. These halos are expected to meet the stellar-to-halo mass relation (SHMR) between the stellar mass of the galaxy and the mass of its dark halo (e.g., \citealp{behroozi13}). This relation is supposed to exist because the transport of baryons inside or outside the halo depends on the mass of the halo; the more massive the halo is, the more it accretes satellites and intergalactic gas. On the other side, the processes expelling the baryons from the halos such as {active galactic nuclei} outflows, supernova explosions, and stellar winds are less effective if the baryons reside in a deeper potential well. This relation can be recovered, for example, by the abundance matching technique based on comparing the halo mass function deduced from cosmological simulations to the observed galaxy stellar mass function while assuming that the most massive galaxies lie within the most massive halos; i.e., the recovered SHMR is a combination of results from observations and simulations. We must remember that the recovered SHMR is not necessarily the SHMR that \lcdm would predict in a perfect simulation, which would include, for example, all baryonic processes correctly. In the current hydrodynamical cosmological simulations  the parameters regulating the baryonic processes have to be tuned so that the galaxy stellar mass function matches the observations \citep{genel14, crain15, pillepich18}. The simulations that are not explicitly tuned to reproduce the stellar mass function do not do so properly \citep{khandai15, kaviraj17}.

Another correlation that the dark matter halos are expected to follow is that between the halo mass and its concentration referred to as the halo  mass-concentration relation (HMCR). The concentration, $c$, of a NFW halo is the virial radius of the halo divided by its scale radius. The {HMCR} is a result deduced from \lcdm cosmological simulations.

\begin{table*}
  \caption{Galaxy sample}
  \label{tab:tabgals}
  \centering
  \begin{tabular}{lccccccccccccc}
 \hline\hline
 Name & $d$ & 1\arcmin &  $\log L$  & $R_\mathrm{e}$ & $n$ & $B-V$  & Type & Env & Prof  & Iso & $\lambda_{R_\mathrm{e}}$ & $\epsilon$ & Rot \\ 
 & [Mpc] & [kpc]  & [$L_\sun$] & [kpc] & & [mag] & & & & & & & \\\hline 

N\,821 & 24$^{+2}_{-2}$ & 7.0 & 10.5 & 4.7$^{14}$ & 4.7$^{14}$ & 0.87 & E6$^{19}$ & F$^{19}$ & $\mid$$^{1}$ & D$^{7, 9}$ & 0.27$^{8}$ & 0.35$^{8}$ & f\\
N\,1023 & 11.4$^{+0.9}_{-0.8}$ & 3.3 & 10.5 & 2.7$^{28}$ & 4.2$^{28}$ & 0.91 & S0$^{8}$ & G$^{28}$ & \textbackslash$^{1}$ & D$^{5}$ & 0.39$^{8}$ & 0.63$^{8}$ & f\\
N\,1399 & 20$^{+2}_{-1}$ & 5.8 & 10.7 & 3.5$^{15}$ & 5.6$^{15}$ & 0.93 & E1pec$^{24}$ & C$^{24}$ & $\cap$$^{1}$ & D$^{10}$ & 0.08$^{11}$ & 0.09$^{11}$ & s\\
N\,1400 & 26$^{+4}_{-4}$ & 7.7 & 10.4 & 3.4$^{14}$ & 4.0$^{14}$ & 0.89 & S0/E0$^{19}$ & G$^{19}$ & $\cap$$^{1}$ & 0$^{2}$ & 0.27$^{8}$ & 0.11$^{8}$ & f\\
N\,1407 & 29$^{+4}_{-3}$ & 8.4 & 11.0 & 9.4$^{14}$ & 8.3$^{14}$ & 0.95 & E0$^{19}$ & G$^{19}$ & $\cap$$^{2}$ & 0$^{2}$ & 0.08$^{8}$ & 0.05$^{8}$ & f\\
N\,2768 & 22$^{+3}_{-2}$ & 6.5 & 10.7 & 8.9$^{16}$ & 3.3$^{16}$ & 0.91 & E6/S0$_{1/2}$$^{19}$ & G$^{19}$ & ?$^{3}$ & D$^{5}$ & 0.25$^{8}$ & 0.57$^{8}$ & f\\
N\,3115 & 9.7$^{+0.4}_{-0.4}$ & 2.8 & 10.2 & 4.8$^{17}$ & 4.4$^{17}$ & 0.90 & S0$^{19}$ & F$^{19}$ & \textbackslash$^{1}$ & D$^{6}$ & 0.58$^{8}$ & 0.49$^{8}$ & f\\
N\,3377 & 11.2$^{+0.5}_{-0.5}$ & 3.3 & 9.9 & 2.9$^{16}$ & 5.0$^{16}$ & 0.82 & E6$^{19}$ & G$^{19}$ & \textbackslash$^{1}$ & D$^{5}$ & 0.52$^{8}$ & 0.33$^{8}$ & f\\
N\,4278 & 16$^{+2}_{-1}$ & 4.7 & 10.2 & 2.5$^{16}$ & 4.8$^{16}$ & 0.90 & E1-2$^{19}$ & G$^{19}$ & $\cap$$^{1}$ & B$^{5}$ & 0.18$^{8}$ & 0.09$^{8}$ & f\\
N\,4365 & 20$^{+2}_{-2}$ & 5.9 & 10.7 & 8.5$^{18}$ & 5.2$^{18}$ & 0.93 & E3$^{19}$ & G$^{19}$ & $\cap$$^{1}$ & B$^{7}$ & 0.09$^{8}$ & 0.24$^{8}$ & s\\
N\,4472 & 16.3$^{+0.8}_{-0.7}$ & 4.7 & 10.9 & 3.9$^{15}$ & 3.0$^{15}$ & 0.93 & E2$^{24}$ & C$^{24}$ & $\cap$$^{1}$ & B$^{7, 9}$ & 0.077$^{9}$ & 0.172$^{9}$ & s\\
N\,4486 & 16$^{+1}_{-1}$ & 4.7 & 10.8 & 5.8$^{18}$ & 2.9$^{18}$ & 0.92 & E0$^{19}$ & C$^{19}$ & $\cap$$^{1}$ & B$^{5}$ & 0.02$^{8}$ & 0.16$^{8}$ & s\\
N\,4494 & 17.1$^{+0.9}_{-0.8}$ & 5.0 & 10.4 & 3.7$^{16}$ & 3.4$^{16}$ & 0.85 & E1-E2$^{19}$ & G$^{19}$ & \textbackslash$^{1}$ & D$^{7}$ & 0.21$^{8}$ & 0.14$^{8}$ & f\\
N\,4526 & 17$^{+2}_{-1}$ & 4.9 & 10.4 & 2.7$^{28}$ & 3.6$^{28}$ & 0.89 & S0$^{8}$ & C$^{28}$ & ? & B$^{5}$ & 0.45$^{8}$ & 0.76$^{8}$ & f\\
N\,4649 & 17$^{+1}_{-1}$ & 4.9 & 10.8 & 5.1$^{15}$ & 3.6$^{15}$ & 0.93 & E2/S0$^{8}$ & C$^{24}$ & $\cap$$^{1}$ & B$^{7, 9}$ & 0.127$^{9}$ & 0.156$^{9}$ & f\\
N\,5128 & 4.2$^{+0.3}_{-0.3}$ & 1.2 & 10.5 & 6.2$^{14}$ & 4.0$^{14}$ & 0.87 & S0pec/Epec$^{25}$ & G$^{26}$ & ?$^{12}$ & ?$^{13}$ & 0.15$^{13}$ & 0.05$^{27}$ & f\\
N\,5846 & 25$^{+2}_{-2}$ & 7.2 & 10.7 & 8.1$^{16}$ & 3.9$^{16}$ & 0.94 & E0$^{19}$ & G$^{19}$ & ?$^{4}$ & B$^{5}$ & 0.03$^{8}$ & 0.08$^{8}$ & s\\\hline
R17x & 19$^{+2}_{-2}$ & 5.5 & 10.2 & 1.5 & 2.6 & 0.90 & - & - &  & - & - & - & -\\
R17y & 19$^{+2}_{-2}$ & 5.5 & 10.2 & 1.7 & 2.0 & 0.90 & - & - &  & - & - & - & -\\
R17z & 19$^{+2}_{-2}$ & 5.5 & 10.2 & 2.0 & 2.0 & 0.90 & - & - &  & - & - & - & -\\

 \hline
  \end{tabular}
  \tablefoot{
 $\boldsymbol{d}$ -- Galaxy distance and its uncertainties according to \citet{tonry01}. 
 \textbf{1\arcmin} -- The distance corresponding to 1\arcmin\ if the galaxy is found at the distance of $d$. 
 $\boldsymbol{L}$ -- Galaxy $B$-band luminosity calculated from $d$ and the apparent magnitude listed in the HyperLeda database\footnote{\url{http://leda.univ-lyon1.fr/}} \citep{hyperleda}.
 $\boldsymbol{R_\mathrm{e}}$ -- S\'ersic effective radius the galaxy according to the cited works assuming the galaxy distence $d$.  
 $\boldsymbol{n}$ -- S\'ersic index of the galaxy. 
 $\boldsymbol{B-V}$ -- Color index according to the HyperLeda database.
 \textbf{Type} -- Galaxy morphological type. 
 \textbf{Env} -- Galaxy environment (F = field, G = group, C = cluster).
 \textbf{Prof} -- Galaxy central photometric profile ($\cap$ is core, \textbackslash\ is power law, $\mid$ is intermediate, and "?" is unknown).
 \textbf{Iso} -- Isophotal shape ("D" indicates disky isophotes, "B" indicates boxy isophotes, "0" indicates pure ellipses, and "?" indicates unknown because of observational problems). 
 $\boldsymbol{\lambda_{R_\mathrm{e}}}$ -- Degree of rotational support of the galaxy (\equ{lam}). For NGC\,5128, $\lambda_{R_\mathrm{e}}$  was estimated from Fig.~14 in \citet{coccato09} as an approximate average between the last measured data point for the stellar kinematics and from the planetary nebula kinematics at $1\,R_\mathrm{e}$.
 $\boldsymbol{\epsilon}$ -- Galaxy ellipticity.
 \textbf{Rot} -- Galaxy rotator type ("s" indicates slow and "f" indicates slow rotators) as determined using the rotator criterion number.
 \textbf{References}:
(1)~\citet{lauer07}; (2)~\citet{spolaor08}; (3)~\citet{lauer05}; (4)~\citet{rest01}; (5)~\citet{emsellem07}; (6)~\citet{nieto91}; (7)~\citet{bender89}; (8)~\citet{arnold14}; (9)~\citet{emsellem11}; (10)~\citet{kissler97}; (11)~\citet{scott14}; (12)~\citet{israel98}; (13)~\citet{coccato09}; (14)~\citet{saxton10}; (15)~\citet{dullo14}; (16)~\citet{atlas3d17}; (17)~\citet{hu08}; (18)~\citet{kormendy09}; (19)~\citet{pota13}; (24)~NASA Extragalactic Database\footnote{\url{https://ned.ipac.caltech.edu/}}; (25)~\citet{harris10}; (26)~\citet{devac75}; (27)~\citet{cappellari09}; (28)~\citet{alabi17}.  }
\end{table*}

The MOND paradigm consists in a change of the laws of inertia or gravity in weak gravitational fields. In other words, according to MOND we encounter the missing mass problem mainly because the laws of general relativity are not valid for weak gravitational fields just as the laws of Newtonian dynamics are not valid for strong gravitational fields. In MOND, test particles move approximately obeying the equation
\begin{equation}
\vec{a_\mathrm{N}} = \vec{a}\ \mu\left( \frac{a}{a_0}\right),
\label{eq:mond}
\end{equation}

where $\vec{a}$ is the actual acceleration of the particle and $\vec{a}_\mathrm{N}$ is the acceleration predicted by the Newtonian dynamics on the basis of the distribution of the observable matter. The interpolation function $\mu$ switches gradually between the Newtonian regime ($\mu(x)\approx 1$) for strong gravitational fields ($x\gg1$) and the deep-MOND regime ($\mu(x)\approx x$) for weak gravitational fields ($x\ll1$). The transition between these regimes occurs at the critical acceleration of $a_0 \approx 1.2\times10^{-10}$\,m\,s$^{-2}$. The expected deviations from \equ{mond} depend on the particular MOND theory, mass distribution, and orbital shape  \citep{bm84, milg94b, ciotti06, brada95,  qumond}. 

For every galaxy, we can construct {an empirical radial acceleration relation (RAR) between the observed acceleration, $a$, and $a_\mathrm{N}$ \citep{mcgaugh16}}. Observations show that most late-type galaxies (LTGs) follow the RAR predicted by MOND in \equ{mond} (e.g., \citealp{begeman91, sanders96, deblok98, milg07, gentile11, angus12, mcgaugh16, lelli17, milg17, li18}). The other attempts to solve the missing mass problem have to explain this empirical fact as well. The attempts to do so in the \lcdm framework employ other empirical relations  whose validity is not an inevitable result of hydrodynamic simulations such as the SHMR (see, e.g., \citealp{cintio15,navarro17}). The form of RAR and the validity of MOND are much less explored in ETGs than in LTGs.

This is because investigating the gravitational fields of ETGs is more complicated than in LTGs. In the \lcdm context, a specific problem is the fact that most of the visible objects in galaxies extend toward galactocentric radii substantially smaller than the characteristic radii of their dark halos and therefore the parameters of the profiles are difficult to constrain, especially given that the mass of stars in a galaxy is also uncertain.  When testing MOND, we encounter a similar problem. We need tracers of gravitational potential at large radii where MOND predicts measurable deviations from the Newtonian gravity without dark matter, i.e., where the gravitational accelerations are much below $a_0$. Otherwise we can just test whether the data are consistent with no deviation. When investigating the gravitational fields of the ETGs, we usually rely on the Jeans analysis of the radial velocities of some kinematic tracers whose motion are fully governed by the gravitational field, such as stars, planetary nebulae, globular clusters (GCs), and satellites. However the shape of the trajectories is unknown and this introduces uncertainties in the measurement; some constraints can be inferred from the higher order moments of the tracer velocity distribution (see \sect{gcss}). Neighboring galaxies can disturb the tracers from virial equilibrium. It is also {possible} {to} use stellar shells to investigate the gravitational fields of ETGs up to large radii {\citep{ebrova12,  bilcjp, bil15b}} but faint shells are difficult to detect. Another method relies on the hydrostatic equilibrium between the pressure of the hot interstellar gas and the gravitational field. The obstacle in this approach is that the equilibrium might not be perfect, for example, because of a recent activity of the galactic nucleus. In addition, the gravitational field cannot usually be investigated up to the radii where the gravitational acceleration is expected to be substantially weaker than $a_0$. \citet{milg12} tested MOND successfully down to very low accelerations in two ETGs with unusually high amounts of hot gas. Some ETGs also contain rotating \ion{H}{I} disks. \citet{weijmans08} used these to verify MOND in one ETG down to the acceleration comparable to $a_0$ and \citet{lelli17} extended this results to a sample of 16 ETGs. Investigations of gravitational fields using gravitational lensing have the advantage of not relying on the uncertain assumptions such as the hydrostatic equilibrium or the shape of the orbits of kinematic tracers. The strong gravitational lensing allows the determination of the dynamical mass enclosed below a certain galactocentric radius. If a galaxy follows MOND, we can never probe the regions where the accelerations are substantially lower than $a_0$ using this method \citep{milg12}. \citet{tian17} used strong lensing to verify MOND down to the acceleration comparable with $a_0$ in a few tens of ETGs. Weak gravitational lensing can be used to probe the gravitational field up to large galactocentric distances but at the price of galaxy stacking. If some of the galaxies have an unusual gravitational field and for example they do not agree with MOND, then we will not learn about their existence. \citet{milg13} verified the consistency of MOND with weak galaxy-galaxy lensing up to very large radii in red galaxies (presumably ETGs) and blue galaxies (presumably LTGs).

In summary, the tests of MOND in ETGs were generally positive. Only the works based on the Jeans modeling of the radial velocities of various tracers in the ETGs gave controversial results \citep{tiret07b,angus08dwarf,richtler08,samur08,samur09,richtler11,tortora14,samurovicn821,samur14,dabringhausen16,samur16,samur16b,samur17,samur18}. Recent examples include the work by \citet{janz16}, who concluded that the spatially resolved kinematics of their 14 fast rotator ETGs is in tension with MOND. On the contrary, \citet{rong18} found that the spatially resolved kinematics of the fast rotator ETGs in the MaNGA survey are consistent with MOND but they identified an inconsistency with the slow rotators. Finally, \citet{lelli17} claimed that the RAR of the LTGs works in all galaxies universally, including the slow rotators.

While Jeans analysis has its down sides, it still enables, as {one of few methods}, investigating the gravitational fields of individual ETGs beyond a few effective radii of the starlight of the galaxy if the GCs are used as the kinematic tracers. Observing the radial velocities of GCs is a demanding task requiring a lot of observing time at large telescopes. In the present work, we collect a sample of 17 ETGs that includes nearly all galaxies for which over 100 GC radial velocities have been measured. The GC systems often extend over 10 effective radii of the galaxy. Together with the real galaxies, we study in the same way the GCs of a galaxy formed in a hydrodynamical \lcdm simulation. Our main motivation is to test the above predictions of \lcdm and MOND, i.e., whether the GC kinematics can be modeled well given the assumed gravitational law and observational constraints, and whether the SHMR, HMCR, and RAR hold true in our galaxies. If we find that these predictions are not met, we will attempt to find an explanation, which is usually an influence of the environment.

This paper is organized as follows. We present our sample of real and simulated galaxies  in \sect{sample} where we also describe the methods used to obtain the stellar mass-to-light ratios from galaxy colors. In \sect{gcss}, we derive the observational characteristics of the studied GC systems. In \sect{jeans} we describe how we computed the theoretical velocity-dispersion profile for a given gravitational field of the galaxy. Our maximum a posteriori method to model the kinematics of the GCs is presented in \sect{fits} and the results to which it led in \sect{res}. We deal with the predictions of the velocity dispersion profiles of the GC systems by the \lcdm and MOND paradigms based on independent estimates of the free parameters in \sect{pred}. We also investigated in \sect{rar} the RARs using the GC systems. All results are discussed in \sect{disc}; specifically, we discuss the possible reason of deviations from the \lcdm predictions in \sect{lcdm} and from the MOND predictions in \sect{mond}. We summarize our work and build the final picture following from it in \sect{sum}.

Throughout the paper, we denote the decadic logarithm by $\log$ and the natural logarithm by $\ln$. {For consistency with \citet{behroozi13} whose results we are building on,} we adopt a flat cosmology with the parameters $H_0=70$\,km\,s$^{-1}$, $\Omega_\mathrm{m} =  0.27$, $\Omega_\mathrm{b} =0.046$, $\sigma_8 =  0.8$ and $n_\mathrm{s} = 0.96$. This is consistent with the WMAP5 \citep{komatsu09} and the WMAP7+BAO+$H_0$ \citep{komatsu11}  cosmologies.

\begin{table*}
  \caption{Properties of globular cluster systems}
  \label{tab:tabgcss}
  \centering
  \begin{tabular}{lcccccccccrr}
 \hline\hline
 Name & $r_\mathrm{min}$  &  $r_\mathrm{max}$& $N$ & $N_\mathrm{bin}$ & $\rho_0$  & $x_\mathrm{br}$  & $a$ & $b$ & $v_\mathrm{rot}$  & $s_3$ & $s_4$ \\ 
 &  [arcmin] & [arcmin] & & & [arcmin$^{-2}$] & [arcmin] & & & [km\,s$^{-1}$] & & \\ \hline

N\,821 & 0.20 & 4.99 & 68 & 5 & 1.20 & 2.27 & -1.6 & -3.4 & 0$^{28}$ & $-0.3 \pm 0.3$  & $-0.5 \pm 0.6$ \\
N\,1023 & 0.35 & 7.86 & 113 & 6 & 1.97 & 2.50 & -1.9 & -3.8 & 119$^{28}$ & $-0.2 \pm 0.2$  & $-1.4 \pm 0.5$ \\
N\,1399 & 1.10 & 18.23 & 790 & 7 & 1.28 & 7.34 & -1.1 & -4.7 & 0$^{20}$ & $-0.12 \pm 0.09$  & $-0.6 \pm 0.2$ \\
N\,1400 & 0.34 & 9.20 & 68 & 8 & 1.48 & 3.09 & -2.4 & -4.5 & 10$^{19}$ & $0.1 \pm 0.3$  & $-1.3 \pm 0.6$ \\
N\,1407 & 0.26 & 14.51 & 374 & 7 & 1.57 & 5.43 & -1.4 & -4.0 & 39$^{19}$ & $-0.0 \pm 0.1$  & $-0.4 \pm 0.3$ \\
N\,2768 & 0.20 & 9.16 & 107 & 8 & 2.11 & 3.30 & -2.2 & -5.0 & 57$^{19}$ & $0.0 \pm 0.2$  & $-0.8 \pm 0.5$ \\
N\,3115 & 0.44 & 7.70 & 150 & 7 & 1.54 & 3.38 & -1.5 & -4.3 & 100$^{19}$ & $0.2 \pm 0.2$  & $-0.7 \pm 0.4$ \\
N\,3377 & 0.32 & 9.38 & 122 & 7 & 2.11 & 1.72 & -1.4 & -3.4 & 18$^{19}$ & $-0.1 \pm 0.2$  & $-0.8 \pm 0.4$ \\
N\,4278 & 0.24 & 8.28 & 269 & 9 & 2.88 & 3.28 & -1.6 & -4.1 & 28$^{19}$ & $-0.0 \pm 0.1$  & $-0.7 \pm 0.3$ \\
N\,4365 & 0.16 & 11.66 & 244 & 7 & 2.68 & 4.99 & -2.0 & -4.7 & 26$^{19}$ & $-0.1 \pm 0.2$  & $-0.5 \pm 0.3$ \\
N\,4472 & 0.43 & 9.48 & 263 & 7 & 0.454 & 5.14 & -0.50 & -5.4 & 53$^{21}$ & $0.2 \pm 0.2$  & $-0.4 \pm 0.3$ \\
N\,4486 & 0.62 & 29.99 & 634 & 15 & 0.927 & 3.13 & -0.14 & -3.4 & 25$^{19}$ & $0.0 \pm 0.1$  & $-0.1 \pm 0.2$ \\
N\,4494 & 0.30 & 7.73 & 105 & 7 & 1.91 & 2.05 & -1.4 & -4.1 & 62$^{19}$ & $0.0 \pm 0.2$  & $-0.9 \pm 0.5$ \\
N\,4526 & 0.34 & 6.74 & 107 & 7 & 1.84 & 2.46 & -2.0 & -3.4 & 142$^{28}$ & $-0.1 \pm 0.2$  & $-1.0 \pm 0.5$ \\
N\,4649 & 0.17 & 21.45 & 423 & 15 & 3.17 & 6.09 & -1.8 & -5.6 & 141$^{22}$ & $-0.0 \pm 0.1$  & $-0.6 \pm 0.2$ \\
N\,5128 & 1.11 & 42.48 & 530 & 20 & 0.111 & 9.43 & -0.40 & -4.3 & 33$^{23}$ & $0.0 \pm 0.1$  & $-0.5 \pm 0.2$ \\
N\,5846 & 0.26 & 8.83 & 205 & 10 & 1.06 & 5.14 & -1.4 & -4.2 & 5$^{19}$ & $-0.2 \pm 0.2$  & $-0.1 \pm 0.3$ \\\hline
R17x & 0.03 & 4.02 & 200 & 10 & 116 & 0.364 & -0.51 & -3.8 & 32 & $-0.0 \pm 0.2$  & $-0.8 \pm 0.3$ \\
R17y & 0.03 & 3.87 & 199 & 10 & 19.8 & 0.535 & -1.6 & -3.8 & 38 & $0.1 \pm 0.2$  & $-0.9 \pm 0.3$ \\
R17z & 0.03 & 5.27 & 200 & 10 & 17.3 & 0.620 & -1.6 & -3.9 & 41 & $-0.0 \pm 0.2$  & $-0.7 \pm 0.3$ \\

 \hline
  \end{tabular}
  \tablefoot{
 $\boldsymbol{r_\mathrm{min}}$ -- Galactocentric radius of the innermost analyzed GC.
 $\boldsymbol{r_\mathrm{max}}$ -- Galactocentric radius of the outermost analyzed GC. 
 $\boldsymbol{N}$ -- Number of analyzed GCs.
 $\boldsymbol{N_\mathrm{bin}}$ -- Number of bins used to fit the surface number density profile.
 $\boldsymbol{\rho_0}$, $\boldsymbol{x_\mathrm{br}}$, $\boldsymbol{a}$, $\boldsymbol{b}$ -- Parameters of the surface number density fit (\equ{sdprof}).
 $\boldsymbol{v_\mathrm{rot}}$ -- Projected systemic rotational velocity of the GC system.  
 $\boldsymbol{s_3}$, $\boldsymbol{s_4}$  -- Skewness and kurtosis, respectively, of the total GC velocity distribution. See \sect{gcss} for details.
 \textbf{References}:  
 (19)~\citet{pota13}; (20)~\citet{dirsch03}; (21)~\citet{cote03}; (22)~\citet{hwang08}; (23)~\citet{woodley10}; (28)~\citet{alabi17}.}
\end{table*}

\begin{table}
  \caption{$M/L$s from the stellar population synthesis models}
  \label{tab:tabmls}
  \centering
  \begin{tabular}{lccccc}
 \hline\hline
 Name &  $M/L_\mathrm{BC}$ & $M/L_\mathrm{P}$ & $M/L_\mathrm{IP1}$ & $M/L_\mathrm{IP2}$ & $M/L_\mathrm{IP3}$  \\\hline  
N\,821 & 5.6 & 6.2 & 3.5 & 3.9 & 6.8\\
N\,1023 & 6.5 & 7.1 & 4.2 & 4.8 & 8.2\\
N\,1399 & 6.9 & 7.6 & 4.6 & 5.2 & 9.0\\
N\,1400 & 6.0 & 6.6 & 3.9 & 4.3 & 7.4\\
N\,1407 & 7.4 & 8.1 & 5.0 & 5.7 & 9.8\\
N\,2768 & 6.4 & 7.0 & 4.2 & 4.7 & 8.2\\
N\,3115 & 6.2 & 6.8 & 4.0 & 4.5 & 7.8\\
N\,3377 & 4.8 & 5.2 & 2.9 & 3.1 & 5.4\\
N\,4278 & 6.2 & 6.8 & 4.0 & 4.5 & 7.8\\
N\,4365 & 6.9 & 7.6 & 4.6 & 5.2 & 9.0\\
N\,4472 & 6.9 & 7.6 & 4.6 & 5.2 & 9.0\\
N\,4486 & 6.7 & 7.3 & 4.4 & 5.0 & 8.5\\
N\,4494 & 5.3 & 5.8 & 3.2 & 3.6 & 6.2\\
N\,4526 & 6.0 & 6.6 & 3.9 & 4.3 & 7.4\\
N\,4649 & 6.9 & 7.6 & 4.6 & 5.2 & 9.0\\
N\,5128 & 5.6 & 6.2 & 3.5 & 3.9 & 6.8\\
N\,5846 & 7.1 & 7.8 & 4.8 & 5.5 & 9.4\\\hline
R17x,y,z & 6.2 & 6.8 & 4.0 & 4.5 & 7.8\\

 \hline
  \end{tabular}
  \tablefoot{ 
 $\boldsymbol{M/L_\mathrm{BC}}$ -- SPS \citet{bruzual03} model with  \citet{salpeter55}  IMF.
 $\boldsymbol{M/L_\mathrm{P}}$ -- SPS PEGASE model with Salpeter IMF  for $Z=0.02$.
 $\boldsymbol{M/L_\mathrm{IP1}}$ -- The exponential star formation history  model with the \citet{kroupa01} IMF \citep{into13}.
 $\boldsymbol{M/L_\mathrm{IP2}}$ -- The disk galaxy model based on the \citet{kroupa01} IMF \citep{into13}.
 $\boldsymbol{M/L_\mathrm{IP3}}$ -- The disk galaxy model based on the Salpeter IMF \citep{into13}. All values are given in solar units. See \sect{mls} for details.}
\end{table}

\section{Galaxy sample and used data}\label{sec:sample}
\subsection{Real galaxies}\label{sec:sampreal}
In this paper we analyze the following 17 nearby ETGs: NGC\,821, NGC\,1023, NGC\,1399, NGC\,1400, NGC\,1407, NGC\,2768, NGC\,3115, NGC\,3377, NGC\,4278, NGC\,4365, NGC\,4472, NGC\,4486, NGC\,4494, NGC\,4526, NGC\,4649, NGC\,5128, and NGC\,5846. We study their dynamics based on their observed GCs, which we do not split into red and blue subpopulations; rather we work with the whole sample of objects. The velocity errors for the objects are typically below 20\,km\,s$^{-1}$ and they rarely exceed 30\,km\,s$^{-1}$.  Since one of the main tasks of this study is to analyze the problem of the missing mass it was important that the available data extend out to several effective radii ($R_\mathrm{e}$) for each ETG for which large deviations from Newtonian dynamics without dark matter are expected. Thus, the galactocentric distance of the outermost point expressed in the units of $R_\mathrm{e}$ varies from 4.5 (for NGC\,3115) to 30.2 (for NGC\,1399).  The galaxies were chosen so as to have at least $100$ confirmed GCs, where NGC\,821 and NGC\,1400 are the exception, with 68 objects in each galaxy and the largest number of GCs are available in the case of NGC\,1399 for which 790 objects were used. Such numbers of tracers are not sufficient to establish the exact anisotropy in a given bin accurately because as shown in \citet{merritt97} at least several hundred tracers per bin are necessary for this purpose. Therefore, our estimates of the $s_3$ and $s_4$ parameters below are to be understood merely as an indicator of anisotropies.  All type of ETGs are included in the sample, from round (for example, NGC\,5846) to flattened (for example, NGC\,3377) ellipticals to lenticulars (for example, NGC\,3115). The selected galaxies belong to various environments: there are 2 field galaxies (NGC\,821 and NGC\,3115), 6 cluster galaxies, and 9 galaxies that belong to groups. We also note that our sample features both fast and slow rotators, and the fast rotators are more numerous (12 fast versus 5 slow rotators). In all our calculations we took into account the systemic rotation, which varies from approximately zero (such as NGC\,821, NGC\,1399, and NGC\,5846) to 140 km s$^{-1}$ (NGC\,4526 and NGC\,4649). Their colors vary from those that resemble their spiral counterparts (NGC\,3377) to red (NGC\,1407).  The observational data for our galaxies are taken from several publicly available databases and the details are as follows. The main part of our sample, namely the galaxies NGC\,821, NGC\,1023, NGC\,1400, NGC\,1407, NGC\,2768, NGC\,3115, NGC\,3377, NGC\,4278, NGC\,4365, NGC\,4486 (=M\,87), NGC\,4494 NGC\,4526, and NGC\,5846 come from the SLUGGS (SAGES Legacy Unifying Globulars and Galaxies  Survey, where SAGES is the Study of the Astrophysics of Globular Clusters in Extragalactic Systems) sample\footnote{\tt http://sluggs.swin.edu.au} \citep{sluggs}. The data for the remaining four galaxies are available in the following papers:  \citet{schuberth10} for NGC\,1399, \citet{cote03} for NGC\,4472 ($=$M49),  \citet{lee08} for NGC\,4649 ($=$M60), and  \cite{woodley10} for NGC\,5128 ($=$Centaurus A). The basic observational data for the galaxies in the sample is available in \tab{tabgals} and the information related to GCs is available in \tab{tabgcss}.

\subsection{Artificial galaxies}\label{sec:sampart}
Together with the real galaxies, we analyzed the GC system of the  galaxy from the simulation by \citet{renaud17}.We used these data to generate artificial observations, which helped us to check the correctness of our methods and interpret the results. This simulation is a zoom-in on a Milky Way-like galaxy in the \lcdm framework with Planck cosmology (see also \citealp{li17, kim18}). The spatial resolution of this simulation was $\sim200$\,pc and the dark matter particles had a mass of $2.1\times10^6$\,M$_\sun$. The simulation includes injection of energy and momentum, and chemical enrichment (in Fe and O) from SN type II and Ia, stellar winds, and radiative pressure. It also implements a metallicity-dependent cooling and heating from UV background.

The radial profile of the cumulated baryonic mass of the simulated galaxy has a distinct plateau between around 10-100\,kpc where it maintains the value of  $4.8\times10^{10}$\,M$_\sun$, which we identify as the baryonic mass of the galaxy. The galaxy has the $r_{200}$ radius below which the average dark matter density is 200 times the Universe critical density of 184\,kpc. We fitted the cumulated-mass profile of the dark matter particles with a NFW profile (\equ{nfw}) up to the radius of $r_{200}$. We obtained 
$\log(\rho_\mathrm{s}\,[ M_\sun\,\mathrm{kpc}^{-3}]  ) = 6.72$ and $\log(r_\mathrm{s}\,[\mathrm{kpc}] ) = 1.29$. Such a NFW halo has the $M_{200}$ mass of $7\times 10^{11}$\,M$_\sun$ and a concentration of $c=9.4$.

We created three artificial galaxies from this single simulated galaxy by making its projection along the Cartesian axes of the simulation. We then treated the projections as separate galaxies, which we labeled as R17x, R17y, and R17z (or collectively as R17x,y,z). The actual distance of these three galaxies was assumed to be 21\,kpc, which is a typical distance to our real galaxies. We however pretended in the following that the observed distance to these galaxies is 19\,kpc in order to simulate the observational errors. The difference is the typical 1$\sigma$ error given by \citet{tonry01} for their measurements at this distance. The artificial galaxies were all assumed to have an actual $M/L = 4.7$, a typical $M/L$ predicted by the SPS models for our real galaxies. We generated images of the artificial galaxies with the resolution of 0.25\arcsec/pixel and the field of view (FOV) of $12\arcmin\times12\arcmin$. These images were then fitted by a S\'ersic profile using Galfit \citep{Penggalfit}. The resulting structural parameters are presented in \tab{tabgals}. The SPS $M/L$s for the R17x,y,z galaxies  were copied from NGC\,3115 because it is a fast rotator with a similar stellar mass. We identified the GC particles in the same way as \citet{renaud17}, i.e., as the stellar particles formed before 10\,Gyr. {We randomly {chose} 200 GCs} so that their maximum galactocentric radius was $15\,R_\mathrm{e}$ to mimic the data for the real galaxies. We then created for the RG17x,y,z, galaxies artificial catalogs of the apparent GC positions and radial velocities. These data were subsequently treated in the same way as the data for the real galaxies. Finally, we obtained the systemic rotation velocity of the GC systems using the standard method described, for example, in \citet{richtler04}.  

The RARs of the R17x,y,z galaxies are plotted  in the panels (f) of Figures~\ref{fig:g17}--~\ref{fig:g19} by the dotted yellow lines. These were obtained in the following way. We calculated a nearly true Newtonian gravitational acceleration in each of these galaxies from the stellar distribution obtained by fitting its artificial image and from the correct values (without the intentionally added error) of the galaxy distance and $M/L$. In order to obtain the dynamical acceleration of the galaxy, we added to the baryonic acceleration the force from the above-described {NFW halo}. We can see that the galaxies follow well the RAR of the LTGs depicted by the gray regions in the figures.

\subsection{Stellar population synthesis models}
\label{sec:mls}
In our paper we rely on the study of \citet{bell01} to calculate the 
theoretical $M/L$s based only on the visible matter using 
the colors corrected for Galactic extinction from the HyperLEDA 
database {\citep{hyperleda}}. Our  estimates of theoretical values for the galaxies in the 
sample are presented in \tab{tabmls} and they 
represent stellar $M/L$s of the galaxies in our 
sample. The $M/L$s of the galaxies in our sample in the 
$B$ band for a given metallicity ($Z=0.02$) were calculated by applying 
the fitting formulas from \citet{bell01}. This value of the metallicity 
was based on the work of \cite{casuso96}, who used the Lick index Mg$_2$, 
as an indicator of metallicity. The  Mg$_2$ index for all our galaxies 
is between $\sim 0.25$ and $\sim 0.30$ (from HyperLEDA) and from figure 
3 of Casuso et al., we can conclude that $Z=0.02$ is an accurate 
approximation\footnote{The only exception is the case of NGC~5128 for 
  which the measurement of this Lick index is not available and thus the 
  value of $Z=0.02$ was assumed.} .

In this paper we use several stellar population synthesis (SPS)  models 
to determine the {mass contribution of the stellar} component. More 
precisely, we use five different models with several initial mass 
functions (IMFs) and our results are given in 
\tab{tabmls}. We used the \citet{bruzual03} model with Salpeter IMF and the PEGASE model 
\citep{fioc97} again with Salpeter IMF.  We also relied on the models of 
\cite{into13}  and our estimates of the mass-to-light ratio based on 
their paper were derived for composite stellar populations (convolutions of simple stellar populations (SSPs), and  SSPs of 
different age and metallicity, according to a given star formation and 
chemical evolution). The estimates in columns 4 and 5 are based on the 
models with the \cite{kroupa01} IMF, and the values in the sixth column 
come from disk galaxy models based on the Salpeter IMF.

{\begin{figure*}
 \centering
 \includegraphics[width=17cm]{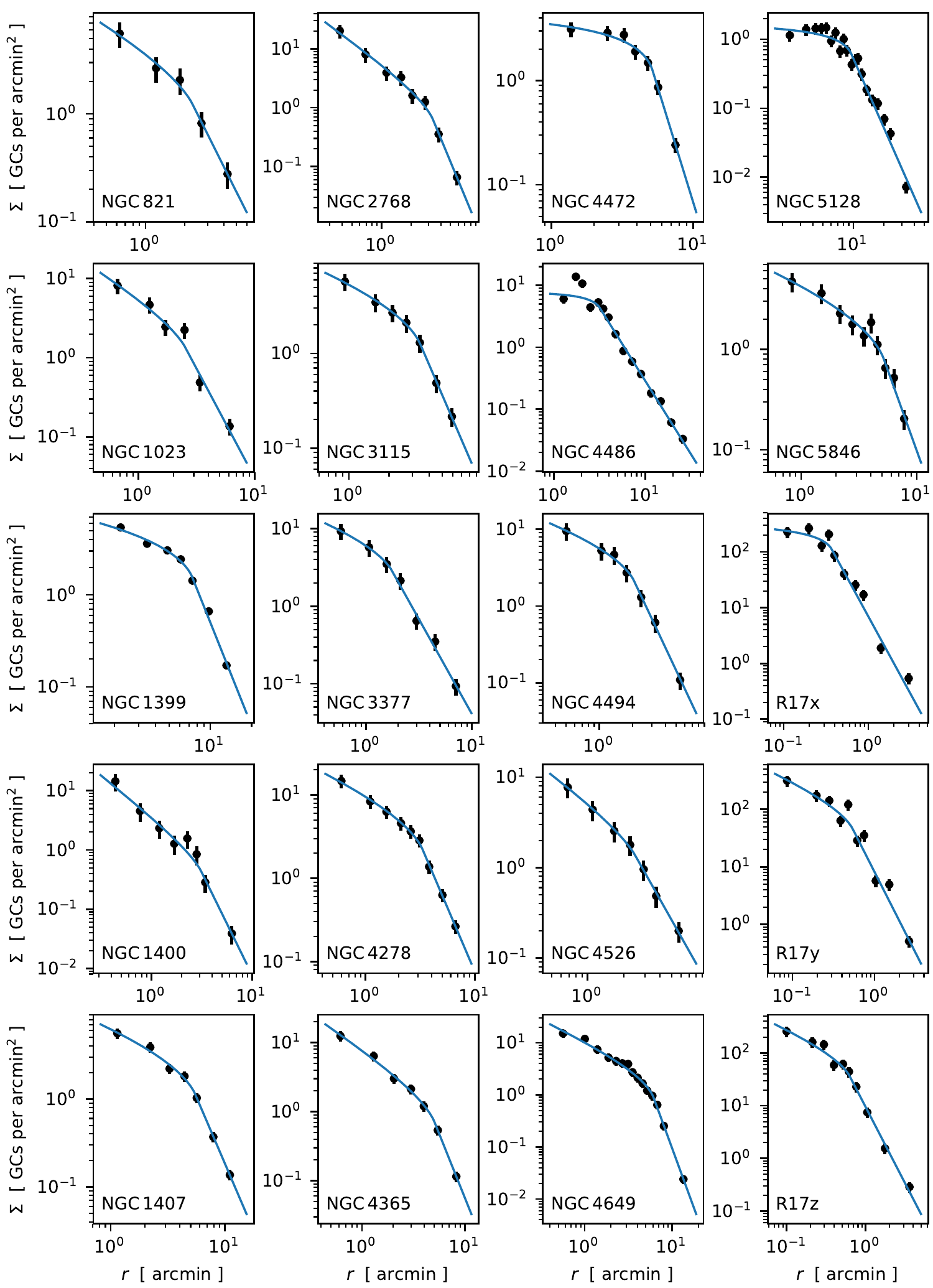}
 \caption{{Surface-density profiles of the investigated GC systems. Points with error bars are measured data. Solid lines are fits by a broken power-law volume density profile (\equ{sdprof}).} }
 \label{fig:sdprof}
  \end{figure*}}

\section{Properties of the globular cluster systems}
\label{sec:gcss}
When solving the Jeans equation (\equ{jeans}), the radial profile of the tracer number density is needed. We noted that for our galaxies the radial profiles of their GC surface number densities follow approximately a broken power law (see \fig{sdprof}). We thus assumed that the GC volume number density has a form of
\begin{equation}
\begin{aligned}
\rho(r) & = \rho_0\ r^a &\quad\textrm{ for }\quad r<r_\mathrm{br},\\
\rho(r) & = \rho_0\ r_\mathrm{br}^{a-b}\ r^b &\quad\textrm{ for }\quad r\geq r_\mathrm{br}.
\end{aligned}
\label{eq:sdprof}
\end{equation} 
The free parameters of this model, the power-law exponents $a$ and $b$, the normalization $\rho_0$, and the break radius $r_\mathrm{br}$, were obtained by fitting the data. For every galaxy we divided its GCs into several radial bins and calculated the surface density of GCs in each of {the bins}. We placed the bins so that each of them contained approximately the same number of GCs. The total number of bins was chosen individually for every galaxy by visual inspection so that both linear parts of the surface density profile were resolved by at least three points and, at the same time, the surface number density profile did not appear too noisy. We assumed that the survey that detected the GCs did so in a homogeneous way on our selected radial range.  It is difficult to detect the GCs in the galaxy cores because of the galaxy glare. The innermost GCs were thus excluded if the surface-density profile appeared to deviate from the double power law profile in the galaxy center. In a few cases, we also noted a decline from the linear trend of the GC surface density profile at large radii, which we interpreted as a consequence of an incomplete GC survey in these distant areas. In these cases we excluded the most distant GCs until the surface density profile formed a single broken line. The volume number density given by \equ{sdprof} was converted to the surface number density numerically using the Abel transform. The parameter $b$ was restricted to be $b<-3$ to ensure a finite number of GCs in the galaxies. In \tab{tabgcss} we list for every galaxy the minimum and maximum radius of the GCs used (for all purposes in our paper), the number of such GCs, the number of bins used for obtaining the surface-density profiles, and the  parameters of the best fits of the density profile. We note that the break radii expressed in kiloparsecs correlate with the stellar mass of the galaxy. This is similar to the behavior of effective radii of GC systems reported by \citet{forbes17}. This suggests that the breaks in the surface density profiles of the GC systems are physically real and do not result from some observational systematics. The systemic rotation velocities of the GC systems compiled from literature are included in \tab{tabgcss}.

\begin{figure}
  \resizebox{\hsize}{!}{\includegraphics{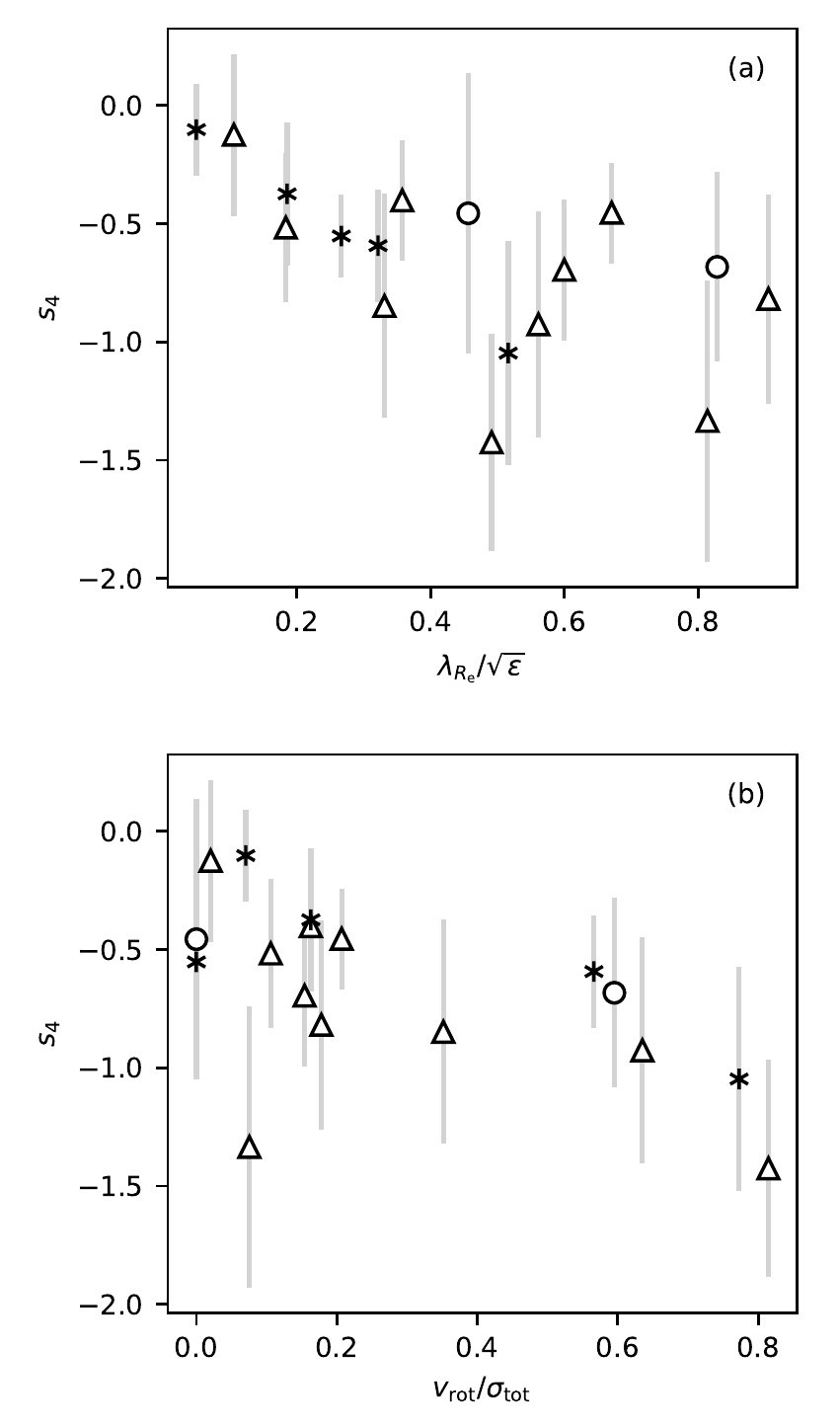}}
  \caption{Panel \textbf{(a):} Kurtosis parameter $s_4$ anticorrelates with the rotator criterion number of the host galaxy's stars. Panel \textbf{(b)}:\ This parameter also anticorrelates with the average rotation velocity of the GC system normalized by the total velocity dispersion of the GC system galaxies. This suggests that the $s_4$ parameter is driven by systemic rotation of the GC system. The host galaxies are distinguished by their environment (circles: field, triangles: group, asterisk: cluster).} 
  \label{fig:s4comb}
\end{figure}

The black points in panels (a) of \allgals show the projected radii and line-of-sight velocities of the GCs with the subtracted average radial velocities of the whole GC systems. We also plotted in those figures the corresponding velocity dispersion profiles (the green lines) and their uncertainty limits (the light green bands).
The velocity dispersion and its uncertainty was calculated as
\begin{equation}
\sigma_\mathrm{los, obs} = \sqrt{\frac{1}{N-1}\sum_{i=1}^N(v_i-v_\mathrm{sys})^2}; \quad \Delta\sigma_\mathrm{obs} = \frac{\sigma_\mathrm{los, obs}}{\sqrt{2(N-1)}},
\label{eq:s2}
\end{equation}
where $N$ denotes the number of GCs in the bin, $v_i$ the radial velocity of the $i$-th GC, and $v_\mathrm{sys}$ the average line-of-sight velocity of all GCs in the galaxy. {The bins had a variable width. Around each of the linearly spaced points where we wanted to calculate the velocity dispersion, we chose the bin edges such that they were symmetric around the point, there were at least 20 GCs between them, and the bin width that they defined was minimal but within the range of 0.5-4\,$R_\mathrm{e}$.}

The panels (h) and (i) of \allgals show the profiles of the $s_3$ and $s_4$ parameters (skewness and kurtosis, respectively) calculated in the same bins. These parameters are useful for constraining the anisotropy parameter $\beta$ in the Jeans equation; see \sect{jeans}. The parameters quantify the deviation of the distribution from a Gaussian for which they are zero. A negative, zero, or positive kurtosis indicates tangential, isotropic, or radial orbits, respectively \citep{gerhard93}. The kurtosis is however influenced by a systemic rotation of the GC system \citep{dekel05}. {These parameters} can be calculated as \citep{joanes98}
\begin{equation}
s_3 = \frac{N}{(N-1)(N-2)}\sum_{i=1}^N\left(\frac{v_i-v_\mathrm{sys}}{\sigma_\mathrm{los, obs}}\right)^3 \pm \sqrt{\frac{6}{N}}
\label{eq:s3}
\end{equation}
and
\begin{equation}
\begin{split}
s_4 = & \frac{N(N+1)}{(N-1)(N-2)(N-3)}\sum_{i=1}^N\left(\frac{v_i-v_\mathrm{sys}}{\sigma_\mathrm{los, obs}}\right)^4  \\
& -\frac{3(N-1)^2}{(N-2)(N-3)} \pm \sqrt{\frac{24}{N}}.
\end{split}
\label{eq:s4}
\end{equation}

We also calculated these parameters for the whole GC systems assuming the velocity dispersion profile described above. The results are presented in \tab{tabgcss}. For all galaxies, the $s_4$ parameter came out negative. This hints that the GCs are on tangential orbits. There are two pieces of evidence that this is a consequence of ordered rotation of the GC systems: First, the values of the $s_4$ parameters anticorrelate with the average rotation velocity of the GC system, $v_\mathrm{rot}$, normalized by the total velocity dispersion of the GC system of the galaxy, $\sigma_\mathrm{tot}$, as we can see in \fig{s4comb}b. Second, the value of the $s_4$ parameters anticorrelates with the rotator criterion number of the stars of the host galaxy  (\sect{lcdm}) so that the GC systems of fast-rotator galaxies have more negative values of the $s_4$ parameter than the slow-rotator galaxies (\fig{s4comb}a) and it was shown in the literature that the rotation of GCs, especially the red, correlates with the rotation of the stars of the host galaxy \citep{pota13}.

\section{Solving the Jeans equation}
\label{sec:jeans}
Our study is based on comparing the observed radial velocities of the GCs to the theoretical models of line-of-sight velocity dispersion. The models of the latter were based on the spherically symmetric Jeans equation corrected approximately for systemic rotation \citep{binney-tremaine08,hui95,peng04}, i.e.,\begin{equation}
\frac{1}{\rho}\frac{\mathrm{d}(\rho\sigma_r^2)}{\mathrm{d}r} + 2\,\frac{\beta(r)}{r}\sigma_r^2 ~=~ a(r)+\frac{v_\mathrm{rot}^2}{r}.
\label{eq:jeans}
\end{equation}
In this equation, $r$ stands for the galactocentric distance and $\rho(r)$ the number density of the GCs at $r$. The quantity $\sigma_r(r)$ expresses the velocity dispersion in the direction toward the galaxy center. The value of velocity dispersion in the tangential direction, $\sigma_\mathrm{t}(r)$, enters the equation through the anisotropy parameter defined as $\beta(r) = 1-\sigma_\mathrm{t}^2/\sigma_r^2$. This means that the GC orbits are predominantly radial if $0<\beta\leq1$, whereas for $-\infty\leq\beta<0$ the GC orbits are predominantly tangential. The $\sigma_r$ profile depends on the gravitational acceleration $a(r)$ (having a negative sign). The term with the systemic rotation velocity $v_\mathrm{rot}$ takes into account that the centrifugal force acts against the direction of gravity.

The radial velocity dispersion $\sigma_r$ is not directly available from observations. Instead, we observe the line-of-sight velocities. The line-of-sight velocitity dispersion $\sigma_\mathrm{los}$ at some projected galactocentric radius $R$ is connected with $\sigma_r$ through the equation (e.g., \citealp{binney82})
\begin{equation}
 \sigma_\mathrm{los}^2(R) = \frac{\int_R^\infty \sigma_r^2(r)\, \left[ 1-(R/r)^2\beta\right]\, \rho(r)\, \left( r^2-R^2\right)^{-1/2}\,r\,\mathrm{d}r  }{\int_R^\infty \rho(r)\, \left(r^2-R^2 \right)^{-1/2}\, r\, \mathrm{d}r }.
\label{eq:siglos}
\end{equation}

Having a model of the gravitational potential, the corresponding $\sigma_\mathrm{los}$ profile can be determined using Eqs.~\ref{eq:jeans} and \ref{eq:siglos}. This requires solving a differential equation and calculating an integral at every $R$ of interest, which is computationally demanding. Therefore we used the single-integral expression {that was published by \citet{mamon05} for a few important profiles of the $\beta$ parameter (their Eqs. A15 and A16).}

In order to test that our code calculates the profiles of $\sigma_\mathrm{los}(R)$ correctly, we compared the{ outputs of our code} to the known analytical solutions to Eqs.~\ref{eq:jeans} and~\ref{eq:siglos}: the isotropic Hernquist sphere \citep{hernquist90}; many forms of a power-law gravitational potential with a power-law tracer-density profile and a constant $\beta$; and the isotropic isothermal sphere \citep{binney-tremaine08}.

The velocity dispersion calculated using Eqs.~\ref{eq:jeans} and~\ref{eq:siglos} does not account for the fact that the observed velocity dispersion is increased by the systemic rotation of the GC system. We thus included the systemic rotation into the modeled line-of-sight velocity dispersion applying the usual approximation

\begin{equation}
\sigma_\mathrm{los, mod}(R) = \sqrt{\sigma_\mathrm{los}^2+v_\mathrm{rot}^2},
\label{eq:mod}
\end{equation}
which is the final expression we were using. We adopted the systemic rotation velocities of the GC systems for every galaxy from literature, see  \tab{tabgcss}. We assumed that the GC systems rotate around an axis perpendicular to the line of sight and that the rotation velocities of all GCs around that axis are equal (see \citealp{cote01} for a discussion of this common assumption).

The anisotropy parameter $\beta$ remains unknown but certain restrictions follow from the  values of the kurtosis parameter $s_4$  discussed in \sect{gcss}. The proximity of the kurtrosis parameter for all our galaxies to zero (\sect{gcss}) led us to consider near-zero anisotropies. For every galaxy, we constructed models with the following types of anisotropy profile:
\begin{itemize}
\item The isotropic profile following $\biso = 0$ at all radii
\item The mildly tangential profile with  $\bneg = -0.5$ at all radii
\item The ``literature'', mildly radial profile $\blit$, given by the prescription,
\begin{equation}
\blit(r) = \frac{0.5\,r}{r+1.4\,R_\mathrm{e}},
\end{equation}
where $R_\mathrm{e}$ is the effective radius of the starlight of the galaxy (see below). With this profile, orbits are isotropic at the galaxy center and become tangential  with $\beta = 0.5$ for radii large compared to $R_\mathrm{e}$. Such a profile is motivated by simulations of galactic mergers in \lcdm (see \citealp{mamon05} for details).
\end{itemize}

We considered the MOND and \lcdm models. The choice of the gravity law then determines  the acceleration field $a$ in \equ{jeans}. The MOND acceleration $a$ was determined from \equ{mond}, in which we employed the interpolating function
\begin{equation}
\mu(x) = x/(1+x).
\end{equation}
The acceleration $a_\mathrm{N}$ in \equ{mond} was calculated in the Newtonian way from the distribution of the visible matter as
\begin{equation}
a_\mathrm{N} = -GM_*(<r)/r^2,
\end{equation}
where $M_*(<r)$ is the stellar mass cumulated below the radius $r$ ; i.e., we neglected, for example, the possible presence of interstellar gas. We assumed that the surface brightness of a galaxy follows the S\'ersic profile \citep{sersic}. This profile is characterized by the total luminosity $L$, the characteristic radius $R_\mathrm{e}$, and the S\'ersic parameter $n$. We gathered these numbers for the photometric $B$ band from literature; see \tab{tabgals}. The mass-to-light ratio, $M/L$, is necessary to convert the luminosity surface density to the matter surface density. This ratio was one of the free parameters of our models. The mass cumulated interior to a given radius was obtained numerically by integrating the formula approximating the density profile of a S\'ersic sphere by \citet{sersdeproj} with the update by \citet{sersdeprojupdate}. 

Our \lcdm models contained, apart from the stellar component, a  NFW dark matter halo \citep{nfw}. The mass of such a halo enclosed interior the radius $r$ is
\begin{equation}
M_\mathrm{DM}(<r) = 4\pi\rho_s r_s^3\left\lbrace \ln\left[ \left(r_s+r\right) /r_s\right]  - r/\left( r+r_s\right)  \right\rbrace , 
\label{eq:nfw}
\end{equation}
where $r_s$ is the scale radius of the halo and $\rho_s$ its scale density. Those were again treated as free parameters of the models. The gravitational field in the \lcdm models  is written as
\begin{equation}
a = -G\left[ M_*(<r)+M_\mathrm{DM}(<r)\right] /r^2.
\end{equation}
We identify in this paper the halo mass with the $M_\mathrm{vir}$ mass cumulated below the $r_\mathrm{vir}$ radius. {These quantities are defined according to \citet{bryan98} so that we are consistent with the HMCR and SHMR definitions used by \citet{diemer15} and \citet{behroozi13}, respectively.}

\section{Fitting the galaxy models}
\label{sec:fits}
We compared the consistency of the observed line-of-sight velocities of the observed and simulated GCs with the expectations of \lcdm and MOND in several ways. The approach described in this section consists in finding the best-fit parameters of the MOND and \lcdm models of the galaxies, the results of which are then interpreted in \sect{disc}. We determined the best values for the free parameters of the models using {the maximum a posteriori} approach{, in which we maximize the product of the likelihood of obtaining the data if the parameters are given and the probability density of the parameters.} Hereafter, when referring to GC radial velocities, we mean their measured radial velocities for which the average velocity of all the GCs was subtracted. We assumed that at a given galactocentric projected radius the distribution of GC radial velocities  is Gaussian, has a zero mean and the velocity dispersion is determined by \equ{mod}. Our priors on the free parameters were based on the independent estimates of the galaxy distances, stellar mass-to-light ratios, and the dark matter scaling relations, as we explain below. Let us denote the vector of the free parameters whose values are to be determined by the fit by  $\vec{p} = (p_1, p_2, p_3,\ldots)$. Our next assumption was that the prior distributions of all the free parameters are Gaussian so that the distribution for the parameter $p_j$ has a mean of $p_{j, \mathrm{exp}}$ and a standard deviation of $\Delta p_j$. {Let us further denote by $\sigma_{\mathrm{mod},i}(\vec{p})$  the velocity dispersion implied by the parameters $\vec{p}$ at the radius of the $i$-th GC and by $v_i$ the measured radial velocity of this GC. According to the maximum a posteriori method, we then have to find the parameters maximizing the function} 
\begin{equation}
\mathcal{F}(\vec{p}) = \prod_{i= 1}^{N_\mathrm{data}}\frac{1}{\sqrt{2\pi\sigma_{\mathrm{mod},i}^2}}e^{\frac{-v_i^2}{2\sigma_{\mathrm{mod},i}^2}}\prod_{j=1}^{N_\mathrm{par}}\frac{1}{\sqrt{2\pi\Delta p_j^2}}e^{\frac{-(p_{j, \mathrm{exp}}-p_j)^2}{2\Delta p_j^2}}.
\label{eq:lik}
\end{equation}
{We denoted the resulting parameters as {$\hat{p_j}$}.} 

The free parameters are described below. All four of these parameters were used in the \lcdm models while only the first two in the MOND models. Specifying these parameters is equivalent to specifying the galaxy distance, $M/L$, halo mass, and concentration, but the parameters we used are more advantageous for incorporating the prior knowledge and subsequently for interpreting the results. The free parameters were chosen so that the mean of the {prior distribution of each $p_{j, \mathrm{exp}}$ is zero}.
\begin{enumerate}
  \item The parameter $p_d$ expresses the deviation of the fitted distance $d$ from the distance $d_0$ measured by \citet{tonry01} on the basis of the surface-brightness fluctuations (see \sect{sample}) as
  \begin{equation}
  p_d = \log\frac{d}{d_0}.
  \end{equation}
  The standard deviation of the prior distribution of this parameter, $\Delta p_d$, follows straightforwardly from the uncertainty of the distance modulus given by \citet{tonry01}. 
  
  \item We similarly introduced the parameter 
  \begin{equation}
  p_{M/L} = \log \frac{M/L}{M/L_0}
  \end{equation}
to account for the uncertainty in the $B$-band $M/L$. Our prior knowledge of $M/L$ follows from the star population synthesis models in \tab{tabmls}. The expected value of $M/L$, $M/L_0$, was chosen as $\log M/L_0 = (\log M/L_\mathrm{min} +\log M/L_\mathrm{max})/2$, where $M/L_\mathrm{min}$ and $M/L_\mathrm{max}$ are, respectively, the minimum and maximum $M/L$s listed for the particular galaxy in \tab{tabmls}. We considered the standard deviation of the prior distribution for $p_{M/L}$ to be  $\Delta p_{M/L} = \log M/L_\mathrm{max}-\log M/L_0$. 
  
  \item The next parameter quantifies the deviation of the galaxy from the mean SHMR. We employed the formula for the mean SHMR by \citet{behroozi13}. We adopted the same cosmological parameters as they did. \citet{behroozi13} concluded that the scatter of the stellar mass $M_*$ for a halo of a given mass $M_h$ is $\Delta \log M_*(M_h) = 0.218$ independently of $M_h$.  Since our parameters $p_{M/L}$ and $p_d$ imply the value of $M_*$, we had to invert this relation numerically and recover the scatter of $M_h$ at a given $M_*$, $\Delta \log M_h(M_*)$. It is then advantageous to quantify the deviation of the galaxy from the SMHR by the parameter
 \begin{equation}
 p_{SH} = \frac{1}{\Delta \log M_h(M_*)}\log\left[ \frac{M_h}{M_h(M_*)}\right].
 \end{equation}
The expression $M_h(M_*)$ means the halo mass given by the \citet{behroozi13} SHMR formula and the stellar mass.
The dispersion of the prior distribution of the parameter $p_\mathrm{SH}$ is $\Delta p_{SH} = 1$.

  \item The halo mass computed from the previous parameters implies a preferred halo concentration via the HMCR. For a given halo mass, we determined the concentration predicted by the HMCR using the COLOSSUS package for Python \citep{colossus} assuming the HMCR function derived by \citet{diemer15}. We benefited from the ability of COLOSSUS to calculate the HMCR function for any set of cosmological parameters. For a given halo mass $M_h$ we defined 
  \begin{equation}
  p_c = \log\left(\frac{c}{c(M_h)} \right).
  \end{equation} 
  \citet{diemer15} found for isolated halos the scatter in the HMCR relation of 0.16\,dex independently of $M_h$ or redshift. An isolated halo is such that its center does not lie in the virial radius of another, larger halo.The dispersion of the prior distribution of the parameter $p_c$ is then $\Delta p_c = 0.16$. 
\end{enumerate}

The 1$\sigma$ uncertainty limits of the fitted value of the parameter $p_i$ can be determined by minimizing and maximizing $p_i$ over the region of the parameter space where 
\begin{equation}
\ln\mathcal{F}(\vec{\hat{p}}) -\ln\mathcal{F}(\vec p)\leq 1/2.
\label{eq:region}
\end{equation}

In order to calculate the 1$\sigma$ uncertainty limits on some quantity $A$ which is a function of $\vec p$, we minimized or maximized $A(\vec p)$ over the same region.

In Appendix~\ref{app:test}, we present a test of a correct recovering the free parameters from artificial data using these methods.

We rated the goodness of the fits {in} several ways: using the $\chi^2$ confidence, the corrected Akaike's information criterion (AICc; \citealp{AIC, AICc}) and the logarithm of Bayes's factor (LBF; \citealp{bayfact}). 

The $\chi^2$ statistics can be used to exclude a model as the only criterion out of the three. It is based on the fact that, for the correct model, the quantity 
\begin{equation}
X^2 = \sum_{i= 1}^{N_\mathrm{data}}\frac{v_i^2}{\sigma_{\mathrm{mod},i}^2} + \sum_{j=1}^{N_\mathrm{par}}\frac{(p_{j, \mathrm{exp}}-p_j)^2}{\Delta p_j^2}
\label{eq:chisq}
\end{equation}
follows the $\chi^2$ statistical distribution with $k$ degrees of freedom. The number of the degrees of freedom is the number of the fitted data points minus the number of the fitted parameters. In our case, $k$ is therefore the number of the GCs. The models for which $X^2$ lies in the tails of the corresponding $\chi^2$ distributions are considered excluded.  We provide in \tabsfit the value of the confidence (conf.) of each model. It is the probability that a random number from the $\chi^2$ distribution has a larger deviation from the median of the distribution than $X^2$. We considered the models with a confidence below 5\% excluded; this is equivalent to a chi-square test with a 5\% significance. This $\chi^2$ confidence allows an easier interpretation of the agreement of the model with the data than the traditional reduced $\chi^2$. If we had correct models for all our galaxies, then the fraction of galaxies with a confidence below $x$ would be $x$.

The AICc and LBF criteria serve for determining which of several competing models matches the given observing data best. These criteria take into account the fact that the model with more free parameters is more likely to provide a better fit to the data. The AICc can be calculated as
\begin{equation}
\mathrm{AICc} = 2X^2+2k+\frac{2k^{2}+2k}{n-k-1}, 
\label{eq:aicc}
\end{equation}
where $X^2$ is defined as above,  $k$ stands for the number of the fitted parameters (two for MOND or four for \lcdm), and $n$ for the number of data points (in our case the number of the GCs plus $k$).  The model with the lowest AICc is preferred. This formula was derived with the assumptions that the model is linear in its free parameters and that the residuals from the fit follow a Gaussian distribution. {{The AICc criterion} can thus give misleading results if systematic errors affect the measurement.}

The LBF is defined as
\begin{equation}
\mathrm{LBF} = \log\int \mathcal{L}\left[\mathrm{data}|M(\vec{p})\right] f(\vec{p})\, \mathrm{d}\vec{p}.
\label{eq:lbc}
\end{equation}
{In this expression} we integrate over the whole space of the free parameters. The term $f(\vec{p})$ expresses the prior probability distribution function of $\vec{p}$ and $\mathcal{L}\left[\mathrm{data}|M(\vec{p})\right]$ the probability that the data is observed given that the model $M$  holds true and has the free parameters $\vec{p}$. If the portion of the parameter space that describes the data well is small, then the LBF comes out small. The LBF is a possible formalization of Occam's razor. When comparing several models of the given data, that with the highest LBF is preferred. This parameter moreover does not require determining the number of free parameters of the model, which might be unclear in some cases (see \citealp{freeparam} for examples). The price to pay is that the LBF is harder to calculate.  We evaluated the integral in \equ{lbc} using the Monte Carlo method. We generated $10^3$ random realizations of $\vec{p}$ according to the prior distribution and {then calculated $10^\mathrm{LBF}$ as the average of the corresponding values of $\mathcal{L}$, i.e., the first product in \equ{lik}.}

\begin{figure}
  \resizebox{\hsize}{!}{\includegraphics{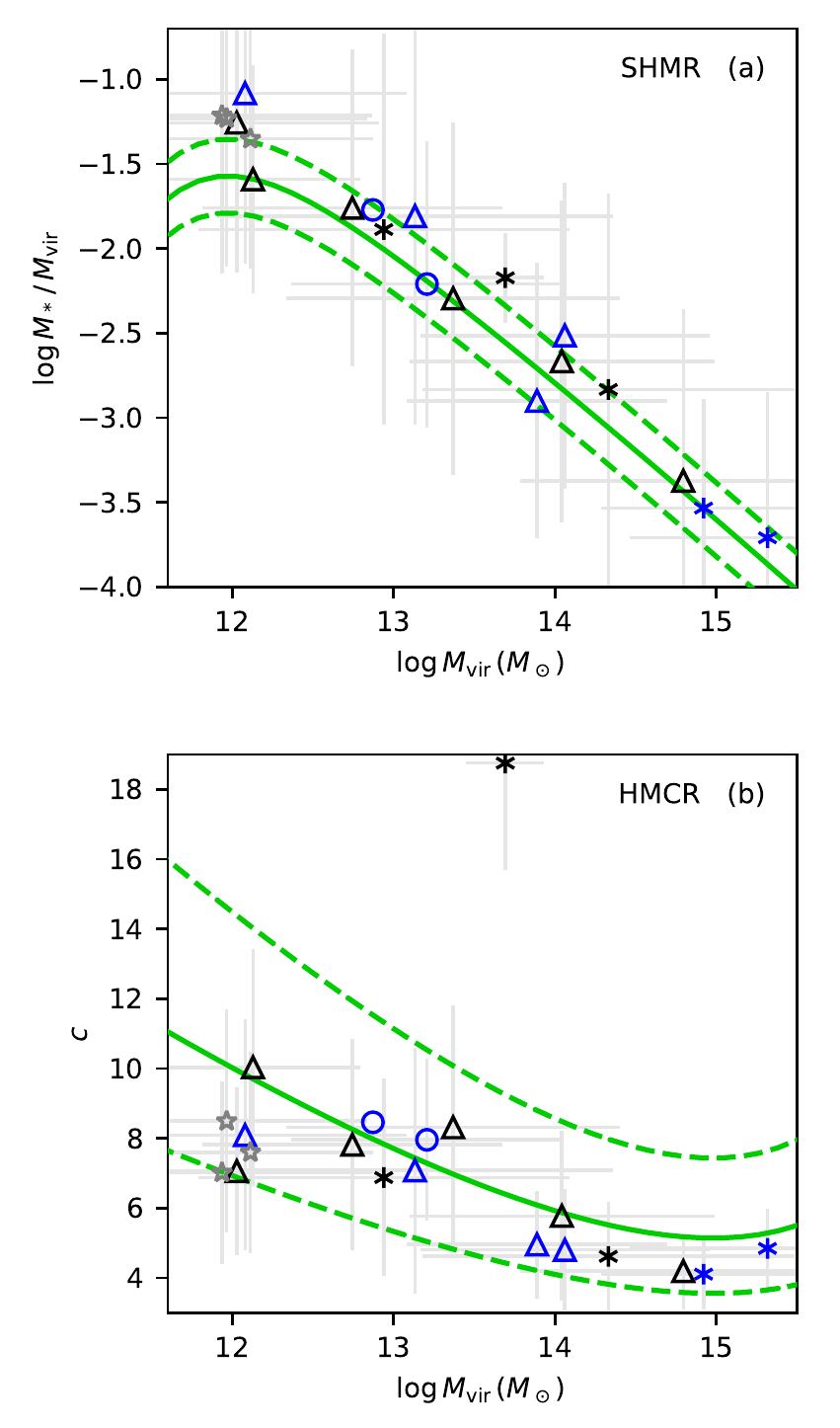}}
  \caption{Panel \textbf{(a)} Stellar-to-halo mass relation. Symbols: recovered relation between the halo mass $M_\mathrm{vir}$ and the baryonic fraction in the halo $M_*/M_\mathrm{vir}$. Black symbols indicate real galaxies distinguished by environment (circles: field, triangles: group, asterisk: cluster). {Blue symbols indicate the real galaxies dominating their environment clearly.} Gray open stars indicate the galaxies R17x,y,z from a \lcdm simulation.  We show the parameters found for the anisotropy profile providing the best agreement with the GC data. The full green line indicates the theoretical SHMR found by \citet{behroozi13} based on the abundance matching technique. The dashed green lines denote the 1$\sigma$ scatter region. The two real galaxies in the top left corner are NGC\,1023 and NGC\,4494. All individual galaxies agree with the prediction within the error bars but the reconstructed SHMR is offset. {The collective offset does not seem to be  substantially smaller for the dominant galaxies.} Panel \textbf{(b)} Halo mass-concentration relation. The green lines  indicate the theoretical HMCR by \citet{diemer15}. The most outlying galaxy is NGC\,4486. Again, the individual galaxies agree with the prediction but they are collectively offset from it{, including the dominant galaxies}.} 
  \label{fig:dmrel}
 \end{figure}

\subsection{Resulting fits, comparing the \lcdm and MOND models, dark halo scaling relations}\label{sec:res}

\begin{table*}
  \caption{Comparison of the best MOND and \lcdm fits}
  \label{tab:comp}
  \centering
  \begin{tabular}{l|lc|lc|c||lc|lc|c}
 \hline\hline
 \multirow{2}{*}{Name}& \multicolumn{2}{c|}{\lcdm} & \multicolumn{2}{c|}{MOND} & \multirow{2}{*}{$\Delta$AICc} & \multicolumn{2}{c|}{\lcdm} & \multicolumn{2}{c|}{MOND} & \multirow{2}{*}{$\Delta$LBF}\\
 &$\beta$ &  AICc &  $\beta$ &  AICc &  &  $\beta$ & LBF & $\beta$ & LBF &  \\
 \hline

N\,821 & lit & -51.2 & lit & -56.4 & 5.2 & lit & 10.8 & lit & 11.4 & -0.55\\
N\,1023 & lit & -106 & lit & -109 & 4.0 & lit & 22.8 & lit & 23.0 & -0.17\\
N\,1399 & neg & 431 & neg & 556 & \textbf{-120} & neg & -95.0 & neg & -195 & \textbf{100}\\
N\,1400 & neg & -76.0 & neg & -80.1 & 4.2 & neg & 16.5 & neg & 16.8 & -0.35\\
N\,1407 & neg & 12.8 & neg & 30.0 & \textbf{-17} & neg & -3.92 & neg & -7.54 & \textbf{3.6}\\
N\,2768 & lit & -84.4 & lit & -91.5 & 7.0 & lit & 18.1 & lit & 19.1 & -1.0\\
N\,3115 & lit & -106 & neg & -110 & 3.6 & lit & 22.5 & neg & 22.6 & -0.098\\
N\,3377 & lit & -215 & iso & -219 & 3.5 & lit & 46.2 & iso & 46.3 & -0.12\\
N\,4278 & iso & -147 & neg & -145 & \textbf{-1.7} & iso & 31.1 & neg & 30.4 & \textbf{0.73}\\
N\,4365 & lit & 14.3 & neg & 35.7 & \textbf{-21} & lit & -4.02 & neg & -8.85 & \textbf{4.8}\\
N\,4472 & neg & 170 & neg & 199 & \textbf{-29} & neg & -38.2 & neg & -45.7 & \textbf{7.5}\\
N\,4486 & neg & 535 & neg & 593 & \textbf{-58} & neg & -121 & neg & -142 & \textbf{21}\\
N\,4494 & iso & -178 & iso & -180 & 1.5 & iso & 38.5 & lit & 38.0 & \textbf{0.45}\\
N\,4526 & neg & -52.8 & neg & -57.3 & 4.5 & neg & 11.3 & neg & 11.7 & -0.37\\
N\,4649 & iso & 41.2 & neg & 43.6 & \textbf{-2.4} & neg & -9.98 & neg & -10.6 & \textbf{0.66}\\
N\,5128 & neg & -419 & neg & -378 & \textbf{-41} & neg & 89.4 & neg & 80.7 & \textbf{8.7}\\
N\,5846 & neg & 16.5 & neg & 31.5 & \textbf{-15} & neg & -4.30 & neg & -7.63 & \textbf{3.3}\\\hline
R17x & neg & -267 & lit & -268 & 1.6 & neg & 57.6 & lit & 57.3 & \textbf{0.28}\\
R17y & lit & -257 & lit & -260 & 2.9 & lit & 55.4 & lit & 55.4 & \textbf{0.037}\\
R17z & lit & -277 & lit & -280 & 3.5 & lit & 59.9 & lit & 59.9 & -0.0072\\

 \hline
  \end{tabular}
  \tablefoot{Lower AICc value and higher LBF value indicate the better fit. In the columns $\Delta$AICc and $\Delta$LBF,  the cases where the \lcdm models performed better than the MOND models with respect to the given criterion are highlighted by boldface. We note that the three last galaxies are not real; they come from a simulation (see \sect{sample}). The number of galaxies for which the MOND models are preferred is comparable to that of the galaxies where the \lcdm models are preferred. If the values of the preference are compared, the \lcdm models come out more successful.  }
\end{table*}

\begin{figure*}
  \centering
  \includegraphics[width=17cm]{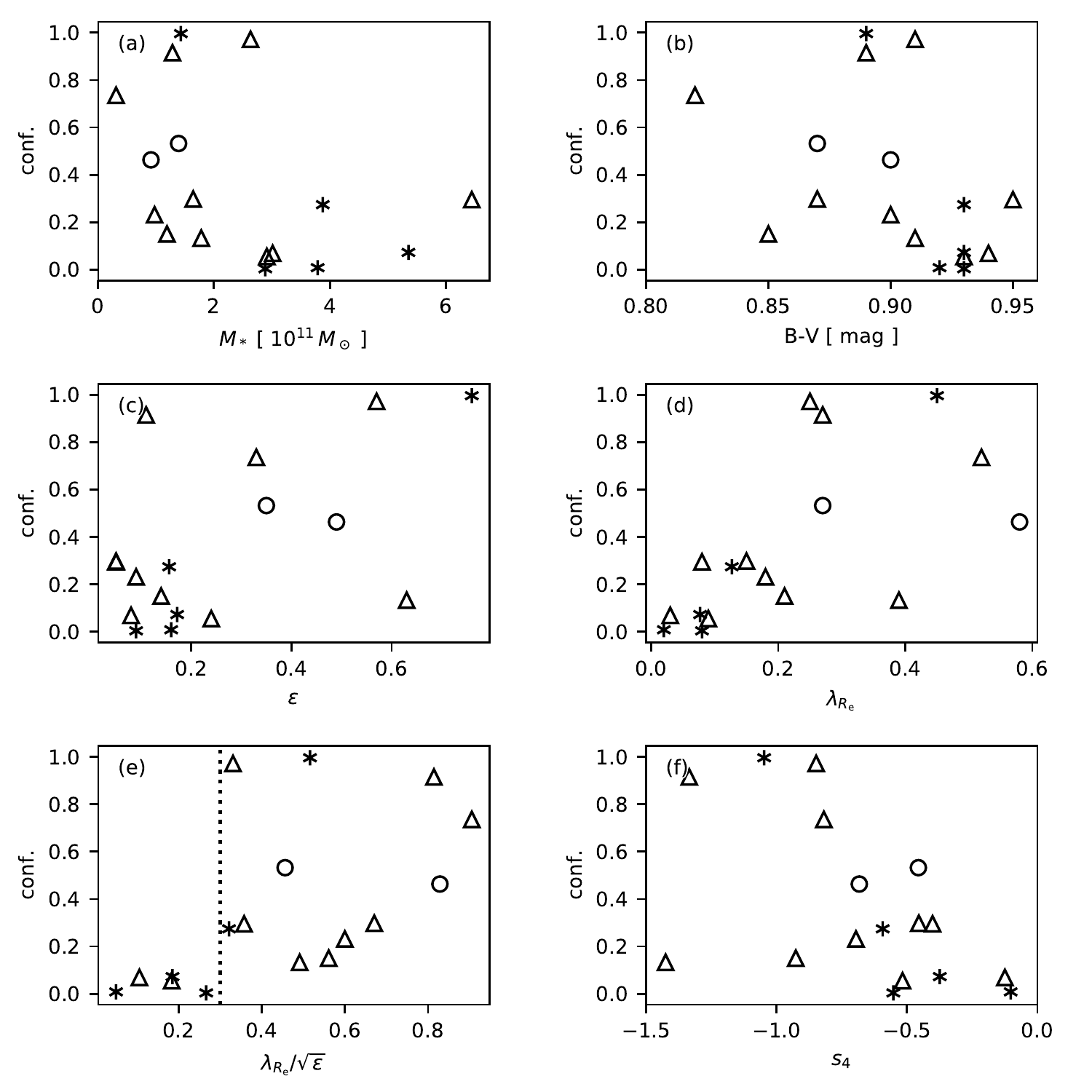}
  \caption{{Panels show how the confidence of the best-fit MOND model depends on various properties of the galaxy}:  \textbf{(a)}: stellar mass of the galaxy (derived for $d_0$ and $M/L_0$ defined in \sect{fits}); \textbf{(b)}: color index $B-V$ of the galaxy; \textbf{(c)}: galaxy ellipticity $\epsilon$; \textbf{(d)}: degree of rotational support $\lambda_{R_\mathrm{e}}$ (\equ{lam}); \textbf{(e)} -- the rotator criterion number $\lambda_{R_\mathrm{e}}/\sqrt{\epsilon}$, the vertical dotted line separates the slow and fast rotators; \textbf{(f)} : the $s_4$ kurtosis parameter of the GC system.  The marker shapes are the same as in \fig{dmrel}. {For the correct models of GC kinematics, the distribution of the confidence value would be homogeneous because of statistical noise.} The success of the MOND fits decreases with the characteristics of the galaxy approaching to those found in galaxy cluster centers.}
  \label{fig:mondconf}
\end{figure*}

Using the methods described above, we derived for every model the estimates and uncertainties of the quantities $M/L$, $d$, $\log\rho_\mathrm{s}$ and $\log r_\mathrm{s}$. These values are stated for the \lcdm and MOND models in Tables~\ref{tab:tablcdm} and~\ref{tab:tabmond}, respectively. In these tables we also provide the corresponding values of halo concentration $c$,  $r_\mathrm{vir}$ halo radius and  $M_\mathrm{vir}$ mass, stellar mass $M_*$, {dark matter fractions ($M_\mathrm{DM}/M_\mathrm{tot}$) interior to 1 and $5\,R_\mathrm{e}$,  $f^\mathrm{DM}_{1R_\mathrm{e}}$ and $f^\mathrm{DM}_{5R_\mathrm{e}}$}, $\chi^2$ of the fit, and the statistical measures of the goodness of the fit, i.e., confidence, AICc, and LBF.

We can see that the \lcdm models can describe all the galaxies successfully (the $\chi^2$ confidence is always higher than 5\%). The worst cases are NGC\,1023 and NGC\,4494 with {a confidence of approximately 38\%}. These galaxies require a weaker gravitational field than expected by the priors, as follows from panels (b)-(e) of \fig{g1} and \fig{g14}. The \lcdm models of these galaxies might require taking into account the rotation in the plane of the sky; see \sect{disc}.

The MOND models were not successful for two galaxies of our sample (NGC\,1399 and NGC\,4486)  meaning that the confidence of their best-fit models does not exceed 5\% for any of the considered anisotropy profiles. We discuss the possible explanations is \sect{disc} (additional mass, another anisotropy parameter, and dynamical heating by neighboring galaxies).

\tab{comp} provides a comparison of the MOND and \lcdm models for our galaxies in the terms of the AICc and LBF criteria. From all the considered anisotropy profiles, we show for every galaxy only the model that provides the best value of AICc or LBF. According to both AICc and LBF criteria, the MOND models are more successful for 9 out of the 17 real galaxies. If a \lcdm model is preferred in some galaxy, then this preference is usually stronger than in the galaxies where a MOND model is preferred. 
The MOND models prefer mostly the $\bneg$ profile,  while the \lcdm prefer the $\blit$ and $\bneg$ profiles in a comparable number of the real galaxies. {Expectedly, the simulated galaxies R17x,y,z prefer mostly  the $\blit$ profile inspired by \lcdm simulations.}

Our results also allow us to construct the SHMR and HMCR from the GC kinematics, which can be compared with theoretical expectations of the \lcdm paradigm.  The comparisons are presented in the Panels (a) and (b) of \fig{dmrel}, respectively. The SHMR is presented in an equivalent form as the fitted baryonic fraction of the galaxy (i.e., the stellar mass divided by the halo mass) as a function of the halo mass. We can see that both the baryonic fraction and halo concentration indeed correlate tightly with the halo mass. We caution that the small scatter can be partly a consequence of our method, which disfavors deviations of the best-fit parameters from the theoretical correlations. The green lines denote the theoretical correlations and their scatter -- that for the SHMR comes from \citet{behroozi13} and that for the HMCR from \citet{diemer15}. Our recovered correlations deviate somewhat from the theoretical models. {There are 14 galaxies out of the 17 lying above the mean theoretical SHMR. Using the binomial distribution, we can calculate that the probability that at least 14 galaxies out of 17 lie above the mean relation is 0.6\%. For the HMCR, only 5 galaxies lie above the mean theoretical relation. Similarly, the probability that less than 5 of our galaxies lie above the theoretical relation is 2\%. The deviations from both theoretical relations are thus statistically significant.} 

{As explained below \equ{chisq},} if we had correct models of the GC kinematics for all our galaxies, then the fraction of galaxies with a confidence below $x$ would be $x$ because of the statistical noise. From this point of view, the \lcdm models were too successful. From the models of 17 real galaxies,  {15 of them could be fitted with the confidence value over 70\%. As above, we can calculate  that the probability that at least 15 out of 17 galaxy fits have a confidence over 70\% is $1\times 10^{-4}$\%.} This probably means that the \lcdm models are very flexible and fit even random noise in the data. On the other hand, the MOND models rather underfit the data since 11 out of the 17 galaxies have a confidence under 30\% and the probability of obtaining {at least this} number of galaxy fits with a confidence below 30\% is 0.32\%. 

{We determined the characteristics of the galaxies with a low confidence of the MOND fits using the diagrams in \fig{mondconf} showing the confidence of the best-fit MOND model versus the given characteristic of the galaxy. If we had correct models of the data, the points would have the uniform distribution of the confidence. If the points concentrate toward a low value of confidence, this signifies an underfitting, i.e., the model does not capture the physical nature of the modeled system well. The concentration of the points toward high values of confidence rather  signifies an overfitting of the data by the model.} From these plots we can see that the confidence of the MOND fits decreases with an increasing stellar mass (as calculated assuming $p_d = p_{M/L} = 0$) for an increasing color index $B-V$, a decreasing ellipticity,  a decreasing degree of rotational support of the galaxy $\lambda_{R_\mathrm{e}}$  (\sect{lcdm}), a decreasing rotator criterion number $\lambda_{R_\mathrm{e}}/\sqrt{\epsilon}$  (\sect{lcdm}), and an increasing total kurtosis parameter of the GC system (as defined in \sect{gcss}).  The decrease of the MOND confidence appears gradual for the degree of rotational support, $\lambda_{R_\mathrm{e}}$, and the kurtosis parameter of the GC system, $s_4$. There seems to be a sudden drop of this confidence for the stellar mass above {$M_* = 2.5\times 10^{11}\,M_\sun$}, the color index above $B-V=0.92$, galaxy ellipticity below $\epsilon = 0.1$, and the galaxy rotator criterion number below $\lambda_{R_\mathrm{e}}/\sqrt{\epsilon}=0.3$. Interestingly, there seems to be a sudden drop  of the confidence of the best-fit MOND models at the value of the {rotator criterion number}  separating the slow and fast rotators, which is indicated by the vertical dotted line in the panel (e) of \fig{mondconf}. {The probability that all the 5 slow rotator best-fits have a confidence below 8\% by coincidence is $3\times 10^{-4}$\%.}

As a check of the correctness of our method, we analyzed in the same way the artificial galaxies R17x,y,z.  {The values of $M/L$, $d$, $\log r_\mathrm{s}$ and $\log \rho_\mathrm{s}$ recovered using the GC kinematics (\tab{tablcdm}) are all consistent with the true values (\sect{sample}) within the $1-2\sigma$ uncertainty limits.} This demonstrates that our method can recover the real parameters reasonably, at least for this particular simulated galaxy.

\section{MOND and \lcdm predictions of the velocity-dispersion profiles}\label{sec:pred}

In this section we compare the observed GC kinematics with the predictions of the \lcdm and MOND paradigms. {They were made} without any fitting, just on the basis of independent estimates of the galaxy distances, luminosities, $M/L$s, and the GC projected systemic rotation velocities. We assumed that the GC systems are kinematically isotropic. For the \lcdm models, we further assumed that the galaxies follow the mean SHMR and HMCR relations. In other words, we set all of the parameters $p_i$ defined in \sect{fits} to zero. 

The results are plotted in the panels (a) of \allgals as the dotted lines: the blue lines correspond to the \lcdm predictions and the red to the MOND predictions. We note several interesting points: 1) MOND usually underpredicts the observed velocity dispersion while \lcdm usually overpredicts it. 2) \lcdm always predicts a higher velocity dispersion than MOND. 3) The MOND predictions are almost always closer to the observed velocity-dispersion profile than the \lcdm predictions {or the goodness of the predictions appears comparable}, i.e., MOND has a better predictive ability.

\begin{figure}
  \resizebox{\hsize}{!}{\includegraphics{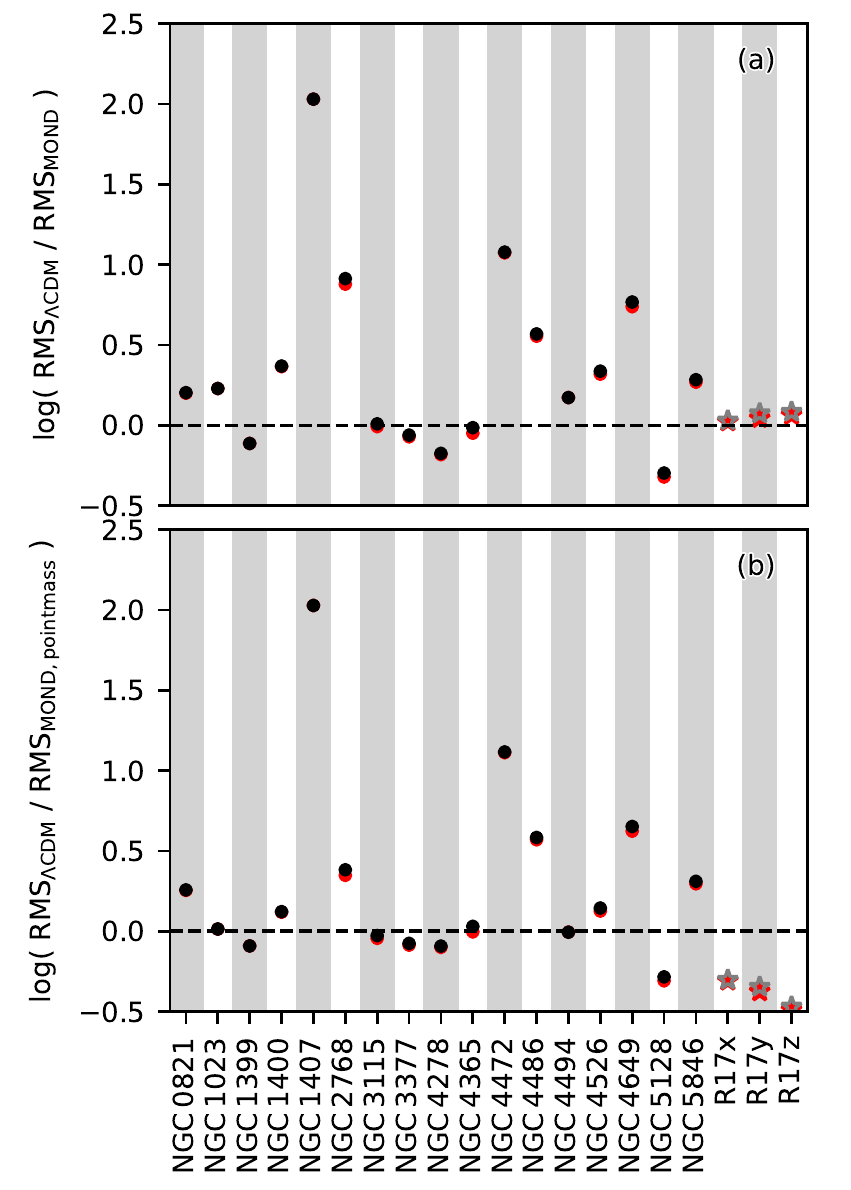}}
  \caption{{Panel \textbf{(a)}: Comparison of the success of the predictions of the GC velocity dispersion profiles by  MOND and \lcdm. These predictions are based on independent estimates of the free parameters. The prediction assuming isotropy for both MOND and \lcdm models are shown in black. The gray open stars represent the artificial galaxies R17x,y,z. It can be seen that MOND has a greater predictive ability. The same is true if the anisotropy model for the \lcdm models is chosen out of the profiles $\biso$, $\bneg$, and $\blit$ so that it gives the lowest RMS, which is shown by the red points. Panel \textbf{(b)}: The same but for the MOND prediction, the real galaxy density profiles were substituted by point masses. Despite this disadvantage, the MOND predictions still perform better, even if we again chose the anisotropy profile giving the best RMS for the
\lcdm.}} 
  \label{fig:preddevs}
\end{figure}

The last point is quantified in \fig{preddevs}a where we calculated the RMS deviation of the predicted velocity dispersion profile, $\sigma_\mathrm{los, pred}$, from the observed velocity dispersion profile $\sigma_\mathrm{los, obs}$ obtained in the way described in \sect{gcss} (i.e., from the green curves in \allgals) as  $\mathrm{RMS} = \left\lbrace n^{-1}\sum_{i=1}^n[\sigma_\mathrm{los, pred}(x_i) - \sigma_\mathrm{los, obs}(x_i)]^2\right\rbrace^{1/2}$. {The ratio of the RMS of the \lcdm and MOND predictions is indeed mostly greater than one, which is shown by the black points. The red points demonstrate that the situation did not change appreciably even if we chose as the \lcdm prediction\footnote{Doing so the \lcdm ``prediction'' is not a genuine prediction anymore.} the anisotropy profile out of $\biso$, $\bneg$, and $\blit$ individually for every galaxy so that the RMS deviation is minimized. The MOND predictions are in majority still better than the \lcdm predictions even if the MOND models are put in disadvantage by substituting the real galaxy density distribution by a point mass that is shown in  \fig{preddevs}b by the black points or by the red points if optimized the \lcdm prediction over the anisotropy profile.}

\begin{figure*}
  \centering
  \includegraphics[width=17cm]{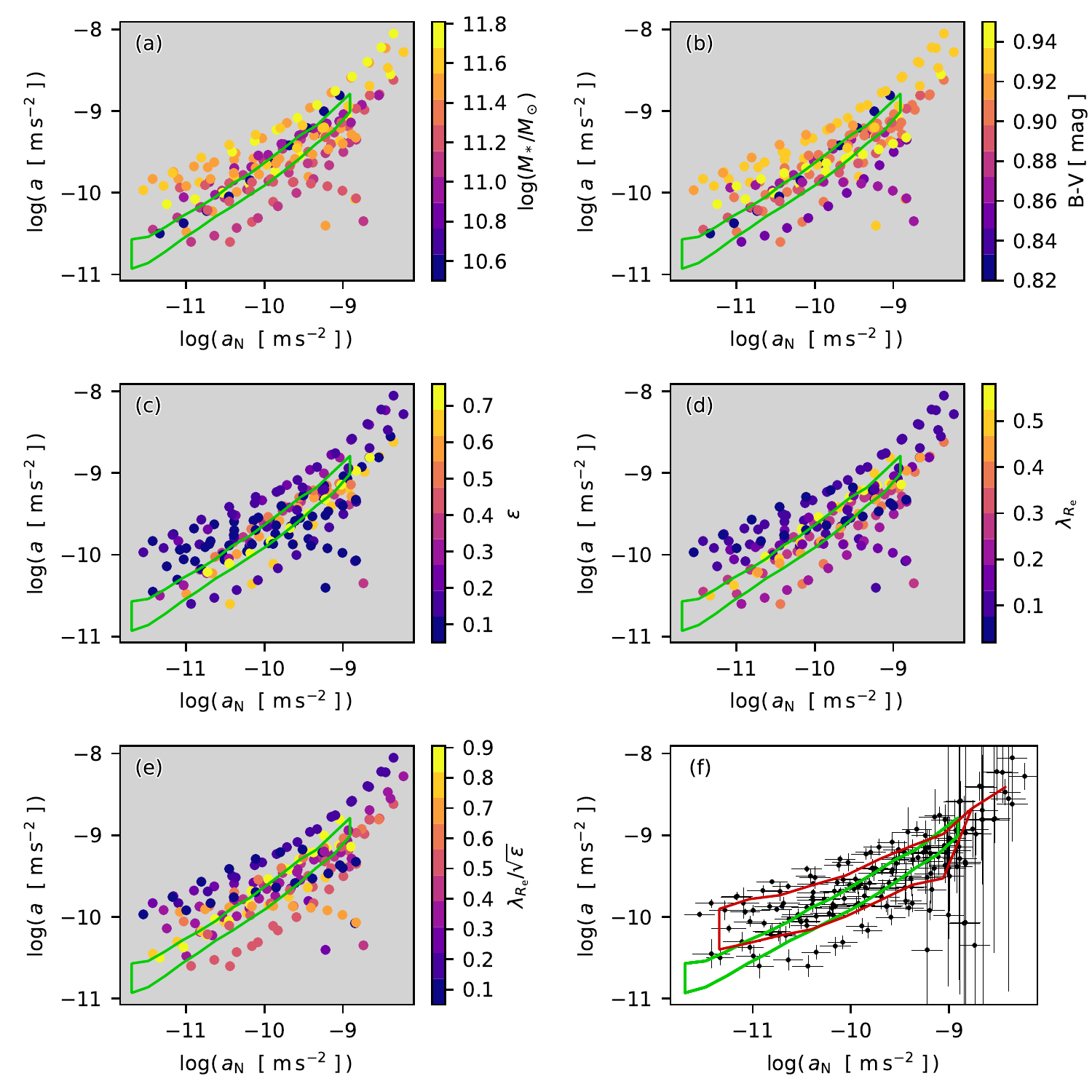}
  \caption{Comparison of the RARs of the individual ETGs (colored points) and to the RAR of the LTGs (green region). The vertical thickness corresponds to the 1\,$\sigma$ scatter (\citealp{mcgaugh16}). The points in the panels (a)--(e) are colored with respect to the properties of their galaxy: (a) -- stellar mass of the galaxy (derived for $d_0$ and $M/L_0$ defined in \sect{fits}); (b) -- color index $B-V$ of the galaxy; (c) -- galaxy ellipticity $\epsilon$; (d) -- degree of rotational support $\lambda_{R_\mathrm{e}}$ (\equ{lam}); (e) -- the rotator criterion number $\lambda/\sqrt{\epsilon}$. Panel (f) indicates the error bars and, as the red region, the 1$\sigma$ intrinsic scatter in the reconstructed collective RAR of the ETGs. The deviation of the ETG RARs from the LTG RAR increases {with their characteristics approaching} to those of galaxies in galaxy cluster centers.}
  \label{fig:rarall}
\end{figure*}

\section{Radial acceleration relations}
\label{sec:rar}
The RAR has been studied extensively in the LTGs but  just a few studies have focused on the ETGs. 
In order to investigate the RAR on our data, we used special models of the gravitational field based on the \lcdm models described in \sect{fits} so that the total gravitational field was approximated by a sum of the contribution from the stars and from a NFW halo. This is meant as a way to parametrize the gravitational field and therefore the values of fitted parameters could come out non-physical in the \lcdm context. This form of gravitational field is justified by the fact that most rotation curves of the LTGs can be fitted well by a NFW halo. These models had three free parameters, the stellar $M/L$, which we allowed to be even negative, and the halo characteristic radius and density, which were both allowed to be only positive. In addition we required that the resulting gravitational field does not repel objects from the galaxy center on the radial range occupied by the GCs. The distance of the galaxies was set to be $d_0$ (defined in \sect{fits}). We checked that the gravitational fields expected by MOND for our galaxies can be approximated precisely in this way; the relative difference in the rotation curves could be made below 2\% on the radial range occupied by the GCs. {We obtained the best-fit parameters using the maximum likelihood method.} This time, we did not use the priors on the free parameters so that we minimized only the first product in \equ{lik} because our goal was to fit the gravitational field as closely as possible. For every galaxy, we considered the three types of the anisotropy profiles  described in \sect{fits}. 

For the best-fitting model, we evaluated the dynamical acceleration at ten galactocentric radii for every galaxy spaced logarithmically on the radial range occupied by the GCs. We found the uncertainty limits on these estimates using the method described in \sect{fits}. The Newtonian accelerations $a_\mathrm{N}$ were calculated assuming $p_{M/L}=p_d=0$. The uncertainty of $a_\mathrm{N}$ is written as
\begin{equation}
\Delta \log a_\mathrm{N} = \sqrt{(\Delta \log M/L)^2+(2\Delta \log d)^2}.
\end{equation}
The results are plotted in the panels (f) of \allgals. We also plot by the gray band the RAR for the LTGs  found by \citet{mcgaugh16}. The thickness of the band corresponds to the 1$\sigma$ scatter found in that study. 

For most galaxies, at least one anisotropy profile allows a match with the LTG RAR within the 2$\sigma$ uncertainty limit. The most discrepant objects are NGC\,1399 and NGC\,4486, which were already found to be inconsistent with MOND in \sect{res}. The objects for which we obtained{ the reconstructed dynamical acceleration, $a$, that is too low}, NGC\,1023 and particularly NGC\,4494, might be quickly rotating in the plane of the sky.  {Another explanation is suggested by the simulated galaxies. Their reconstructed RARs deviate in a similar fashion. We know how quickly these {galaxies} rotate in all perpendicular projections.} {Since this rotation is not sufficient to explain the deviations and since the galaxies follow the LTG RAR (\sect{sampart}),} these galaxies provide examples of how  errors in the $M/L$ and galaxy distance can affect the appearance of the RAR. 

When all our RAR data points are put together, they broadly follow the RAR of the LTGs but with more scatter and the mean dynamical acceleration is higher at a given Newtonian acceleration; see \fig{rarall}. {Its panel (f) indicates the error bars of the measured points obtained as the minimum and maximum $a$ allowed by models with the three anisotropy profiles. We depicted in red the region covering the 1$\sigma$ intrinsic scatter in the total RAR of ETGs determined using a maximum likelihood approach assuming that the average RAR does not change substantially over a 0.33\,dex range in $a_\mathrm{N}$ and that the total scatter is a combination of a Gaussian intrinsic scatter and a Gaussian scatter in the estimation of $a$.
 The 1$\sigma$ scatter region of the LTG RAR (intrinsic scatter plus measurement error) as given by \citet{mcgaugh16} is denoted by the green boundary.
The points in the other panels are colored with respect to various properties of the particular galaxies}: the stellar mass (as determined assuming $p_d = p_{M/L} = 0 $), galaxy color index $B-V$, galaxy ellipticity $\epsilon$, galaxy degree of rotational support $\lambda_{R_\mathrm{e}}$ (\equ{lam}), and the galaxy rotator criterion number $\lambda_{R_\mathrm{e}}/\sqrt{\epsilon}$.  We see {a picture similar to} the case with the fit confidence in \sect{res}: the deviation of the reconstructed RAR from that for the LTGs increases with a higher galaxy mass, redder color, decreasing support by rotation,  and decreasing rotator criterion number. 

We verified the correctness of our method for reconstructing the RAR using the artificial galaxies R17x,y,z. The reconstructed data points indeed agree with the correct RARs (the orange dotted lines in the panels (f) of Figs.~\ref{fig:g17}~--~\ref{fig:g19})  within approximately 2$\sigma$ uncertainty limits.

\section{Discussion}
\label{sec:disc}

\subsection{Implications in the \lcdm context} \label{sec:lcdm}
The GC systems have the advantage that they cover a substantial fraction of the dark halo. This helps when estimating the parameters of the halo. The scale radii of the NFW halos come out for LTGs typically five times their disk scale lengths \citep{navarro17} while our most distant GCs lie further than $5\,R_\mathrm{e}$ (see \allgals). Our best-fit \lcdm models confirm that the scale radii are often comparable to the extent of the GC system; see \tab{tablcdm}. 

We were able to find a statistically successful \lcdm fit for every galaxy in the sample as determined using the chi-square test. The \lcdm predictions on velocity dispersion are usually too high (\sect{pred}). This is also the case of the galaxies NGC\,1023 and NGC\,4494 for which we obtained the worst \lcdm models ({confidence of around 38\%} for both). Statistical error can easily account for this confidence value. Nevertheless, the shape of the best-fit \lcdm models for these galaxies  seems to be too high for the outermost GCs (Figures~\ref{fig:g1} and~\ref{fig:g12}). The reason for this discrepancy might be the rotation of the GC system in the plane of the sky so that the attractive force is overestimated. This suggestion is supported by the fact that NGC\,1023 and NGC\,4494 have very low values of the $s_4$ parameter among the real galaxies in our sample (\tab{tabgcss}) and as we showed in \sect{gcss}, this suggests a systemic rotation of the GC system. We can see in \tab{tabgcss} that the systemic rotation of GC systems can exceed 100\,km\,s$^{-1}$. Future studies can use the rotation velocity in the plane of the sky as a free parameter.

Our galaxies form well-defined SHMR and HMCR relations (\fig{dmrel}). We should remember that our method forces the  best-fit models not to deviate substantially from the theoretical relations. Nevertheless, both of the recovered relations deviate from the theoretical expectations. In particular, most galaxies in the recovered HMCR relation lie below the relation deduced from cosmological dark-matter-only simulations of isolated halos by \citet{diemer15}. We propose a few  possible explanations. Some of our galaxies are not isolated, i.e., their center lies within a virial radius of another, more massive halo. Then this neighboring halo can tidally strip the outer part of the halo of the galaxy of interest and its concentration thereby decreases. {In such a case, we expect that isolated galaxies and the galaxies dominating their environment would be affected the least. In order to test this idea, we chose the galaxies that are clearly dominant objects of their environment by visual inspection of the Digitized Sky Survey images using the Aladin software -- NGC\,821, NGC\,1023, NGC\,1399, NGC\,1407, NGC\,2768, NGC\,3115, NGC\,4472 and NGC\,5128. These galaxies are denoted by the blue symbols in \fig{dmrel}. The offset for the dominant galaxies does not seem to be substantially different than for the non-dominant sample. }  Another explanation might be that the ETGs prefer less concentrated halos than the LTGs that are missing in our sample{, i.e.} the result that our galaxies lie below the theoretical HMCR is a selection effect. This hypothesis could be investigated in existing cosmological hydrodynamical simulations or observationally using the rotation curves of the LTGs. The existing {observational} results suggest that the positions of the halos of the LTGs on the HMCR and SHMR depend on the assumed form of the halo and the priors \citep{katz17,li18b}.  On the other hand, the most luminous galaxies are nearly always ETGs, while our reconstructed HMCR deviates from the theoretical HMCR even for the most massive halos. The deviation of the recovered HMCR from the theoretical HCMR might also be caused by some systematic insufficiency of our \lcdm models such as the plane-of-the-sky rotation of the GC system or some of the mechanisms suggested for the MOND models below. {This kind of explanation seems inevitable for the} extreme deviation of the galaxy NGC\,4486 from the relation formed by the other galaxies. We should keep in mind a possible error in the theoretical HMCR.

We showed the SHMR as the baryonic fraction, $\log(M_*/M_\mathrm{vir})$, as a function of the halo mass in \fig{dmrel}a. The theoretical relation was based on abundance matching \citep{behroozi13}. The recovered relation is linear. Compared to the theoretical relation, the measured baryonic fraction in a halo is higher for a given halo mass and there is no turnover in the recovered relation near the halo mass of $10^{12}$\,M$_\sun$. We recall that  three of the galaxies with $\log(M_*/M_\mathrm{vir})>-1.3$ denoted by the open gray stars are the artificial galaxies R17x,y,z. The remaining two are NGC\,1023 and NGC\,4494. It is interesting that {the RARs of these galaxies most clearly lie}  below the RAR of the LTGs; see panels (f) of \allgals. Moreover, these galaxies have very low values of the $s_4$ parameter (\tab{tabgcss}) among the sample and their chi-square confidences belong to the lowest. Interestingly, the artificial galaxies indicated by gray open stars have nearly the same fitted halo mass and the baryonic fraction, and we checked that these quantities were recovered correctly; the simulation {that the galaxies come from}  was not designed to reproduce the SHMR. The same explanations of the deviation of the recovered HMCR from the theoretical HMCR apply for the SHMR apart from the fact that no assumption on galaxy isolation was made in the derivation of the theoretical SHMR \citep{behroozi13}.

We were interested in whether the velocity dispersion profiles of the artificial galaxies R17x,y,z were affected by the continual mergers that happen in a \lcdm universe. To this end, we calculated the velocity dispersion profiles for the correct values of the galaxy distances, stellar masses, and dark halo parameters (\sect{sampart}). The galaxy S\'ersic indices, effective radii, and GC rotation velocities were left as we derived them from the projected data (Tables~\ref{tab:tabgals} and~\ref{tab:tabgcss}). The results are indicated in panels (a) of \figs{g17}{g19} by the orange dotted lines. A zero anisotropy was assumed. These theoretical profiles are well consistent with the measured profiles. This suggests that the GC kinematics was not affected by the mergers in this case. {The galaxy in the simulation by \citet{renaud17} was relatively isolated and did not experience any major merger for 5\,Gyr before the data were read, which means that the many subhalos moving through the GC system did not affect its kinematics substantially.} 

In the \lcdm context, the apparent validity of MOND in the LTGs is considered a consequence of the profile of the dark halos, SHMR, HMCR, and the {baryonic scaling relations} \citep{cintio15,navarro17}. Interestingly, the MOND prescription (\equ{mond}) has a better predictive ability than the HMCR and SMHR, even if the MOND prediction is put at a disadvantage by substituting the galaxy by a point mass {or optimizing the \lcdm ``prediction'' over the anisotropy profile} (\sect{pred}). It might be again because we used a HMCR for isolated halos or because the halo shape also depends on the density distribution of the galaxy.  The second point is probable because the MOND prescription works for the LTGs and there are pairs of LTGs with the same luminosity but very different rotation curves {(the ``Tully-Fisher pairs'; see, e.g., Sect.~4 of  \citealp{deblok98})}.

We note that the degree of success of a MOND model of a GC system correlates with how much slow or fast rotator the host galaxy is. The slow and fast rotating ETGs can be distinguished observationally using integral field units. In these observations, we can introduce the degree of rotational support interior to $1\,R_\mathrm{e}$ as
\begin{equation}
\lambda_{R_\mathrm{e}} = \frac{\sum_i F_i R_i |V_i|}{\sum_i F_i R_i\sqrt{V_i^2+\sigma_i^2}},
\label{eq:lam}
\end{equation}
where we sum over the spaxels interior to $1\,R_\mathrm{e}$ and $F_i$ means the light flux from the spaxel, $R_i$ its galactocentric radius, $V_i$ the systemic velocity in the spaxel, and $\sigma_i$ the velocity dispersion in the spaxel. The $\lambda_{R_\mathrm{e}}$ parameter measures the degree of rotational support of the galaxy. The slow rotators are defined as the galaxies having the rotator criterion number $\lambda_{R_\mathrm{e}}/\sqrt{\epsilon}$ below 0.31, where $\epsilon$ is the ellipticity of the galaxy \citep{emsellem11}. Fast and slow rotators are two types of galaxies that have many contrasting properties. Most ETGs are fast rotators (86\% according to \citealp{emsellem11}). We can use the rotator criterion number $\lambda_{R_\mathrm{e}}/\sqrt{\epsilon}$ to quantify how fast or slow rotator a galaxy is. We found that this parameter correlates with the degree of success of the best MOND model of the galaxy measured by the chi-square confidence; see the panel (e) of \fig{mondconf}. \citet{janz16} noted that the dynamical masses of fast rotators determined from the GC kinematics using the tracer-mass-estimator method agree better with MOND than the dynamical masses of slow rotators. Similarly, \citet{rong18} found that the stellar kinematics of fast rotator ETGs is consistent with MOND while that of the slow rotators is less so. On the other hand, other observational evidence supports the validity of MOND even in slow rotators as we argue in \sect{mond}. Our results suggest that in the \lcdm context some connection between the degree of rotational support of the galaxy and its dark halo exists. \citet{tenneti18} detected a tight RAR for the rotationally supported galaxies in the MassiveBlack-II hydrodynamic cosmological simulation, which however differs from the RAR observed for the LTGs. It would be interesting to see whether the velocity-dispersion-supported galaxies in such simulations follow the same RAR and possibly spot the origin of the dependence on the galaxy rotation. Also, some of the explanations we suggest below for the MOND context could be true. However, most of these explanations cause an increase in velocity dispersion while the \lcdm models in most galaxies already overpredict the velocity dispersion.

\begin{figure}
  \resizebox{\hsize}{!}{\includegraphics{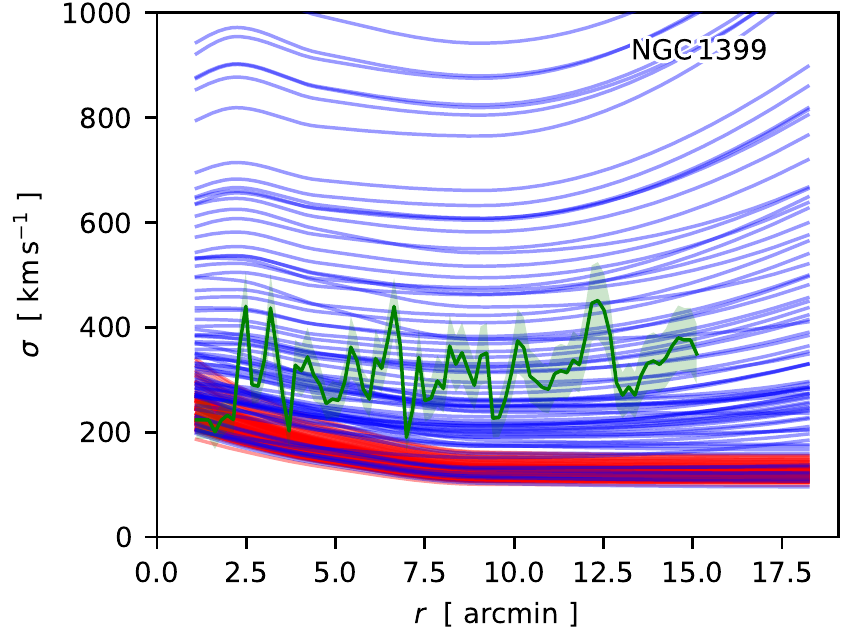}}
  \caption{Demonstration that GC velocity dispersion profiles can be fitted easier by a \lcdm model than by a MOND model even if the models omit some important aspect (e.g., halo stripping or perturbed GC kinematics). Blue lines indicate 100 random \lcdm models of the velocity dispersion profile of NGC\,1399, according to the statistical distribution of the free parameters described in \sect{fits}. Such free parameters are well consistent with all the observational and theoretical constraints we have on the galaxy. Red lines indicate the same for the MOND models. Green line indicates the observed velocity dispersion profile of NGC\,1399.} 
  \label{fig:shoot}
\end{figure}

\subsection{Implications in the MOND context}\label{sec:mond}
Our MOND models were not successful in two galaxies as determined by the chi-square test (NGC\,1399 and NGC\,4486, see \sect{res}). The AICc and LBF discriminators prefer the MOND models over the \lcdm models in about a half of the galaxies but the preference is usually stronger for the \lcdm models {(\tab{comp})}. It is hard to decide whether this means that the \lcdm paradigm is preferred over MOND based on our data alone because the assumed gravity prescription is just one of the uncertain ingredients of the models. The others include {the anisotropy profile or the assumption} of dynamical equilibrium, as described below.

The following example demonstrates that the \lcdm models are more likely to produce a better fit to the data even if some of the model assumptions are broken because of the higher flexibility of these models. We generated 100 radial profiles of the velocity dispersion for NGC\,1399 with $\biso$, where we chose the $p_i$ parameters of the models randomly according to the prior statistical distributions (\sect{fits}). The results are the blue curves in \fig{shoot}. We note that the most of these models follow the SHMR and HMCR well and are consistent with the observational constraints. The red curves in this figure were generated for the MOND models in the same way. The green curve is the observed velocity dispersion profile of NGC\,1399. It is obvious that the \lcdm models can provide a successful fit for a much larger variety of observed velocity dispersion profiles than the MOND models. 

If our MOND models were not successful for the slow rotator ETGs because MOND is invalid, then other tests of MOND in slow rotators based on independent methods should also be negative. The studies of stellar kinematics by Jeans analysis (e.g., \citealp{rong18}) are still similar to our method and could share the same weak points. \citet{rong18} might have come to their conclusion that the dynamical masses for the slow rotators are up to 60\% higher than predicted by MOND because they used axially symmetric models for the galaxies, while most slow rotators exhibit more complex kinematics and misalignments between the kinematic and photometric axes \citep{emsellem07,emsellem11, graham18}. On the contrary, an independent method of measuring the gravitational field based on the requirement of hydrostatic equilibrium of the hot interstellar gas with gravity resulted in successful tests of MOND in slow rotators \citep{milg12, lelli17}. We are not aware of a test of MOND by weak gravitational lensing specifically  in slow rotators. Nevertheless, MOND was tested  successfully in an ``averaged'' sample of red, presumably ETGs \citep{milg13}. While most ETGs are fast rotators, the most massive ETGs are usually slow rotators; i.e., three-quarters above the stellar mass of $10^{11.5}$\,M$_\sun$ according to \citealp{emsellem11} (see also \citealp{veale17}). There does not seem to be any indication of the deviation of the weak lensing data from the MOND prediction for the most massive galaxies (see Fig.~1 of \citealp{milg13}). A dedicated test of MOND in slow rotators based on weak lensing would be welcome (\citealp{rong18} had 144 slow rotators in their sample). Analysis of strong lensing by massive ETGs by \citet{tian17} revealed that their dynamical masses agree with the MOND prediction and the interpolating function found for the LTGs. The small systematic deviation of around 20\% could have easily been  caused by a systematic error in the estimated stellar $M/L$. If MOND does not hold true in velocity-dispersion supported objects, it should be impossible to obtain satisfactory fits to the rotation curves of galaxies with dominant classical bulges. NGC\,7814 has a classical bulge and bulge-to-total ratio of 0.85  \citep{bell17}. The rotation curve of this galaxy was fitted successfully in MOND \citep{angus13}. 

These facts led us to look for ways to explain the low success of some of our MOND models other than excluding MOND. These approaches are described in the following subsections. {The MOND predicted velocity dispersions are usually lower than what is observed.}  The discrepancy probably has several sources. Some mechanisms to increase the velocity dispersion even more are necessary because of the external field {effect:} if the galaxy accelerates in an external gravitational field (such as from the galaxy cluster) its internal gravitational field is reduced in comparison to the galaxy being isolated (e.g., \citealp{milg83a, bm84, milgmondlaws}). In our models, the galaxies were treated as isolated. In the discussion below we assumed that the central galaxies of groups or clusters lie indeed in the geometrical centers of their environments and thus the external field effect is negligible.

\subsubsection{Unaccounted matter}
\label{sec:unacc}
 Our results could be explained by the presence of some mass component that we did not take into account. This contradicts some of the observations mentioned in the previous section but could be a viable option if substantial amounts of this {material} (say more than twice the stellar mass because of the uncertainty in the $M/L$ in the above studies) are present just in a small fraction of galaxies, such as in the two most problematic galaxies NGC\,1399 and NGC\,4486. 
 These are both central cluster galaxies whose GC systems extend over 100\,kpc, therefore we might suspect the contribution of the hot intercluster gas is important. According to the observations of \citet{paolillo02}, the mass of the X-ray emitting hot gas of NGC\,1399 cumulated interior to 15\arcmin\ is $10^{10}-10^{11}$\,M$_\sun$. This is comparable to the stellar mass of the galaxy. Since the velocity dispersion in MOND scales with the fourth root of the baryonic mass, this contribution increases the velocity dispersion by around 20\%, which is not enough. Such an increase would however be substantial for other galaxies, nevertheless exploring the mass of the interstellar gas in these galaxies is beyond the scope of this paper. 

In NGC\,1399, there seems to be several times more invisible matter than stellar matter as we can deduce either from the comparison of the MOND prediction on the velocity dispersion profile and the observed profile or from the comparison of the $M/L$ of the best-fit MOND model of the galaxy and the SPS models. This has been already noted by \citet{richtler08} based on an analysis of the GC system of this galaxy. These authors suggested that this is a small scale version of the remaining missing mass problem of MOND in galaxy clusters. This MOND cluster-scale missing mass seems to form cores in galaxy cluster centers with characteristic radii of a few hundreds of kiloparcsec \citep{sanders03,takahashi07, milgcbdm}. {In the whole extents of galaxy clusters, MOND requires about as much of the invisible matter as baryonic matter. In galaxy cluster centers, the ratio of the invisible to visible mass increases up to the factor of a few.} It is then expected that some of {the invisible mass} is also contained within the extents of the GC systems of the central cluster galaxies and thus the mass deficit in these galaxies does not pose an additional problem to MOND. The trend for slow rotators would be expected because they are mostly found in the cores of their clusters \citep{cappellari13, houghton13, deugenio13, scott14} because the slow rotator fraction increases with stellar mass and the most massive galaxies are found near galaxy cluster centers \citep{brough17, veale17}.

The following proposition adds to the ``unaccounted matter'' in a different way. Massive elliptical galaxies have a bottom-heavy  IMF in their centers \citep{conroy12}. Thus, their $M/L$ are higher than expected from the SPS models based on the IMF in the Milky Way or nearby galaxies. This might alleviate the deviation for our galaxies. \citet{vandokkum17} found that this change of the $M/L$  is by a factor of 2.5 in galaxy centers but it decreases with radius so that the average $M/L$ is enhanced by a factor of  1.3-1.5 at 1\,$R_\mathrm{e}$. The fraction of stellar light enclosed in 1\,$R_\mathrm{e}$ is around 40\% for our galaxies, thereby the total mass is increased by around 20\%.  This would help for some of the galaxies; compare Tables~\ref{tab:tabmls} and~\ref{tab:tabmond}. This correction would probably remove the small deviation in the strong lensing data from the MOND expectations in the work by \citet{tian17}.

\subsubsection{GC systems out of dynamical equilibrium}\label{sec:heating} 
Galaxy groups and clusters are dense environments {where galaxies interact often}. An encounter can dynamically heat the GC population. This would increase the velocity dispersion for a few dynamical times of the GC system until the system relaxes {because of}  phase mixing \citep{mo}. This {explanation of the deviations from our MOND models appears very promising} as we demonstrate on the example of NGC\,4486 (=M\,87), the central galaxy of the Virgo Cluster. The velocity field and galaxy distribution suggest that the center of this cluster is dynamically young and non-relaxed \citep{binggeli87, binggeli93, conselice01, lisker18}. A smaller galaxy cluster centered on M\,86  is passing around the main core of the Virgo Cluster surrounding M\,87. We can roughly estimate how much the velocity dispersion of the GCs around M\,87 increased using the impulse approximation for high-speed encounters; i.e., we assume that the perturber moves with a constant velocity on a linear trajectory and the GCs do not move during the encounter. If the perturbing object has a mass of $M_\mathrm{p}$, a velocity of $v_\mathrm{p}$, and the pericentric distance of the encounter is $b$, then the tidal force increases the velocity of a GC at distance  of $x$ from the center of M\,87 by approximately \citep{mo}
\begin{equation}
\Delta v = \frac{2GM_\mathrm{p}x}{v_\mathrm{p}b^2}.
\label{eq:ia}
\end{equation}
 If we substitute in this equation for $v_\mathrm{p}$ the average velocity dispersion in the Virgo Cluster, 700\,km\,s$^{-1}$ \citep{binggeli93}, the dynamical mass of the M\,86 subclump, $M_\mathrm{p} = 3\times 10^{13}$\,M$_\sun$ \citep{bohringer94}, the current projected distance of M\,87 from M\,86, $d=350$\,kpc, and $x = 50$\,kpc for the characteristic distance of a GC, we obtain $\Delta v = 150$\,km\,s$^{-1}$. This would substantially alleviate the tension between the MOND models and the data{, especially if the pericentric distance was closer than we used in this example.} This is just a rough estimate and \equ{ia} should be adjusted for MOND, but this example already demonstrates{ the potential of galaxy encounters to increase the velocity dispersion of the GCs}. These effects have to be studied using numerical simulations. In MOND this heating mechanism can be important since galaxies can have close encounters without ending in a merger {because the dynamical friction is generally weaker in such configurations  compared to Newtonian gravity with dark matter \citep{nipoti07, tiret08, combtir10,kroupacjp,renaud16,bil18} In other configurations, however the opposite can be the case (see \citealp{nipoti08} for details).} The signs of these encounters on the GC systems probably depend on the particular orbital geometry, masses of the galaxies, and orientation with respect to the observer. A close encounter with a pericenter closer than the extent of the GC system would probably cause asymmetries in the GC spatial distribution and kinematics. During more distant encounters, tidal torques  would cause an increase of angular momentum of the corotating part of the GC population by the mechanism known to form tidal arms in spiral galaxies (see, e.g., \citealp{bournaud10}), which would induce systemic rotation of the GC system. The GC system would relax in a few dynamical times. We can estimate the dynamical time as $2\pi x/\sigma_\mathrm{tot}$, where $\sigma_\mathrm{tot} = 300$\,km\,s$^{-1}$ stands for the typical observed velocity dispersion of the GCs around M\,87 (\fig{g11}). The result is 1\,Gyr. In this time, a galaxy with the average velocity equal to the velocity dispersion in the {Virgo Cluster} passes the path of 700\,kpc. This is more than the observed projected distance to M\,86 and it also is a substantial fraction of the size of the Virgo Cluster, meaning that many other galaxies could have perturbed {the GC system}. Interacting galaxies in MOND can {reach a perhaps surprisingly large separation} in 1\,Gyr, even in sparse environments. In a self-consistent simulation reproducing the orbital history of the Local Group in MOND \citep{bil18}, the Milky Way and Andromeda galaxies receded from the pericentric distance of about 25\,kpc to about 600\,kpc in 1\,Gyr (see Fig.~5 there). Such close encounters probably affect the GC systems of galaxies substantially. Galaxy groups are also possibly an environment susceptible to disturbing the GC systems. They are often less dense than the clusters but the encounters have a lower relative speed.

There is more evidence that the galaxies in our sample experienced interactions apart from the high velocity dispersion of their GCs. We turn again to M\,87 and M\,86. These galaxies have approximately the same luminosity. If the smaller M\,86 subclump increased the velocity dispersion of the GCs of M\,87, then the GCs of M\,86 had to be affected even more. According to \citet{park12}, the velocity dispersion of the GCs of M\,86 is virtually the same as that of M\,87, around 300\,km\,s$^{-1}$. The average recession velocity of the GC system of M\,86 is different from that of the M\,86 core by over 100\,km\,s$^{-1}$, which can be a sign of tidal stripping. The isophotes of  M\,87 and its GC system are elongated approximately toward M\,86 \citep{cote01,pota13}. The galaxy M\,86 moves toward us with respect to the center of M\,87, which is also the case of the GCs at M\,87 at the side closer to M\,86 \citep{cote01,pota13}. Regardless of the hypothesized influence of M\,86, \citet{murphy14} showed that the stellar halo of M\,87 is probably non-relaxed at the distance of 45\,kpc from the galaxy center. Signs of galaxy interactions are detected in the kinematics of GCs and planetary nebulae of M\,87 \citep{romanowsky12, longobardi15}. Some mechanism to explain the high concentration parameter is required for M\,87 in the \lcdm context as well (\fig{g11}).

Numerous signs of encounters exist in NGC\,1399, the second of the two galaxies most problematic for our MOND models. {Many of these signs} have already been summarized in the GC review by \citet{brodie06}. This galaxy is the central galaxy of the Fornax Cluster. It has unusually many GCs for its luminosity while four other nearby galaxies have too few \citep{bassino06a, bekki03}. This suggests that the GCs were transferred from these smaller galaxies to NGC\,1399. Such a transfer can be indeed achieved in simulations \citep{bekki03}. Additionally, the GCs of  NGC\,1399 have asymmetric spatial distribution \citep{bassino06b} and kinematics \citep{richtler04}. Deep optical images of the Fornax Cluster show that the galaxies are connected by bridges of intracluster light and that one of these bridges is followed by an overdensity of the GCs of NGC\,1399 \citep{iodice17}. This proves that the baryonic halo of this galaxy is not relaxed.

Signs of encounters have also been observed in other galaxies where the MOND models did not work very well. The galaxy  NGC\,4365 analyzed in this work still interacts with another galaxy, NGC\,4342 \citep{blom12, blom12b, blom14}. There is an overdensity of GCs between the galaxies, which is aligned to the tidal features in NGC\,4365. It was suggested that the GCs in this overdensity were stripped from NGC\,4365. This would be supported by the fact that NGC\,4365 (=VCC\,731) has an unusually large amount of GCs for its luminosity \citep{peng06}. The {alleged} three-peaked color distribution of the GCs in NGC\,4365 (e.g., \citealp{blom12}) might be a consequence of the presence of GCs coming from NGC\,4342 or another galaxy. The most obvious sign that NGC\,4342 lost its outer parts because of the stripping is its central black hole, which is too massive for the stellar mass of the galaxy. This galaxy is an outlier from the relations between luminosity and effective radius, luminosity and metallicity, and luminosity and the central black hole mass \citep{blom14}. Galaxies with this mass of black hole of $ M_\mathrm{BH} = 10^{8.7}\,M_\sun$ usually have  a total stellar mass of around $10^{11}\,M_\sun$ \citep{reines15,graham16, davis18}. This is a substantial fraction of the stellar mass of NGC\,4365 (approximately $4\times10^{11}$\,M$_\sun$ according to Tables~\ref{tab:tabgals} and~\ref{tab:tabmls}). Interestingly, \citet{bogdan12} excluded the possibility that NGC\,4342 was tidally stripped because the analysis of the X-ray halo implies a high dynamical mass but the dark matter halo should have been stripped preferentially before the stars. This problem is removed by assuming MOND instead of a dark halo.

The fourth most discrepant galaxy, NGC\,5846, possesses an asymmetric X-ray halo with sharp-edged structures that has been suggested to be a result of a galactic interaction \citep{machacek11, paggi17}. A suitable perturber could be the neighboring spiral galaxy NGC\,5850 visible in \fig{g16}. Its strong bar and irregularities in spiral arms indicate that this galaxy interacted recently as well  \citep{higdon98}.

Many signs of a strong interaction in NGC\,5128 (=Cen A), whose RAR deviates noticeably from the RAR of the LTGs (panel (f) of \fig{g15}), are well known. These include the dust band \citep{quillen93}, stellar shells \citep{malin83}, or the twist in the velocity field of the planetary nebulae \citep{peng04}. The rotating disk of satellite galaxies detected at this galaxy by \citet{muller18} might be another sign. The best proposed mechanism to form such structures are galaxy encounters in which the satellites in the disks are tidal dwarf galaxies \citep{kroupa05,pawlowski11,fouquet12,pawlowski15persist,kroupacjp,bil18,banik18}. This would be supported by the fact that the satellites of Cen~A seem to have a too high metallicity for their mass \citep{taylor18} as if it was inherited from their mother galaxy. \citet{malin83} estimated that  the time since the encounter is around 1\,Gyr from the morphology of the dust lane and from the positions of the shells. The dynamical time of the GCs at Cen~A at the galactocentric distance of 30\,kpc is around 1\,Gyr as well.

Galaxy encounters however also increase velocity dispersion in {another way} apart from tidal heating \citep{hilker18}. This mechanism is based on the fact that the baryonic halos of central cluster galaxies are grown by material released from the galaxies that fall on the central galaxy from the outskirts of the galaxy cluster or from the field. Because of the large difference of gravitational potential, the new GCs have a large velocity and increase the velocity dispersion of their new host substantially before the system relaxes dynamically. Employing this mechanism, \citet{hilker18} were able to explain the apparent disagreement of MOND with the stellar kinematics in the central galaxy of the Hydra~I cluster, NGC\,3311. This mechanism {suggests itself particularly for}  the galaxies NGC\,1399 and NGC\,4365  showing multiple signs of capturing GCs of other galaxies.

When a galaxy moves through the galaxy cluster, its outskirts can also get out of  dynamical equilibrium because of the varying external gravitational field via the MOND external field effect. This mechanism was studied recently by simulations \citep{candlish18}. The authors noted that the outskirts of the perturbed galaxy form a one-sided tail pointing in the direction away from the galaxy cluster center.  We can then expect that this mechanism creates lopsided GC systems.

\subsubsection{Other influences}\label{sec:other}
Jeans analysis has a well-known difficulty with the anisotropy parameter. Its value is unknown and can vary within the volume of the galaxy. We noted that the degree of success of the MOND fits correlates with the degree of rotational support of the stellar component of the galaxy and with its rotator criterion number (panels (d) and (e) of \fig{mondconf}). As we concluded in \sect{gcss}, more rotating galaxies have more rotating GC systems.  This suggests that the degree of the rotational support of the GC system drives the success of the MOND fit. Since the support by random motion is weaker in rotating GC systems, a wrong anisotropy parameter or non-ordered motions are less important. This possibly partially explains why the GC systems of the fast rotators could be fitted by our MOND models better and also the results by \citet{rong18} that the kinematics of the stars of the fast rotators agrees better with MOND than that of the slow rotators. We  recalculated our MOND predicted models from \sect{pred} for the $\bneg$, $\blit$, and radial ($\beta = 1$) and nearly tangential ($\beta = -20$) anisotropy parameters. The parameter giving the highest velocity dispersion at large radii came out as nearly tangential. This provided velocity dispersion profiles increased by tens of km\,s$^{-1}$ beyond $2-5\,R_\mathrm{e}$ for all galaxies compared to the profiles calculated for the models with $\biso$. 

Other mechanisms influencing the velocity dispersion include the following: the contamination of the GC population by the GCs moving in the total gravitational field of the cluster; the contamination by the GCs orbiting the satellite galaxies of the investigated galaxies; the presence of binary GCs, just as binary stars increase the observed velocity dispersion in dwarf galaxies (\citealp{dabringhausen16}); the anisotropic spatial distribution of GCs, such as that observed around the Milky Way (e.g., \citealp{pawlowski12,arakelyan18}); the substructures in the GC system resulting from stripping of other galaxies (e.g., like in NGC\,3311, \citealp{hilker18}); or the systematic error introduced by our approximative treatment of the systemic rotation of the GC systems by \equ{mod}.

An error in estimating the galaxy distance using the surface brightness fluctuation method could influence our results. We note that even a several-sigma deviation of the parameter $p_d$ from zero is actually not an extreme deviation in the distance itself. The greatest relative deviation of the best-fit distance to the prior estimate is 25\% for the MOND models  (NGC\,1399) and 11\% for the \lcdm models (NGC\,1407).

We should not forget the modified inertia versions of MOND in which the validity of \equ{mond} might be limited {\citep{milg94b,milgcjp}}. Unfortunately, no fully fledged modified inertia theory of MOND has been constructed yet.

\section{Summary and conclusions}\label{sec:sum}

Gravitational fields of ETGs are difficult to investigate, especially at their outskirts (over a few effective radii, \re, of the galaxy's stars)  where the gravitational fields are expected to deviate substantially from the predictions of Newtonian dynamics without dark matter. The hypotheses proposed to explain the missing mass problem are thus tested poorly in ETGs. In this paper, we aimed to improve the situation and collected archival high-quality measurements of radial velocities of  GCs in a sample of 17 ETGs. There were, in most cases, over 100 GC per galaxy. Such data have the advantage that the GCs often extend over 10\,\re. The GC kinematics was investigated using the Jeans analysis. We focused on testing the predictions of the two most discussed propositions to solve the missing mass problem: the paradigms of the cold dark matter with a cosmological constant (\lcdm) and MOND. 

We investigated the gravitational fields in three ways. 1) We looked for the best-fit parameters of MOND and \lcdm models of the GC kinematics using the Jeans equation and {the maximum a posteriori} fitting method (\sect{fits}). The priors were an estimate of the galaxy distance, its stellar mass-to-light ratio ($M/L$) determined from the galaxy color, an estimate of the galaxy halo mass from the SHMR, and an estimate of the halo concentration from the HMCR. 2) We compared the observed velocity dispersion profiles to the predictions of MOND and \lcdm supposing the galaxies follow the above-mentioned observational constraints and the mean HMCR and SHMR exactly (\sect{pred}). 3) We constructed the RARs of the ETGs using a maximum likelihood approach (\sect{rar}). Together with the real galaxies, we processed in the same way artificial data from a \lcdm simulation of GC formation \citep{renaud17}, which {helped with} the interpretation of the results.

Successful \lcdm fits were found for every galaxy in the terms of the chi-square test with a 5\% significance level. The parameters of the best-fit models are presented in \tab{tablcdm}. The GC systems analyzed in this work are advantageous for determining the halo parameters because the GCs often extend to a galactocentric radii comparable to the scale radii of the NFW dark halos. We compared the fit results with the theoretically predicted SHMR and HMCR (\fig{dmrel}). The reconstructed points indeed show well-defined correlations in the SHMR and HMCR spaces and every individual galaxy agrees with the theoretical relations within the error bars. This might be partly because our fitting method disfavored deviations from the theoretical relations. Nevertheless, our recovered relations are offset from the theoretical relations. As for the HMCR, the real galaxies have a lower concentration for a given halo mass than predicted by \citet{diemer15} for isolated halos. A possible reason might be tidal stripping of the halos since most of our galaxies are not isolated, belonging to galaxy groups or clusters. We then expect the least deviations for the field galaxies and the central galaxies of groups and clusters.  Another explanation might be that the ETGs reside in less concentrated halos than LTGs, i.e., that the offset of the reconstructed HMCR is a selection effect. This could be tested in cosmological hydrodynamical simulations or by a similar analysis of the rotation curves of the LTGs. The available facts nevertheless speak against both of these suggestions (\sect{lcdm}). Comparing the reconstructed SHMR to the theoretical SHMR by \citet{behroozi13}, our galaxies contain too much stellar mass for their halo masses. In addition, the reconstructed SHMR, if expressed as the baryonic fraction (i.e., the stellar mass divided by the halo mass) as a function of the halo mass, is linear and does not show the well-known maximum for the $L_*$ galaxies at the halo mass of around $10^{12}\,$M$_\sun$; however, this conclusion is based on just two data points, NGC\,1023 and NGC\,4494. The reason might again be halo stripping. We have to remember that all free parameters were fitted to the same data, and thus if the priors on one parameter are wrong, then all fitted parameters would be affected. The \lcdm model usually predicts higher-than-observed GC velocity dispersion profiles (\sect{pred}).

All but two galaxies could be fitted well by the MOND models according to the same test. The problematic cases are the central galaxies of the Fornax and Virgo clusters, NGC\,1399 and NGC\,4886, respectively. The success of the MOND fits generally decreases as the characteristics of the galaxies approach those found in the centers of galaxy groups or clusters (\fig{mondconf}) because the GCs have {an observed} velocity dispersion that is too large (\sect{mond}). A similar trend was found for the deviation of the RARs of our ETGs from the RAR of the LTGs, i.e., from the MOND prediction. It would not be surprising if these results were a consequence of a contribution of the additional dark matter that MOND requires in galaxy clusters: this dark matter is known to form cores of characteristic radii of a {few hundreds of} kiloparsecs {\citep{sanders03,takahashi07, milgcbdm}} and the extents of the GC systems are over 100\,kpc for the most problematic galaxies (\sect{unacc}). Additionally, the dynamics of GCs of galaxies in dense environments is likely perturbed by galaxy interactions, which increase the velocity dispersion temporarily (\sect{heating}). Indeed, we found many signatures of recent interactions or out-of-equilibrium GC systems in the most discrepant galaxies. Analytic estimates of the increase of the GC velocity showed that this proposition is promising, but simulations are necessary  for a firmer conclusion. 

We demonstrated that  MOND is able to predict the velocity dispersion profiles in most ETGs better than \lcdm even if the galaxies in the MOND models are approximated by point masses  (\sect{pred}). In the \lcdm context, this probably means that the theoretical SHMR or HMCR we used should be updated for galaxies {similar to those in our sample}, for example, because dark halos are affected by tidal stripping. But then it is remarkable that MOND predicted the necessary direction of the correction.

{Statistical criteria prefer  MOND or \lcdm models in a comparable number of galaxies:} nine cases for MOND versus eight for \lcdm, (see \tab{comp}). The preference is however {stronger} in the cases where \lcdm was preferred; thus \lcdm emerged as the theory more suitable for describing our whole data set. This seems to be in a large part caused by the fact that the \lcdm models are much more flexible than the MOND models. As we demonstrated in \fig{shoot}, if a galaxy is affected by some effect that we did not take into account, such as halo stripping or galaxy interactions, then a \lcdm model is more likely to provide a better fit. This is supported by the fact that the \lcdm fits had a high measure of success in a larger percentage of the galaxies than expected because of the statistical noise and this means that the \lcdm models were even fitting noise in the data.

\begin{acknowledgements}
  We thank Ivana Ebrov\'a, Benedikt Diemer,  and the anonymous referee for their valuable comments.
  
 SS acknowledges the support from the Ministry of Education, Science and Technological Development of the Republic of Serbia through project no.~176021 ``Visible and Invisible Matter in Nearby Galaxies: Theory and Observations''.
  
  FR acknowledges support from the Knut and Alice Wallenberg Foundation.
  
  We acknowledge the usage of the HyperLeda database (\url{http://leda.univ-lyon1.fr}).
  
  This research has made use of "Aladin sky atlas" developed at CDS, Strasbourg Observatory, France. 
  
  The Digitized Sky Surveys were produced at the Space Telescope Science Institute under U.S. Government grant NAG W-2166. The images of these surveys are based on photographic data obtained using the Oschin Schmidt Telescope on Palomar Mountain and the UK Schmidt Telescope. The plates were processed into the present compressed digital form with the permission of these institutions. 
  
\end{acknowledgements}

\bibliographystyle{aa}
\bibliography{citace}

\clearpage
\onecolumn
\begin{landscape}

{
  \begin{longtable}{llcccccccccccccc}
 
\caption{\label{tab:tablcdm} \lcdm fits.}\\ \hline\hline
Name & $\beta$ & $M/L$ & $d$ & $\log \rho_\mathrm{s}$ & $\log \,r_\mathrm{s}$ & $c$ & $\log r_\mathrm{vir}$ & $\log M_\mathrm{vir}$ & $\log M_*$ & $f^\mathrm{DM}_{1R_\mathrm{e}}$ & $f^\mathrm{DM}_{5R_\mathrm{e}}$ & $\chi^2$ & conf. & AICc & LBF \\
& & [$M_\sun / L_\sun$] & [kpc] & [$M_\sun\,$kpc$^{-3}$] & [kpc] & & [kpc] & [$M_\sun$] & [$M_\sun$] & [\%] &[\%] & & [\%] & & \\
\hline
\endfirsthead
\caption{\lcdm fits, continued.}\\\hline\hline
Name & $\beta$ & $M/L$ & $d$ & $\log \rho_\mathrm{s}$ & $\log \,r_\mathrm{s}$ & $c$ & $\log r_\mathrm{vir}$ & $\log M_\mathrm{vir}$ & $\log M_*$ & $f^\mathrm{DM}_{1R_\mathrm{e}}$ & $f^\mathrm{DM}_{5R_\mathrm{e}}$ & $\chi^2$ & conf. & AICc & LBF \\
& & [$M_\sun / L_\sun$] & [kpc] & [$M_\sun\,$kpc$^{-3}$] & [kpc] & & [kpc] & [$M_\sun$] & [$M_\sun$] & [\%] & [\%] & & [\%] & & \\
\hline
\endhead
\hline
\endfoot

N\,821 & iso & $5 \pm 1$  & $24 \pm 2$  & $6.3 \pm 0.5$  & $1.8 \pm 0.3$  & 8.5 & 2.7 & 12.9 & 11.1 & 21 & 71 & 70 & 81 & -51 & 10.7\\
        & neg & $5 \pm 1$  & $24 \pm 2$  & $6.3 \pm 0.5$  & $1.8 \pm 0.3$  & 8.6 & 2.7 & 12.9 & 11.1 & 21 & 70 & 71 & 74 & -50 & 10.6\\
        & lit & $4 \pm 1$  & $24 \pm 2$  & $6.3 \pm 0.3$  & $1.8 \pm 0.3$  & 8.5 & 2.7 & 12.9 & 11.1 & 21 & 71 & 68 & 89 & -51 & 10.8\\
 & & & & & & & & & & & & & & & \\
N\,1023 & iso & $3.5 \pm 0.7$  & $11.0 \pm 0.8$  & $6.2 \pm 0.5$  & $1.5 \pm 0.3$  & 8.0 & 2.4 & 12.1 & 11.0 & 5.2 & 33 & 99 & 37 & -110 & 22.8\\
        & neg & $3.4 \pm 0.7$  & $11.0 \pm 0.8$  & $6.2 \pm 0.5$  & $1.5 \pm 0.3$  & 7.9 & 2.4 & 12.1 & 11.0 & 5.1 & 33 & 98 & 35 & -100 & 22.7\\
        & lit & $3.5 \pm 0.7$  & $11.0 \pm 0.8$  & $6.2 \pm 0.4$  & $1.5 \pm 0.3$  & 8.1 & 2.4 & 12.1 & 11.0 & 5.2 & 33 & 99 & 39 & -110 & 22.8\\
 & & & & & & & & & & & & & & & \\
N\,1399 & iso & $5 \pm 1$  & $20 \pm 1$  & $5.4 \pm 0.2$  & $2.9 \pm 0.2$  & 3.6 & 3.5 & 15.1 & 11.3 & 15 & 69 & 784 & 90 & 440 & -96.5\\
        & neg & $6 \pm 1$  & $20 \pm 1$  & $5.6 \pm 0.2$  & $2.8 \pm 0.2$  & 4.1 & 3.4 & 14.9 & 11.4 & 14 & 68 & 786 & 95 & 430 & -95.0\\
        & lit & $5 \pm 1$  & $19 \pm 1$  & $5.3 \pm 0.2$  & $3.0 \pm 0.2$  & 3.2 & 3.5 & 15.3 & 11.3 & 15 & 71 & 784 & 89 & 460 & -101\\
 & & & & & & & & & & & & & & & \\
N\,1400 & iso & $5 \pm 1$  & $25 \pm 3$  & $6.2 \pm 0.4$  & $1.8 \pm 0.3$  & 7.6 & 2.7 & 12.8 & 11.0 & 13 & 60 & 65 & 87 & -76 & 16.4\\
        & neg & $5 \pm 1$  & $25 \pm 3$  & $6.2 \pm 0.4$  & $1.8 \pm 0.3$  & 7.8 & 2.7 & 12.7 & 11.0 & 13 & 58 & 65 & 89 & -76 & 16.5\\
        & lit & $4 \pm 1$  & $25 \pm 3$  & $6.1 \pm 0.3$  & $1.9 \pm 0.3$  & 7.2 & 2.7 & 13.0 & 11.0 & 14 & 61 & 64 & 82 & -75 & 16.3\\
 & & & & & & & & & & & & & & & \\
N\,1407 & iso & $4 \pm 1$  & $26 \pm 2$  & $5.5 \pm 0.3$  & $2.6 \pm 0.1$  & 4.0 & 3.2 & 14.3 & 11.5 & 28 & 84 & 369 & 90 & 16 & -4.76\\
        & neg & $5 \pm 1$  & $26 \pm 3$  & $5.7 \pm 0.4$  & $2.4 \pm 0.3$  & 4.8 & 3.1 & 14.1 & 11.5 & 28 & 83 & 371 & 95 & 13 & -3.92\\
        & lit & $4.1 \pm 0.9$  & $25 \pm 3$  & $5.4 \pm 0.2$  & $2.7 \pm 0.2$  & 3.3 & 3.2 & 14.5 & 11.5 & 29 & 84 & 368 & 86 & 23 & -6.47\\
 & & & & & & & & & & & & & & & \\
N\,2768 & iso & $5 \pm 1$  & $22 \pm 2$  & $6.2 \pm 0.5$  & $1.9 \pm 0.4$  & 7.5 & 2.7 & 13.0 & 11.4 & 32 & 77 & 109 & 83 & -83 & 17.8\\
        & neg & $6 \pm 2$  & $22 \pm 2$  & $6.2 \pm 0.5$  & $1.8 \pm 0.3$  & 8.0 & 2.7 & 12.7 & 11.4 & 28 & 72 & 112 & 69 & -82 & 17.6\\
        & lit & $5 \pm 1$  & $22 \pm 2$  & $6.1 \pm 0.5$  & $1.9 \pm 0.3$  & 7.1 & 2.8 & 13.1 & 11.3 & 35 & 81 & 107 & 93 & -84 & 18.1\\
 & & & & & & & & & & & & & & & \\
N\,3115 & iso & $6 \pm 1$  & $9.7 \pm 0.4$  & $6.2 \pm 0.4$  & $1.9 \pm 0.3$  & 7.9 & 2.8 & 13.2 & 11.0 & 31 & 81 & 152 & 85 & -110 & 22.5\\
        & neg & $6 \pm 1$  & $9.7 \pm 0.4$  & $6.2 \pm 0.4$  & $1.9 \pm 0.3$  & 8.0 & 2.8 & 13.1 & 11.0 & 29 & 79 & 153 & 79 & -110 & 22.4\\
        & lit & $6 \pm 1$  & $9.7 \pm 0.4$  & $6.2 \pm 0.3$  & $1.9 \pm 0.3$  & 8.0 & 2.8 & 13.2 & 11.0 & 32 & 82 & 151 & 88 & -110 & 22.5\\
 & & & & & & & & & & & & & & & \\
N\,3377 & iso & $4.5 \pm 0.9$  & $11.2 \pm 0.4$  & $6.5 \pm 0.4$  & $1.5 \pm 0.2$  & 10 & 2.5 & 12.1 & 10.6 & 20 & 69 & 125 & 81 & -210 & 46.1\\
        & neg & $5 \pm 1$  & $11.2 \pm 0.4$  & $6.4 \pm 0.4$  & $1.48 \pm 0.07$  & 9.5 & 2.5 & 12.1 & 10.6 & 18 & 66 & 126 & 76 & -210 & 45.8\\
        & lit & $4.3 \pm 0.9$  & $11.2 \pm 0.4$  & $6.5 \pm 0.4$  & $1.5 \pm 0.2$  & 10 & 2.5 & 12.1 & 10.5 & 22 & 70 & 123 & 87 & -220 & 46.2\\
 & & & & & & & & & & & & & & & \\
N\,4278 & iso & $7 \pm 1$  & $16 \pm 1$  & $6.3 \pm 0.4$  & $2.0 \pm 0.3$  & 8.3 & 2.9 & 13.4 & 11.1 & 11 & 59 & 273 & 81 & -150 & 31.1\\
        & neg & $8 \pm 1$  & $16 \pm 1$  & $6.3 \pm 0.4$  & $1.9 \pm 0.3$  & 8.3 & 2.8 & 13.3 & 11.1 & 9.5 & 54 & 277 & 68 & -150 & 30.8\\
        & lit & $6 \pm 1$  & $16 \pm 1$  & $6.2 \pm 0.3$  & $2.0 \pm 0.3$  & 7.7 & 2.9 & 13.5 & 11.0 & 13 & 63 & 270 & 94 & -150 & 31.1\\
 & & & & & & & & & & & & & & & \\
N\,4365 & iso & $7 \pm 2$  & $20 \pm 1$  & $5.6 \pm 0.2$  & $2.7 \pm 0.3$  & 4.4 & 3.3 & 14.7 & 11.5 & 40 & 89 & 246 & 88 & 15 & -4.14\\
        & neg & $9 \pm 2$  & $20 \pm 1$  & $5.6 \pm 0.3$  & $2.6 \pm 0.3$  & 4.3 & 3.3 & 14.6 & 11.6 & 32 & 85 & 252 & 67 & 18 & -4.80\\
        & lit & $6 \pm 1$  & $20 \pm 1$  & $5.6 \pm 0.2$  & $2.7 \pm 0.3$  & 4.2 & 3.4 & 14.8 & 11.4 & 45 & 91 & 242 & 97 & 14 & -4.02\\
 & & & & & & & & & & & & & & & \\
N\,4472 & iso & $5 \pm 1$  & $16.1 \pm 0.7$  & $5.7 \pm 0.2$  & $2.9 \pm 0.2$  & 4.9 & 3.6 & 15.4 & 11.6 & 19 & 72 & 259 & 92 & 180 & -39.5\\
        & neg & $5 \pm 1$  & $16.2 \pm 0.7$  & $5.7 \pm 0.2$  & $2.8 \pm 0.3$  & 4.8 & 3.5 & 15.3 & 11.6 & 17 & 70 & 260 & 95 & 170 & -38.2\\
        & lit & $5 \pm 1$  & $16.1 \pm 0.7$  & $5.8 \pm 0.2$  & $2.9 \pm 0.2$  & 5.0 & 3.6 & 15.5 & 11.6 & 21 & 75 & 259 & 91 & 190 & -41.7\\
 & & & & & & & & & & & & & & & \\
N\,4486 & iso & $5 \pm 1$  & $16 \pm 1$  & $7.0 \pm 0.1$  & $1.8 \pm 0.1$  & 16 & 3.0 & 13.8 & 11.5 & 50 & 88 & 649 & 66 & 540 & -121\\
        & neg & $5 \pm 1$  & $16 \pm 1$  & $7.2 \pm 0.2$  & $1.71 \pm 0.04$  & 19 & 3.0 & 13.7 & 11.5 & 51 & 88 & 653 & 58 & 540 & -121\\
        & lit & $5 \pm 1$  & $16 \pm 1$  & $6.79 \pm 0.08$  & $1.9 \pm 0.1$  & 13 & 3.1 & 13.9 & 11.5 & 47 & 88 & 643 & 77 & 540 & -121\\
 & & & & & & & & & & & & & & & \\
N\,4494 & iso & $2.3 \pm 0.5$  & $16.7 \pm 0.8$  & $6.1 \pm 0.3$  & $1.6 \pm 0.3$  & 7.1 & 2.4 & 12.0 & 10.8 & 13 & 54 & 92 & 39 & -180 & 38.5\\
        & neg & $2.3 \pm 0.5$  & $16.7 \pm 0.8$  & $6.1 \pm 0.4$  & $1.6 \pm 0.2$  & 7.1 & 2.4 & 12.0 & 10.8 & 13 & 53 & 91 & 37 & -180 & 38.4\\
        & lit & $2.3 \pm 0.5$  & $16.7 \pm 0.8$  & $6.1 \pm 0.4$  & $1.6 \pm 0.2$  & 7.0 & 2.4 & 12.1 & 10.8 & 13 & 55 & 93 & 42 & -180 & 38.4\\
 & & & & & & & & & & & & & & & \\
N\,4526 & iso & $4 \pm 1$  & $16 \pm 1$  & $6.1 \pm 0.4$  & $1.9 \pm 0.3$  & 6.9 & 2.7 & 12.9 & 11.0 & 7.6 & 46 & 101 & 75 & -53 & 11.3\\
        & neg & $4 \pm 1$  & $16 \pm 1$  & $6.1 \pm 0.4$  & $1.9 \pm 0.3$  & 6.9 & 2.7 & 12.9 & 11.1 & 7.6 & 46 & 101 & 75 & -53 & 11.3\\
        & lit & $4 \pm 1$  & $16 \pm 1$  & $6.1 \pm 0.4$  & $1.9 \pm 0.3$  & 7.1 & 2.7 & 12.9 & 11.0 & 7.6 & 45 & 102 & 77 & -52 & 11.2\\
 & & & & & & & & & & & & & & & \\
N\,4649 & iso & $5 \pm 1$  & $17 \pm 1$  & $5.7 \pm 0.3$  & $2.5 \pm 0.4$  & 4.6 & 3.2 & 14.3 & 11.5 & 16 & 69 & 422 & 99 & 41 & -9.98\\
        & neg & $6 \pm 1$  & $16.6 \pm 0.6$  & $5.7 \pm 0.4$  & $2.5 \pm 0.4$  & 4.7 & 3.2 & 14.2 & 11.6 & 14 & 64 & 424 & 93 & 41 & -9.98\\
        & lit & $4.7 \pm 0.9$  & $16.4 \pm 0.8$  & $5.6 \pm 0.3$  & $2.6 \pm 0.3$  & 4.4 & 3.3 & 14.5 & 11.4 & 20 & 73 & 422 & 99 & 43 & -10.5\\
 & & & & & & & & & & & & & & & \\
N\,5128 & iso & $2.9 \pm 0.5$  & $4.2 \pm 0.2$  & $5.7 \pm 0.2$  & $2.4 \pm 0.2$  & 4.5 & 3.1 & 14.0 & 11.0 & 43 & 89 & 517 & 71 & -410 & 87.8\\
        & neg & $3.0 \pm 0.5$  & $4.2 \pm 0.2$  & $5.8 \pm 0.3$  & $2.4 \pm 0.2$  & 5.0 & 3.0 & 13.9 & 11.0 & 42 & 89 & 518 & 74 & -420 & 89.4\\
        & lit & $2.8 \pm 0.5$  & $4.2 \pm 0.2$  & $5.6 \pm 0.2$  & $2.5 \pm 0.2$  & 4.1 & 3.2 & 14.2 & 11.0 & 43 & 90 & 516 & 68 & -400 & 85.6\\
 & & & & & & & & & & & & & & & \\
N\,5846 & iso & $5 \pm 1$  & $24 \pm 2$  & $5.7 \pm 0.3$  & $2.5 \pm 0.3$  & 4.9 & 3.2 & 14.2 & 11.3 & 40 & 88 & 201 & 90 & 18 & -4.68\\
        & neg & $6 \pm 1$  & $24 \pm 2$  & $5.9 \pm 0.4$  & $2.3 \pm 0.3$  & 5.8 & 3.1 & 14.0 & 11.4 & 40 & 87 & 204 & 99 & 17 & -4.30\\
        & lit & $5 \pm 1$  & $24 \pm 2$  & $5.7 \pm 0.3$  & $2.5 \pm 0.3$  & 4.5 & 3.2 & 14.3 & 11.3 & 40 & 88 & 200 & 84 & 22 & -5.59\\
 & & & & & & & & & & & & & & & \\\hline
R17x & iso & $3.8 \pm 0.5$  & $18 \pm 2$  & $6.1 \pm 0.4$  & $1.5 \pm 0.3$  & 7.3 & 2.4 & 11.9 & 10.8 & 2.6 & 19 & 192 & 73 & -270 & 57.5\\
        & neg & $3.7 \pm 0.4$  & $18 \pm 2$  & $6.1 \pm 0.4$  & $1.6 \pm 0.3$  & 7.0 & 2.4 & 11.9 & 10.7 & 2.6 & 19 & 191 & 69 & -270 & 57.6\\
        & lit & $4.0 \pm 0.5$  & $18 \pm 2$  & $6.2 \pm 0.4$  & $1.5 \pm 0.3$  & 7.5 & 2.4 & 12.0 & 10.8 & 2.7 & 20 & 193 & 76 & -270 & 57.6\\
 & & & & & & & & & & & & & & & \\
R17y & iso & $3.9 \pm 0.5$  & $18 \pm 2$  & $6.1 \pm 0.4$  & $1.6 \pm 0.2$  & 7.3 & 2.4 & 12.1 & 10.8 & 3.7 & 24 & 191 & 72 & -250 & 55.0\\
        & neg & $3.9 \pm 0.5$  & $18 \pm 2$  & $6.1 \pm 0.4$  & $1.6 \pm 0.2$  & 7.0 & 2.4 & 12.0 & 10.8 & 3.4 & 22 & 191 & 71 & -250 & 53.9\\
        & lit & $3.9 \pm 0.5$  & $18 \pm 2$  & $6.2 \pm 0.4$  & $1.6 \pm 0.2$  & 7.6 & 2.5 & 12.1 & 10.8 & 4.1 & 26 & 191 & 73 & -260 & 55.4\\
 & & & & & & & & & & & & & & & \\
R17z & iso & $3.7 \pm 0.5$  & $18 \pm 2$  & $6.3 \pm 0.4$  & $1.5 \pm 0.2$  & 8.2 & 2.4 & 11.9 & 10.7 & 4.9 & 28 & 191 & 69 & -280 & 59.8\\
        & neg & $3.7 \pm 0.5$  & $18 \pm 2$  & $6.2 \pm 0.4$  & $1.5 \pm 0.3$  & 7.8 & 2.4 & 11.9 & 10.7 & 4.4 & 26 & 190 & 67 & -280 & 59.5\\
        & lit & $3.7 \pm 0.5$  & $18 \pm 2$  & $6.3 \pm 0.4$  & $1.5 \pm 0.2$  & 8.5 & 2.4 & 12.0 & 10.7 & 5.5 & 31 & 192 & 72 & -280 & 59.9\\

\end{longtable}
\tablefoot{$\boldsymbol{\beta}$ -- Assumed anisotropy parameter; see \sect{jeans} for explanation.
  $\boldsymbol{M/L}$, $\boldsymbol{d}$, $\boldsymbol{\rho_\mathrm{s}}$, $\boldsymbol{r_\mathrm{s}}$, $\boldsymbol{c}$, $\boldsymbol{r_\mathrm{vir}}$, $\boldsymbol{M_\mathrm{vir}}$, $\boldsymbol{M_*}$ -- Best-fit parameters of the \lcdm model; see \sect{jeans}. 
  {$\boldsymbol{f^\mathrm{DM}_{1R_\mathrm{e}}}$, $\boldsymbol{f^\mathrm{DM}_{5R_\mathrm{e}}}$ -- Average dark matter fractions below 1 and $5R_\mathrm{e}$, respectively.}
  $\boldsymbol{\chi^2}$, \textbf{conf.}, \textbf{AICc}, \textbf{LBF} -- Statistical measures of the quality of the best fit; see \sect{fits}.  
  }

}
\end{landscape}

\onecolumn
\begin{longtable}{llccccccc}
  
  \caption{\label{tab:tabmond} MOND fits}\\ \hline\hline
  Name & $\beta$ & $M/L$ & $d$ & $\log M_*$ & $\chi^2$ & conf. & AICc & LBF \\ 
  & & [$M_\sun / L_\sun$] & [kpc]  & [$M_\sun$] & & [\%] & & \\
  \hline
  \endfirsthead
  \caption{MOND fits, continued.}\\\hline\hline
  Name & $\beta$ & $M/L$ & $d$ & $\log M_*$ & $\chi^2$ & conf. & AICc & LBF \\ 
  & & [$M_\sun / L_\sun$] & [kpc]  & [$M_\sun$] & & [\%] & & \\
  \hline
  \endhead
  \hline
  \endfoot 
  
 N\,821 & iso & $6 \pm 1$  & $25 \pm 2$  & 11.3 & 75 & 52 & -56 & 11.4\\
        & neg & $6 \pm 1$  & $25 \pm 2$  & 11.3 & 74 & 53 & -56 & 11.4\\
        & lit & $6 \pm 1$  & $25 \pm 2$  & 11.3 & 75 & 52 & -56 & 11.4\\
 & & & & & & & & \\
N\,1023 & iso & $3.2 \pm 0.7$  & $11.0 \pm 0.8$  & 11.0 & 90 & 13 & -110 & 22.9\\
        & neg & $3.2 \pm 0.7$  & $10.9 \pm 0.8$  & 10.9 & 90 & 12 & -110 & 22.8\\
        & lit & $3.3 \pm 0.7$  & $11.0 \pm 0.8$  & 11.0 & 91 & 13 & -110 & 23.0\\
 & & & & & & & & \\
N\,1399 & iso & $42 \pm 5$  & $25 \pm 2$  & 12.5 & 922 & 0.15 & 590 & -226\\
        & neg & $37 \pm 4$  & $25 \pm 2$  & 12.4 & 909 & 0.39 & 560 & -195\\
        & lit & $53 \pm 6$  & $26 \pm 2$  & 12.6 & 945 & 0.021 & 660 & -294\\
 & & & & & & & & \\
N\,1400 & iso & $5 \pm 1$  & $27 \pm 3$  & 11.1 & 70 & 80 & -79 & 16.6\\
        & neg & $5 \pm 1$  & $27 \pm 3$  & 11.1 & 68 & 91 & -80 & 16.8\\
        & lit & $6 \pm 1$  & $28 \pm 3$  & 11.2 & 72 & 64 & -77 & 16.1\\
 & & & & & & & & \\
N\,1407 & iso & $11 \pm 2$  & $35 \pm 3$  & 12.2 & 407 & 23 & 42 & -10.3\\
        & neg & $10 \pm 2$  & $34 \pm 3$  & 12.1 & 402 & 30 & 30 & -7.54\\
        & lit & $12 \pm 2$  & $36 \pm 4$  & 12.2 & 413 & 15 & 61 & -14.5\\
 & & & & & & & & \\
N\,2768 & iso & $6 \pm 1$  & $22 \pm 2$  & 11.4 & 106 & 97 & -91 & 19.0\\
        & neg & $6 \pm 1$  & $22 \pm 2$  & 11.4 & 105 & 95 & -90 & 18.8\\
        & lit & $6 \pm 1$  & $23 \pm 2$  & 11.4 & 108 & 89 & -91 & 19.1\\
 & & & & & & & & \\
N\,3115 & iso & $8 \pm 1$  & $9.8 \pm 0.4$  & 11.1 & 164 & 38 & -110 & 22.5\\
        & neg & $8 \pm 1$  & $9.8 \pm 0.4$  & 11.1 & 162 & 46 & -110 & 22.6\\
        & lit & $8 \pm 1$  & $9.8 \pm 0.4$  & 11.1 & 167 & 32 & -110 & 22.3\\
 & & & & & & & & \\
N\,3377 & iso & $4.7 \pm 0.8$  & $11.3 \pm 0.5$  & 10.6 & 127 & 68 & -220 & 46.3\\
        & neg & $4.7 \pm 0.7$  & $11.3 \pm 0.5$  & 10.6 & 126 & 74 & -220 & 46.1\\
        & lit & $4.7 \pm 0.8$  & $11.3 \pm 0.5$  & 10.6 & 128 & 64 & -220 & 46.3\\
 & & & & & & & & \\
N\,4278 & iso & $12 \pm 2$  & $18 \pm 1$  & 11.4 & 301 & 17 & -140 & 29.3\\
        & neg & $12 \pm 2$  & $18 \pm 1$  & 11.4 & 297 & 23 & -140 & 30.4\\
        & lit & $13 \pm 2$  & $18 \pm 2$  & 11.5 & 308 & 9.7 & -130 & 27.1\\
 & & & & & & & & \\
N\,4365 & iso & $18 \pm 2$  & $23 \pm 2$  & 12.0 & 294 & 3.0 & 44 & -10.6\\
        & neg & $17 \pm 2$  & $23 \pm 2$  & 12.0 & 288 & 5.5 & 36 & -8.85\\
        & lit & $19 \pm 3$  & $23 \pm 2$  & 12.1 & 302 & 1.2 & 59 & -14.0\\
 & & & & & & & & \\
N\,4472 & iso & $20 \pm 2$  & $16.9 \pm 0.8$  & 12.3 & 313 & 3.5 & 220 & -50.9\\
        & neg & $18 \pm 2$  & $16.8 \pm 0.8$  & 12.2 & 305 & 7.3 & 200 & -45.7\\
        & lit & $25 \pm 3$  & $17.0 \pm 0.7$  & 12.4 & 326 & 0.96 & 240 & -61.1\\
 & & & & & & & & \\
N\,4486 & iso & $31 \pm 3$  & $20 \pm 1$  & 12.5 & 738 & 0.50 & 610 & -149\\
        & neg & $29 \pm 3$  & $19 \pm 1$  & 12.4 & 732 & 0.83 & 590 & -142\\
        & lit & $32 \pm 4$  & $20 \pm 1$  & 12.5 & 748 & 0.23 & 630 & -157\\
 & & & & & & & & \\
N\,4494 & iso & $2 \pm 1$  & $16.6 \pm 0.8$  & 10.7 & 83 & 13 & -180 & 38.0\\
        & neg & $2 \pm 1$  & $16.6 \pm 0.8$  & 10.7 & 82 & 11 & -180 & 37.9\\
        & lit & $2.0 \pm 0.4$  & $16.6 \pm 0.8$  & 10.7 & 84 & 15 & -180 & 38.0\\
 & & & & & & & & \\
N\,4526 & iso & $5 \pm 1$  & $17 \pm 1$  & 11.1 & 105 & 96 & -57 & 11.7\\
        & neg & $5 \pm 1$  & $17 \pm 1$  & 11.1 & 104 & 92 & -57 & 11.7\\
        & lit & $5 \pm 1$  & $17 \pm 1$  & 11.1 & 106 & 100 & -57 & 11.6\\
 & & & & & & & & \\
N\,4649 & iso & $9 \pm 1$  & $18 \pm 1$  & 11.8 & 467 & 14 & 50 & -12.0\\
        & neg & $9 \pm 1$  & $17 \pm 1$  & 11.8 & 454 & 27 & 44 & -10.6\\
        & lit & $9 \pm 1$  & $18 \pm 1$  & 11.8 & 487 & 3.2 & 61 & -14.4\\
 & & & & & & & & \\
N\,5128 & iso & $6.9 \pm 0.8$  & $4.5 \pm 0.3$  & 11.4 & 571 & 20 & -360 & 76.6\\
        & neg & $6.3 \pm 0.7$  & $4.5 \pm 0.3$  & 11.4 & 563 & 30 & -380 & 80.7\\
        & lit & $7.6 \pm 0.9$  & $4.6 \pm 0.3$  & 11.5 & 581 & 12 & -330 & 71.1\\
 & & & & & & & & \\
N\,5846 & iso & $17 \pm 3$  & $29 \pm 2$  & 12.0 & 246 & 4.8 & 39 & -9.28\\
        & neg & $16 \pm 3$  & $28 \pm 2$  & 12.0 & 243 & 6.9 & 31 & -7.63\\
        & lit & $18 \pm 3$  & $29 \pm 2$  & 12.0 & 252 & 2.8 & 50 & -11.7\\
 & & & & & & & & \\\hline
R17x & iso & $3.5 \pm 0.5$  & $18 \pm 2$  & 10.7 & 185 & 49 & -270 & 57.2\\
        & neg & $3.4 \pm 0.5$  & $17 \pm 2$  & 10.7 & 184 & 45 & -270 & 57.2\\
        & lit & $3.6 \pm 0.5$  & $18 \pm 2$  & 10.7 & 187 & 54 & -270 & 57.3\\
 & & & & & & & & \\
R17y & iso & $3.6 \pm 0.5$  & $18 \pm 2$  & 10.7 & 185 & 50 & -260 & 54.7\\
        & neg & $3.6 \pm 0.5$  & $18 \pm 2$  & 10.7 & 183 & 46 & -250 & 53.4\\
        & lit & $3.5 \pm 0.5$  & $18 \pm 2$  & 10.7 & 186 & 54 & -260 & 55.4\\
 & & & & & & & & \\
R17z & iso & $3.3 \pm 0.5$  & $18 \pm 2$  & 10.7 & 185 & 48 & -280 & 59.6\\
        & neg & $3.3 \pm 0.5$  & $17 \pm 2$  & 10.7 & 184 & 44 & -280 & 59.0\\
        & lit & $3.3 \pm 0.5$  & $18 \pm 2$  & 10.7 & 187 & 53 & -280 & 59.9\\

\end{longtable}
\tablefoot{$\boldsymbol{\beta}$ -- Assumed anisotropy parameter; see \sect{jeans} for explanation.
  $\boldsymbol{M/L}$, $\boldsymbol{d}$, $\boldsymbol{M_*}$ -- Best-fit parameters of the MOND model; see \sect{jeans}. 
  $\boldsymbol{\chi^2}$, \textbf{conf.}, \textbf{AICc}, \textbf{LBF} -- Statistical measures of the quality of the best fit; see \sect{fits}.  
}
\twocolumn

\begin{appendix}

\section{Test of recovering of free parameters}
\label{app:test}

\begin{figure}[h!]
  \resizebox{\hsize}{!}{\includegraphics{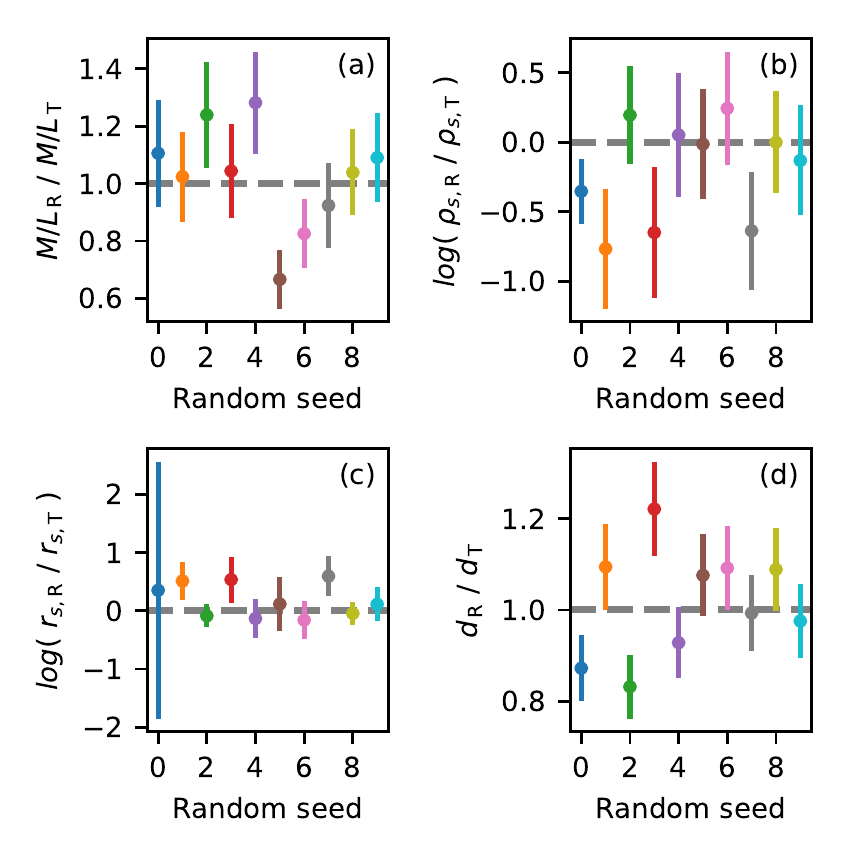}}
  \caption{Test confirming that we can reconstruct the free parameters of the gravitational potential correctly from artificial data. The index ``T'' denotes the true value of the free parameter and ``R'' the reconstructed value.} 
  \label{fig:cfit}
\end{figure}

We tested our method and codes for reconstructing the parameters of the gravitational potential from the GC radial velocities by applying these {codes} on artificial data generated especially for this purpose. We assumed that we have artificial observations of  10 galaxies for which the following independent estimates are available: distance of 20\,Mpc, luminosity of $2\times10^{12}$\,L$_\sun$,  $M/L$ of 5, effective radius of 3\,kpc, and a S\'ersic index of 4. In order to take the observational errors into account in our test, we randomly assigned each artificial galaxy the ``real'' parameters so that the $p_i$ parameters followed the statistical distributions described in \sect{fits} with $\Delta p_{M/L} = 0.15$ and $\Delta p_d = 0.04$. This defines the real parameters of a galaxy: its distance, $M/L$, and the halo parameters. Then we could generate a GC system for every galaxy. Its number density was always proportional to $r^{-3.5}$ and it contained 200 members extending from 0.1 to 5 effective radii of the galaxy. The radial velocities were generated with a Gaussian distribution with a velocity dispersion calculated at every radius according to the solution to \equs{jeans}{siglos}. We assumed a zero anisotropy and no systemic rotation. We applied our code to these data and compared the recovered and real parameters. This is shown in \fig{cfit}. In this figure the index T stands for the true value and the index R for the recovered value. We can see that the recovered values are always correct at least by an order of magnitude and that the estimated uncertainty limits are reasonable.

\section{Results for individual galaxies}
\label{app:plots}
Panels (a): Black points indicate galactocentric radii and absolute values of the radial velocities of the GCs. The systemic velocity of the whole GC system was subtracted. The galactocentric radius is expressed on the horizontal scale in the units of arcminutes, effective radii of the starlight of the galaxy, and kiloparsecs (assuming the mean distance from \tab{tabgcss}). The green line and band represent the running local line-of-sight velocity dispersion and its uncertainty (\sect{gcss}). The thick gray line indicates the line-of-sight velocity dispersion profile implied by the best-fit MOND model. The thick black line  represents the same for \lcdm. The dotted red line indicates the velocity dispersion profile  predicted by MOND (\sect{pred}). The dotted blue line indicates the same for \lcdm. For the galaxies NGC\,1407 and NGC\,4472, the \lcdm prediction lies outside the plot. The orange dotted lines in \figs{g17}{g19} (R17x,y,z)  represent the line-of-sight velocity dispersion profile calculated for a zero anisotropy and the correct parameters that these simulated galaxies have (\sect{res}). The red arrows indicate the radii where the acceleration calculated from the distribution of the stars in the Newtonian way ($a_\mathrm{N}$) reaches the quoted value (assuming the independent estimates of the galaxy distance and mass-to-light ratio).

Panels (b)--(e): Comparison of the $p_i$ parameters  expected for the galaxies from independent estimates  (that we used as priors in \sect{fits}) with the values we obtained by fitting the GC kinematics. The filled small black points with error bars represent the means and dispersions of the priors. The open gray points indicate the parameters of the best-fit MOND models. The open black points denote the same for the \lcdm models.

Panels (f): Gray band  indicates the RAR for the LTGs reported by \citet{mcgaugh16} and its scatter. The colored points with error bars denote the RAR for our ETGs derived in \sect{rar}. The color indicates the assumed anisotropy type (see \sect{jeans}).

Panels (g): Image of the galaxy (prepared using the Aladin software \citep{aladin1,aladin2}; the images come from the Digitized Sky Survey). North is up and east is left. The numbers stated in the middle of the bottom side show the size of the FOV. We chose the FOV so that its size is 3-4 times larger than the galactocentric distance of the furthermost GC. The size was increased (decreased) if there was (was not) another big galaxy nearby. In \fig{g11} (NGC\,4486), nearly all visible objects are galaxies.

Panels (h) and (i) show the radial profiles of the skewness ($s_3$) and kurtosis ($s_4$) parameters and their uncertainties, respectively (see \sect{gcss}). The plotted radial range matches that in the panel (a). The horizontal dashed lines indicate the zero value.

\begin{figure*}
 \centering
 \includegraphics[width=17cm]{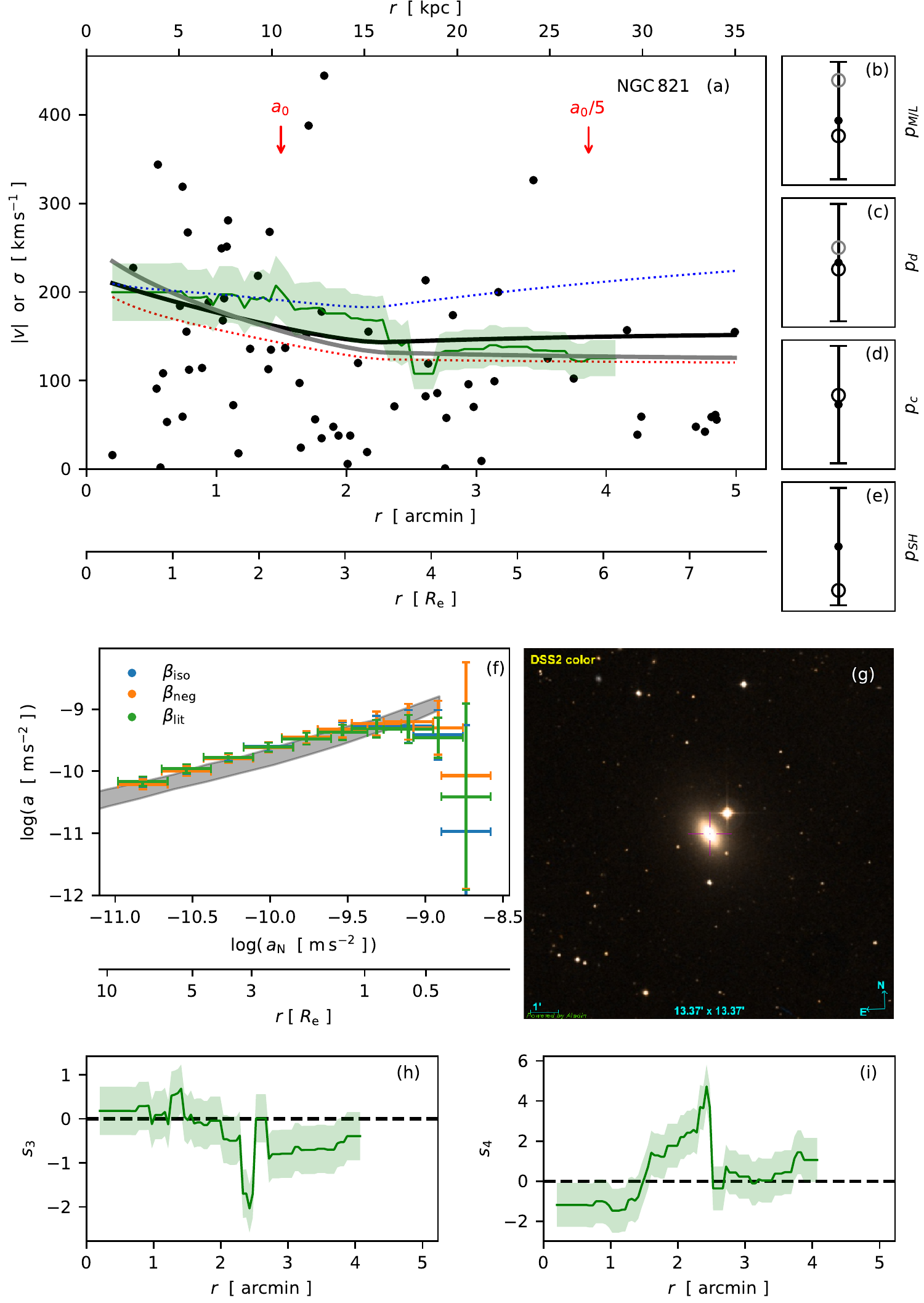}
 \caption{\protect\input{"./tables/latex/galaxycapt0.tex"}}
 \label{fig:g0}
  \end{figure*} 
\begin{figure*}
 \centering
 \includegraphics[width=17cm]{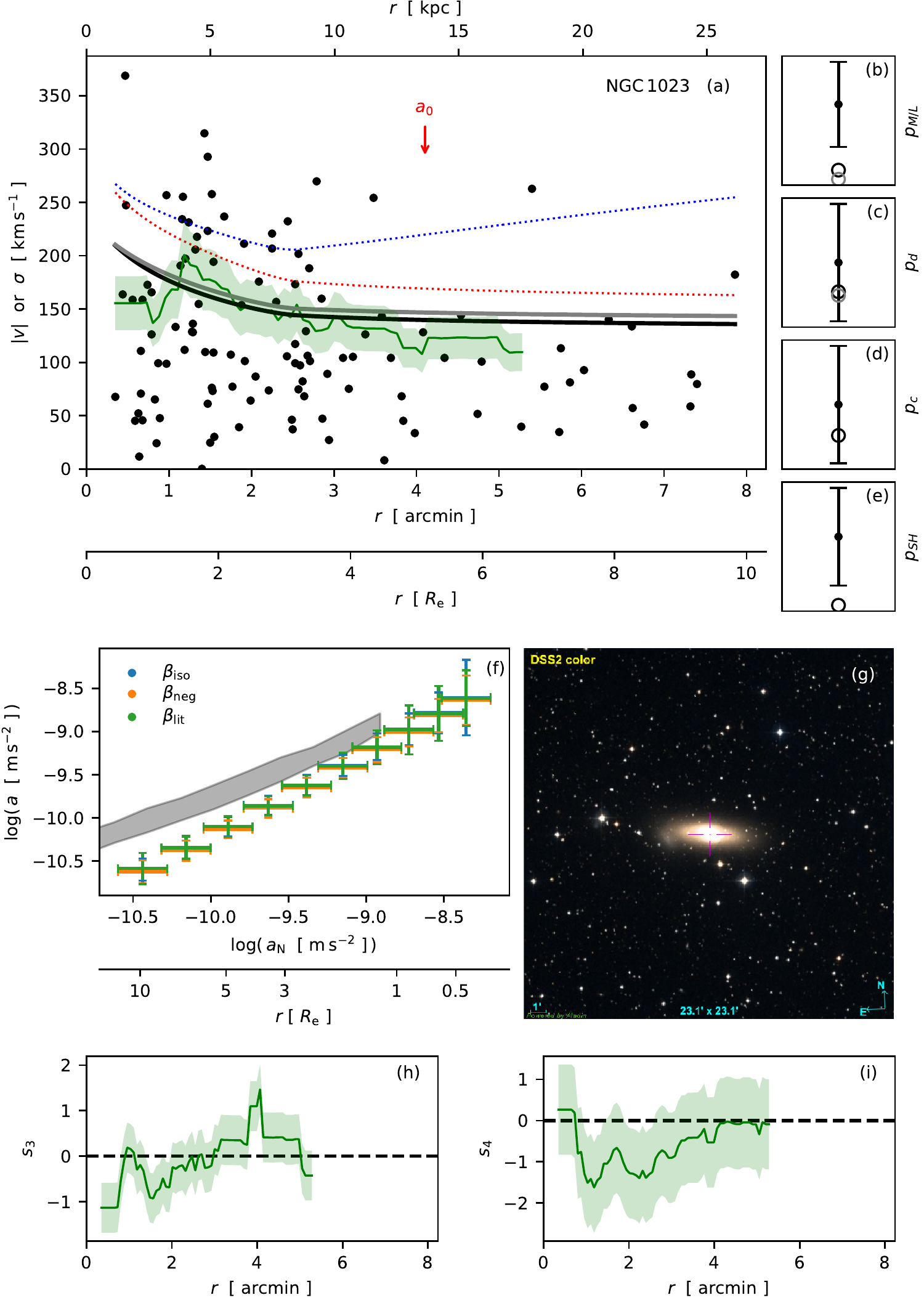}
 \caption{\protect\input{"./tables/latex/galaxycapt1.tex"}}
 \label{fig:g1}
  \end{figure*}
\begin{figure*}
 \centering
 \includegraphics[width=17cm]{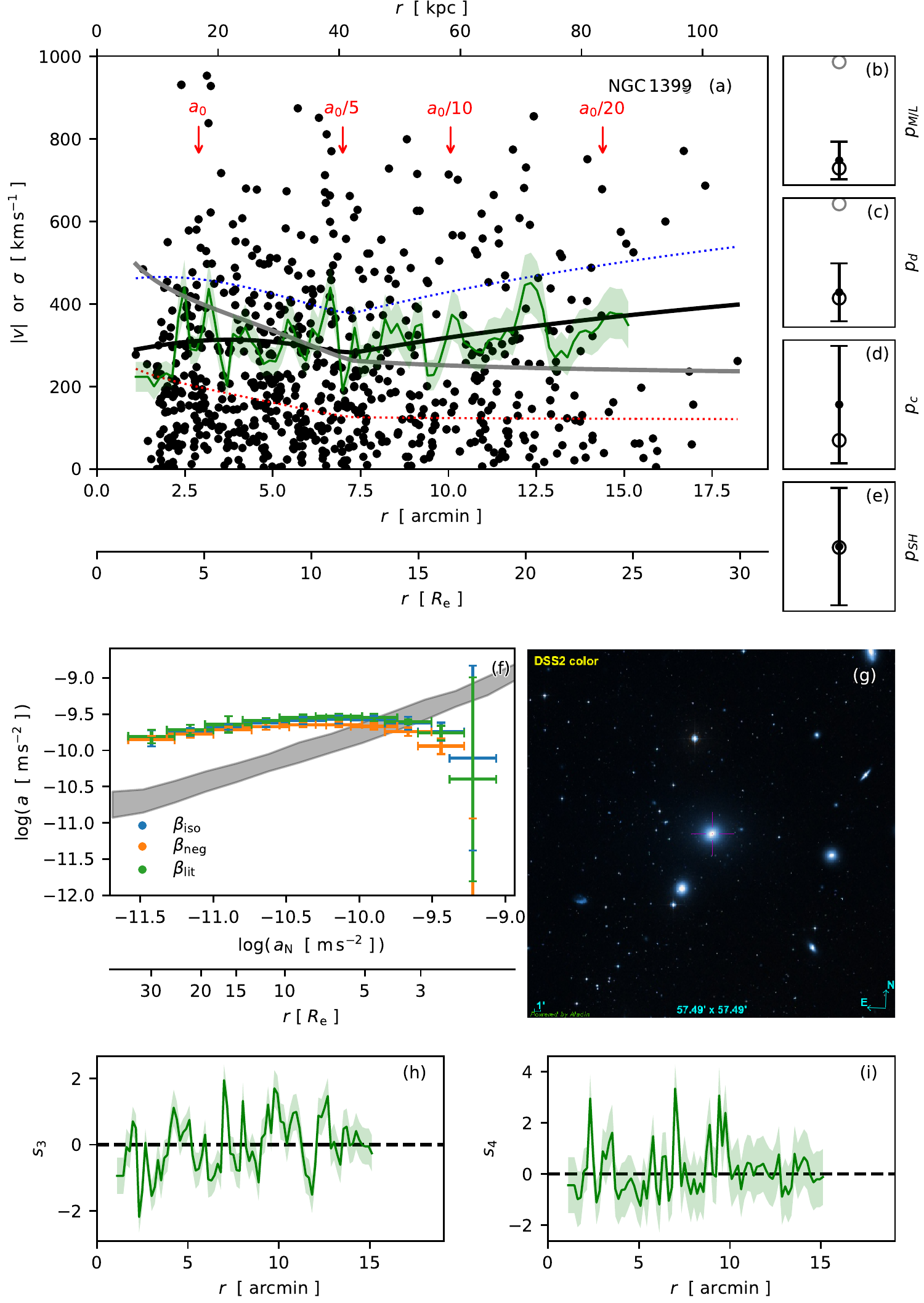}
 \caption{\protect\input{"./tables/latex/galaxycapt2.tex"}}
 \label{fig:g2}
  \end{figure*}
\begin{figure*}
 \centering
 \includegraphics[width=17cm]{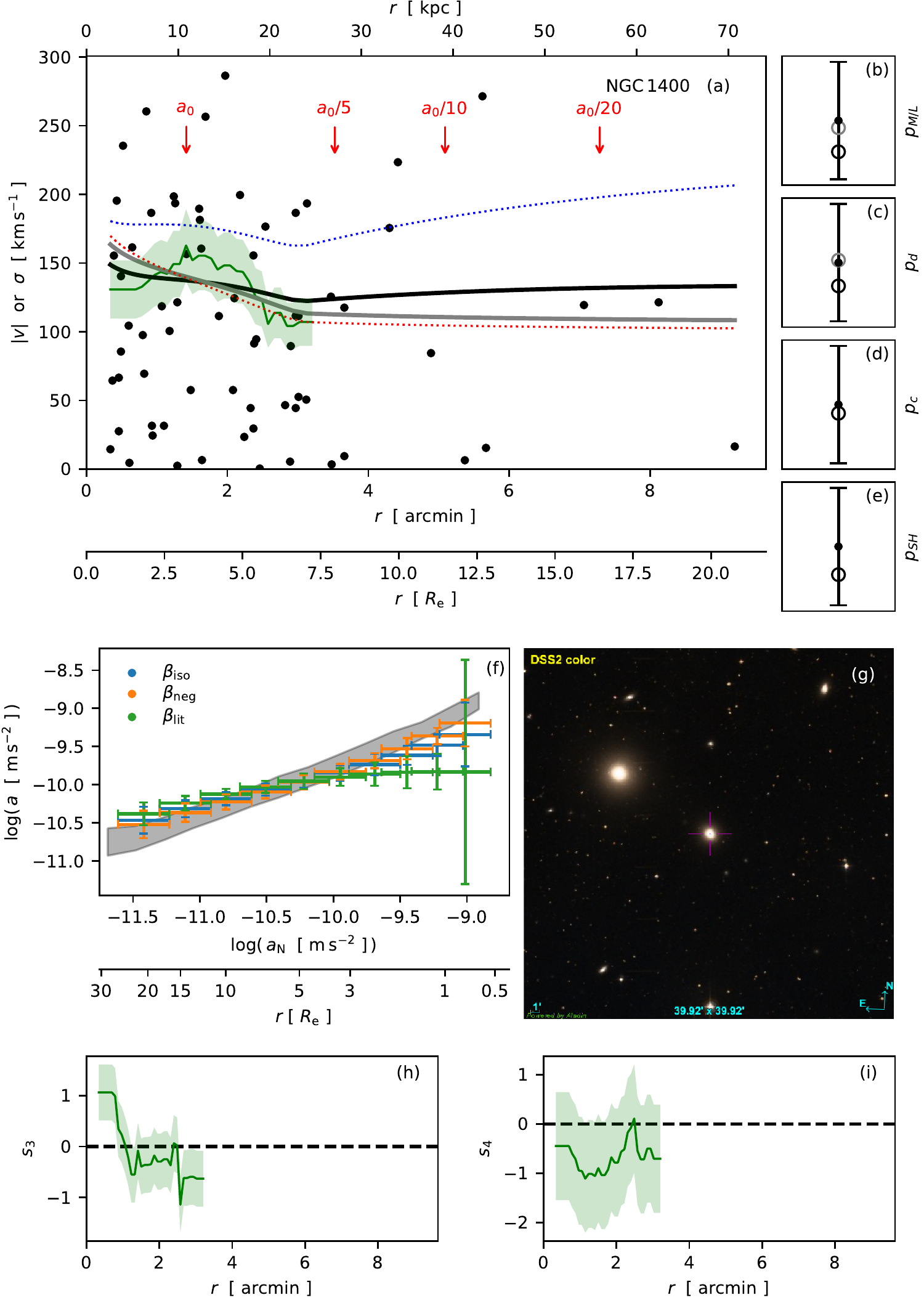}
 \caption{\protect\input{"./tables/latex/galaxycapt3.tex"}}
 \label{fig:g3}
  \end{figure*}
\begin{figure*}
 \centering
 \includegraphics[width=17cm]{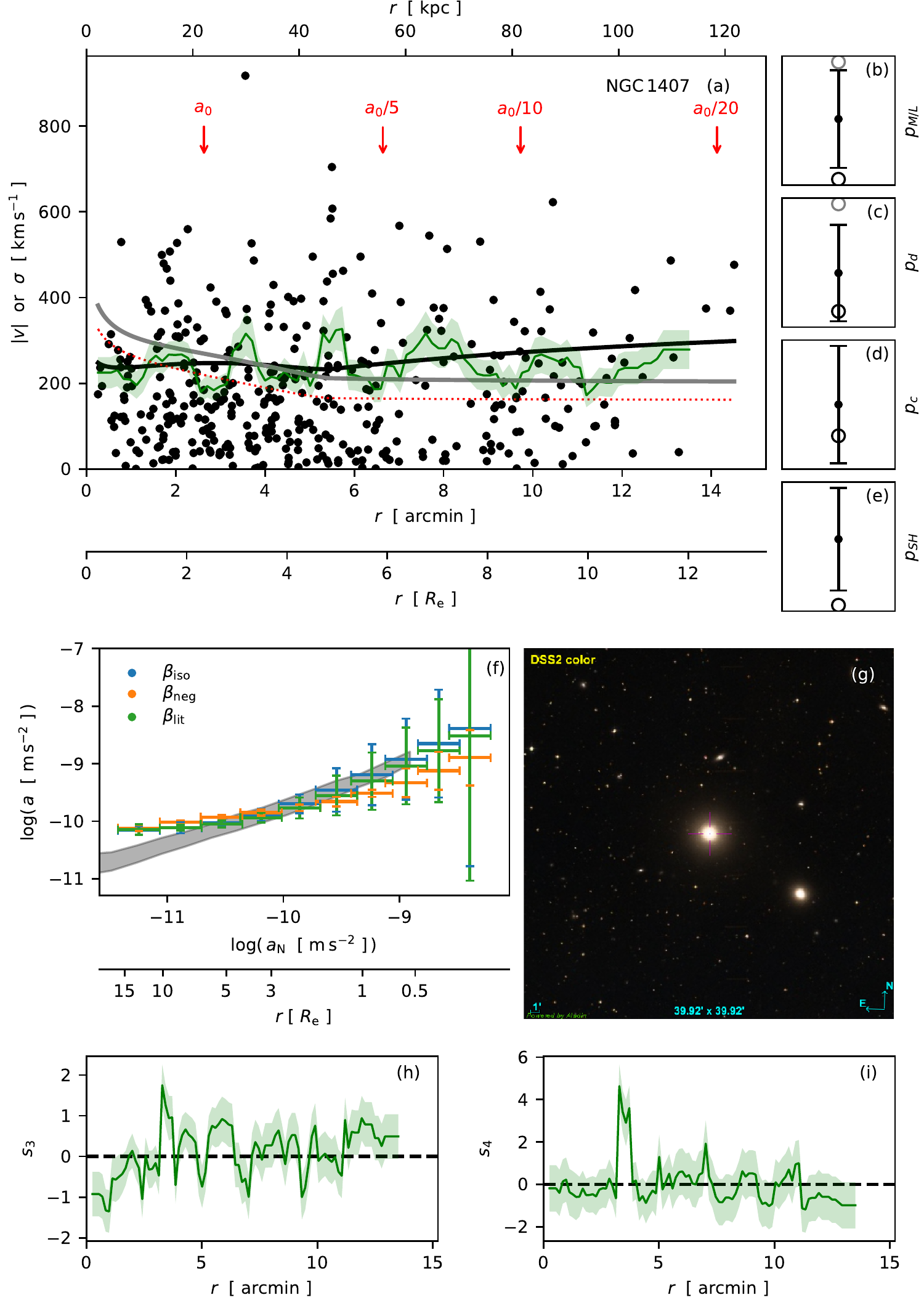}
 \caption{\protect\input{"./tables/latex/galaxycapt4.tex"}}
 \label{fig:g4}
  \end{figure*}
\begin{figure*}
 \centering
 \includegraphics[width=17cm]{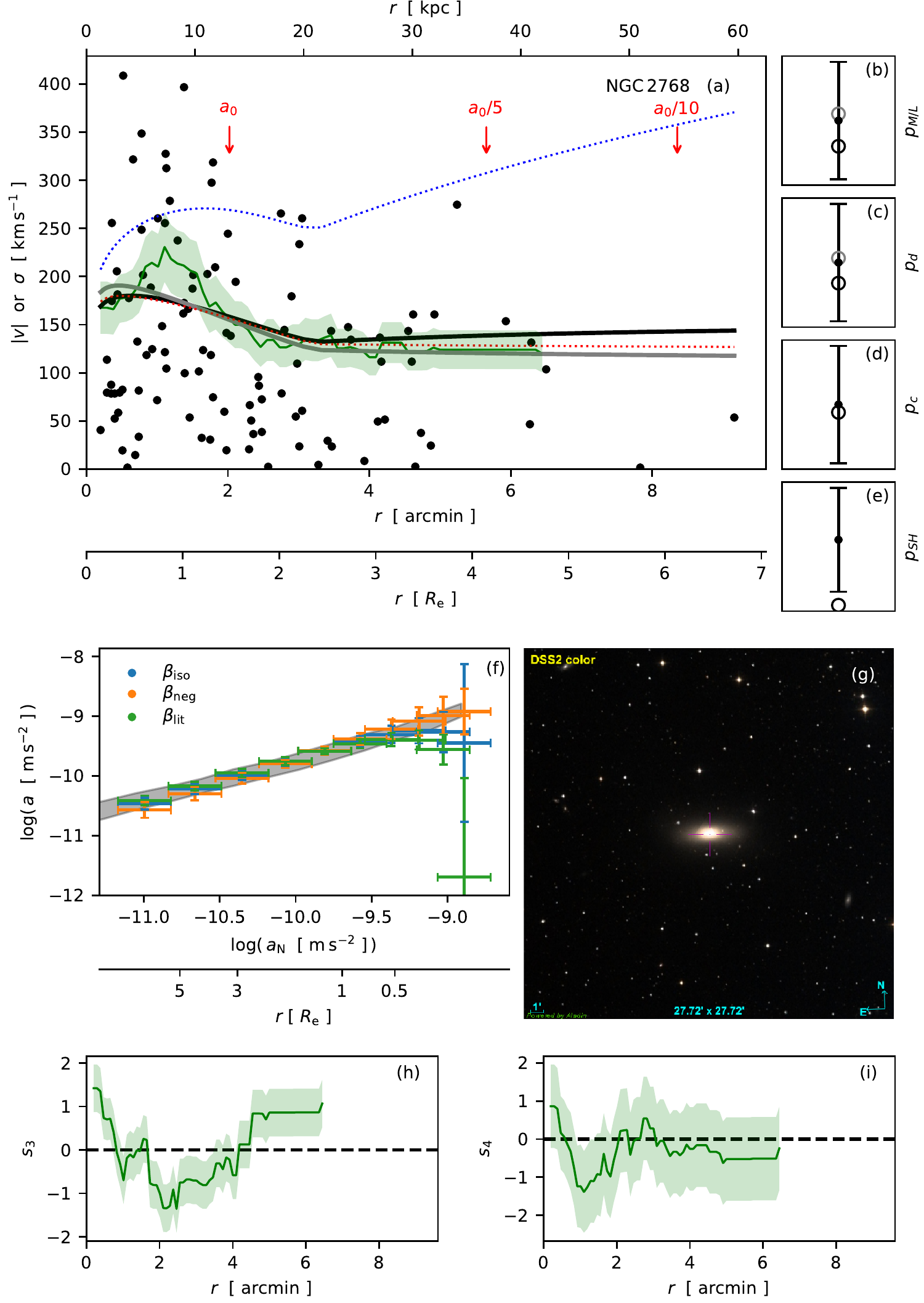}
 \caption{\protect\input{"./tables/latex/galaxycapt5.tex"}}
 \label{fig:g5}
  \end{figure*}
\begin{figure*}
 \centering
 \includegraphics[width=17cm]{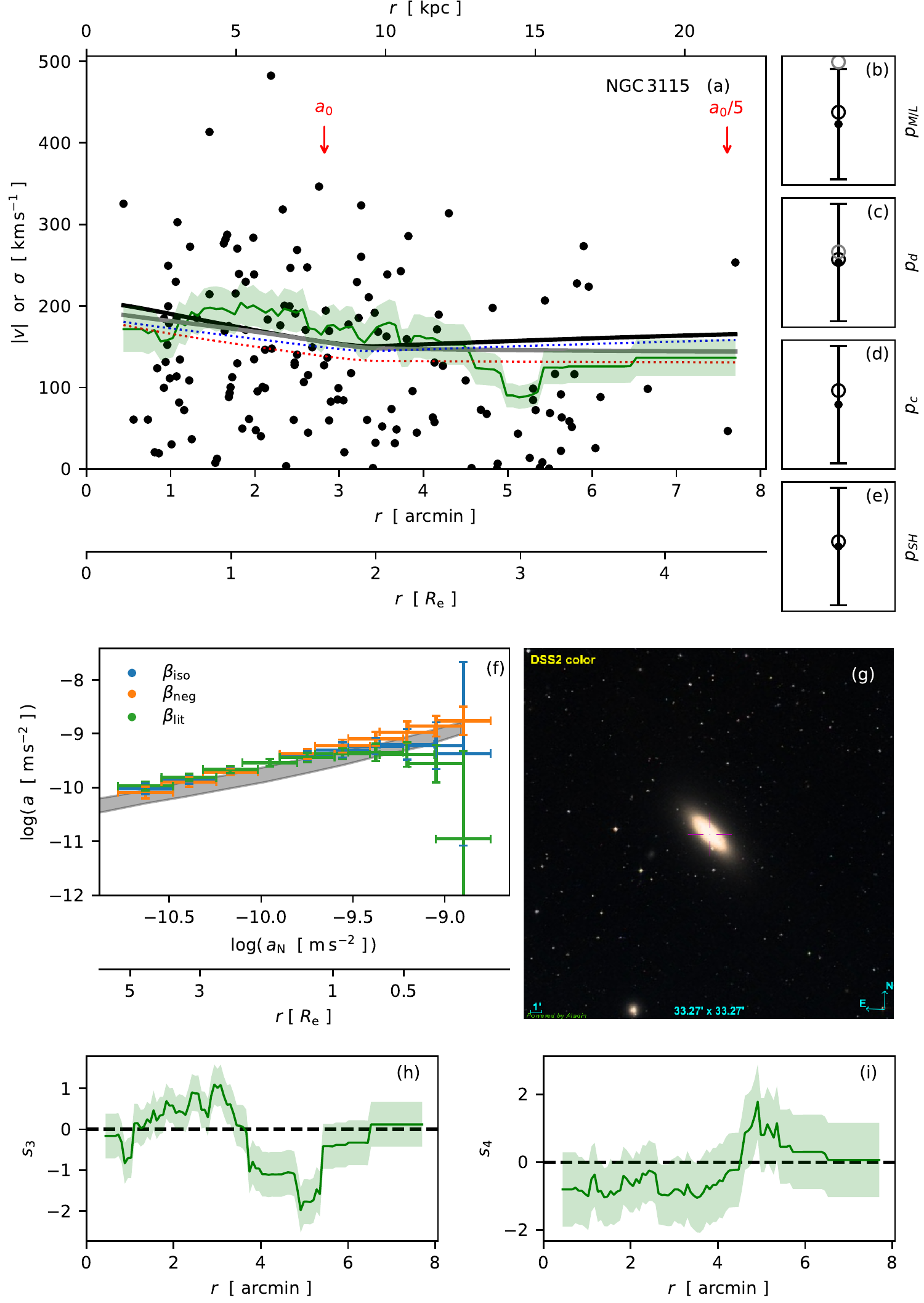}
 \caption{\protect\input{"./tables/latex/galaxycapt6.tex"}}
 \label{fig:g6}
  \end{figure*}
\begin{figure*}
 \centering
 \includegraphics[width=17cm]{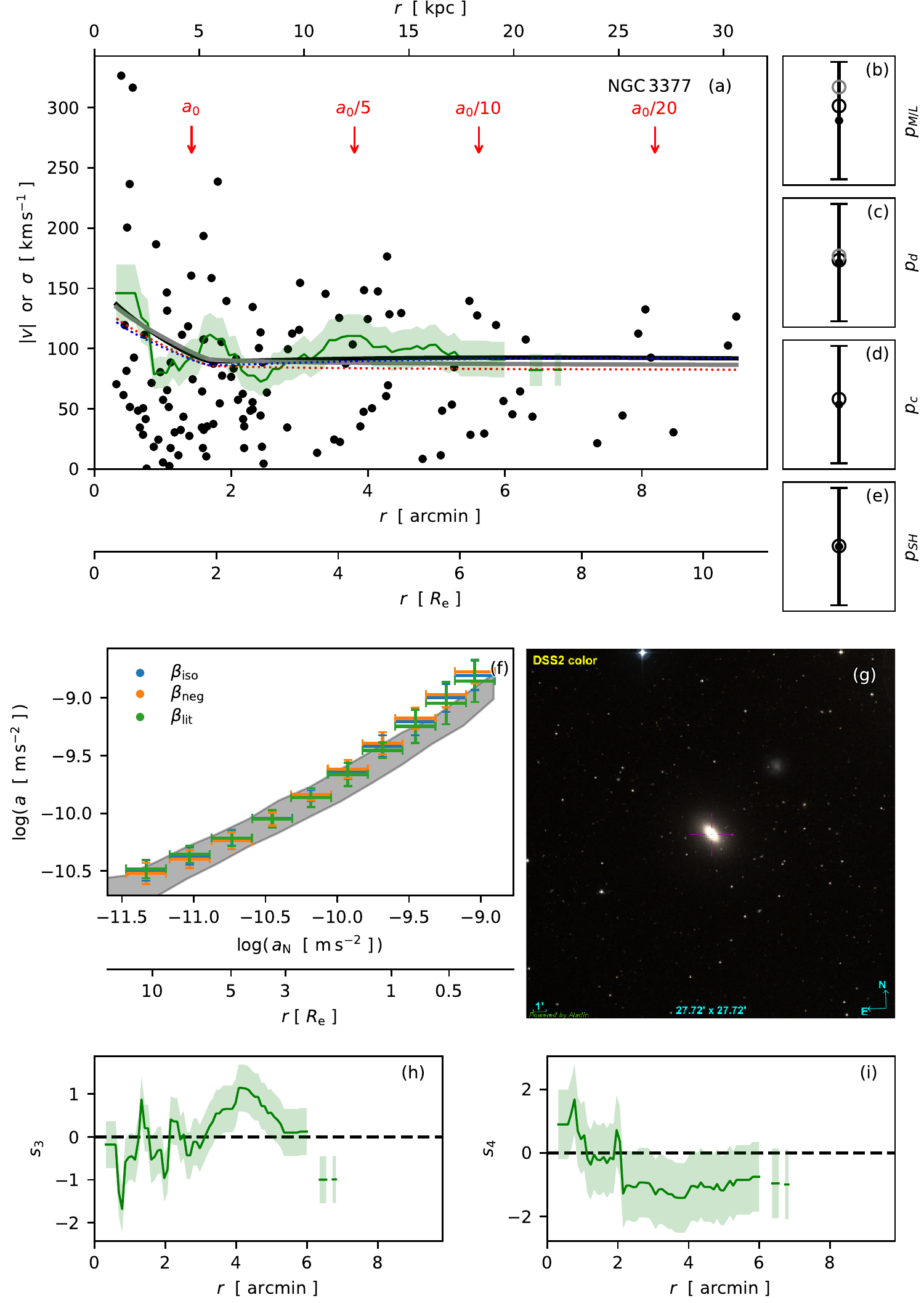}
 \caption{\protect\input{"./tables/latex/galaxycapt7.tex"}}
 \label{fig:g7}
  \end{figure*}
\begin{figure*}
 \centering
 \includegraphics[width=17cm]{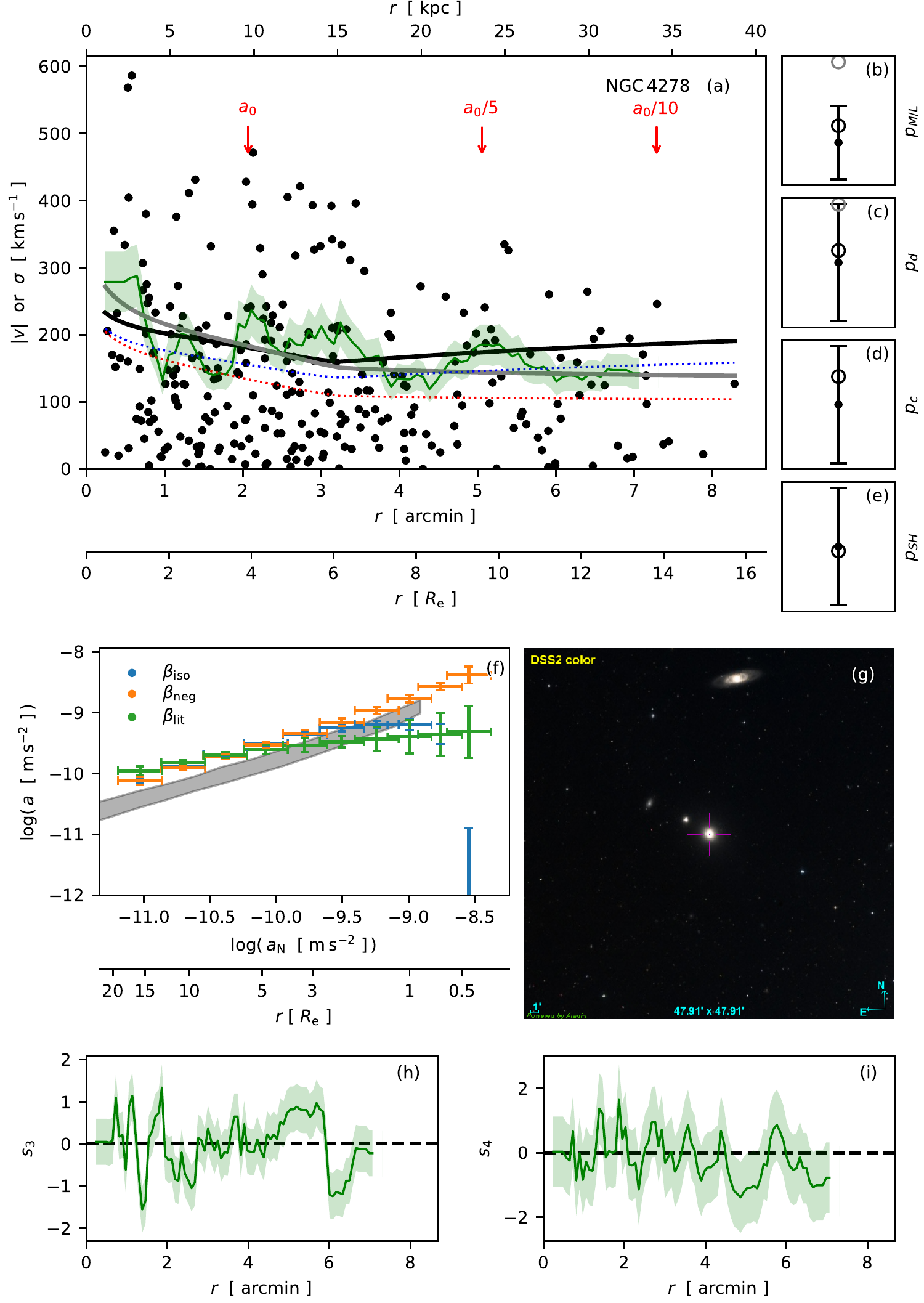}
 \caption{\protect\input{"./tables/latex/galaxycapt8.tex"}}
 \label{fig:g8}
  \end{figure*}
\begin{figure*}
 \centering
 \includegraphics[width=17cm]{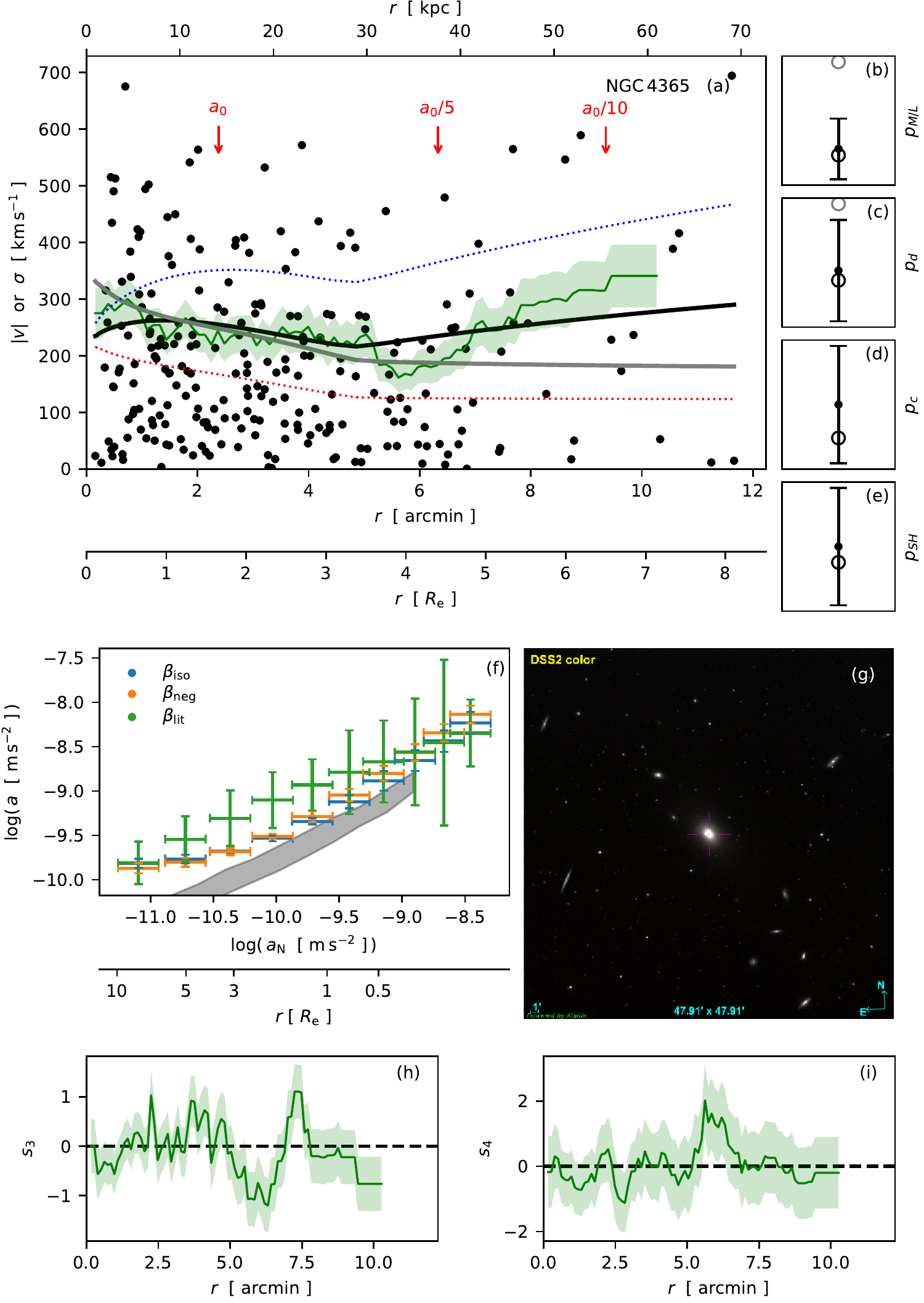}
 \caption{\protect\input{"./tables/latex/galaxycapt9.tex"}}
 \label{fig:g9}
  \end{figure*}
\begin{figure*}
 \centering
 \includegraphics[width=17cm]{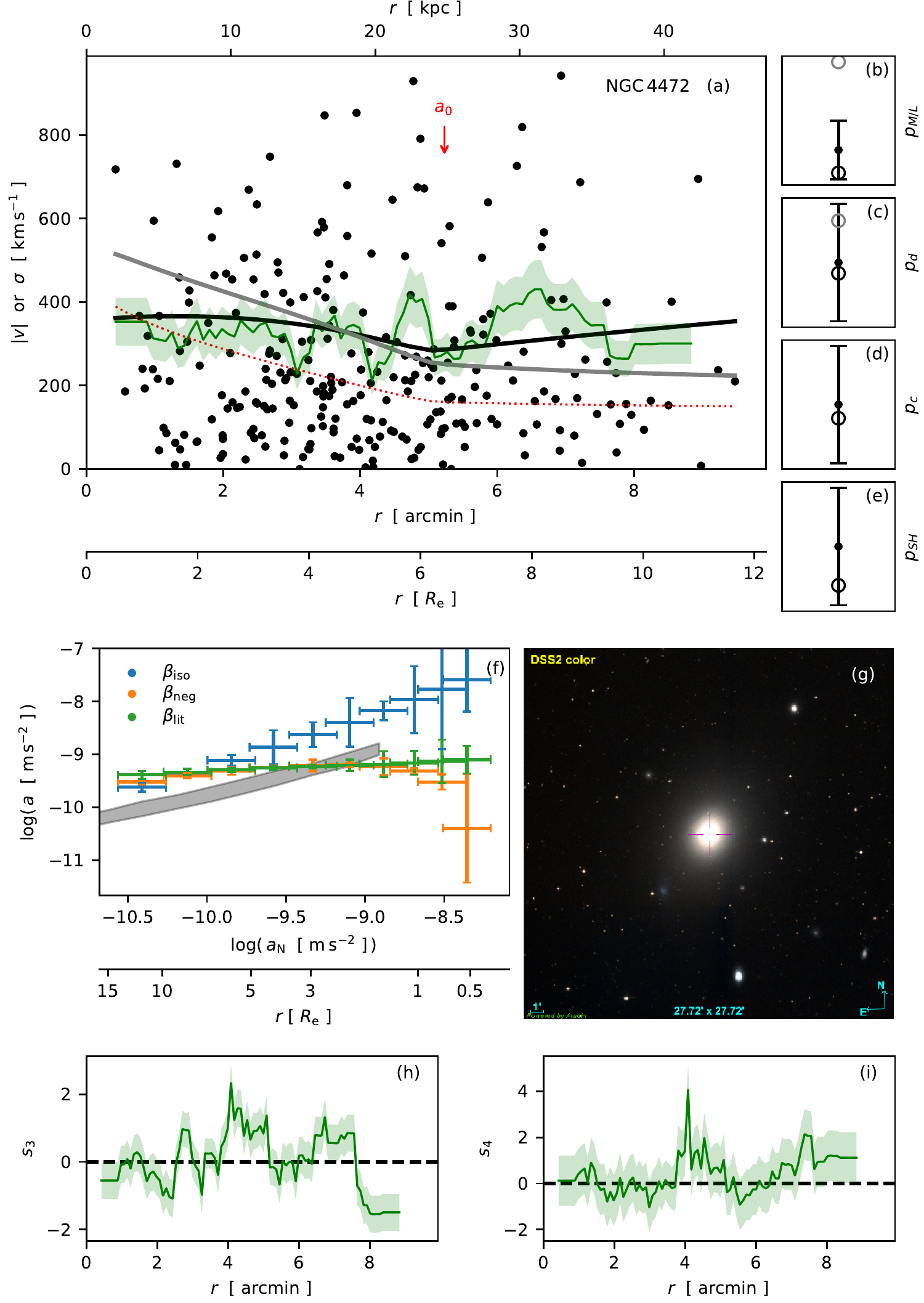}
 \caption{\protect\input{"./tables/latex/galaxycapt10.tex"}}
 \label{fig:g10}
  \end{figure*}
\begin{figure*}
 \centering
 \includegraphics[width=17cm]{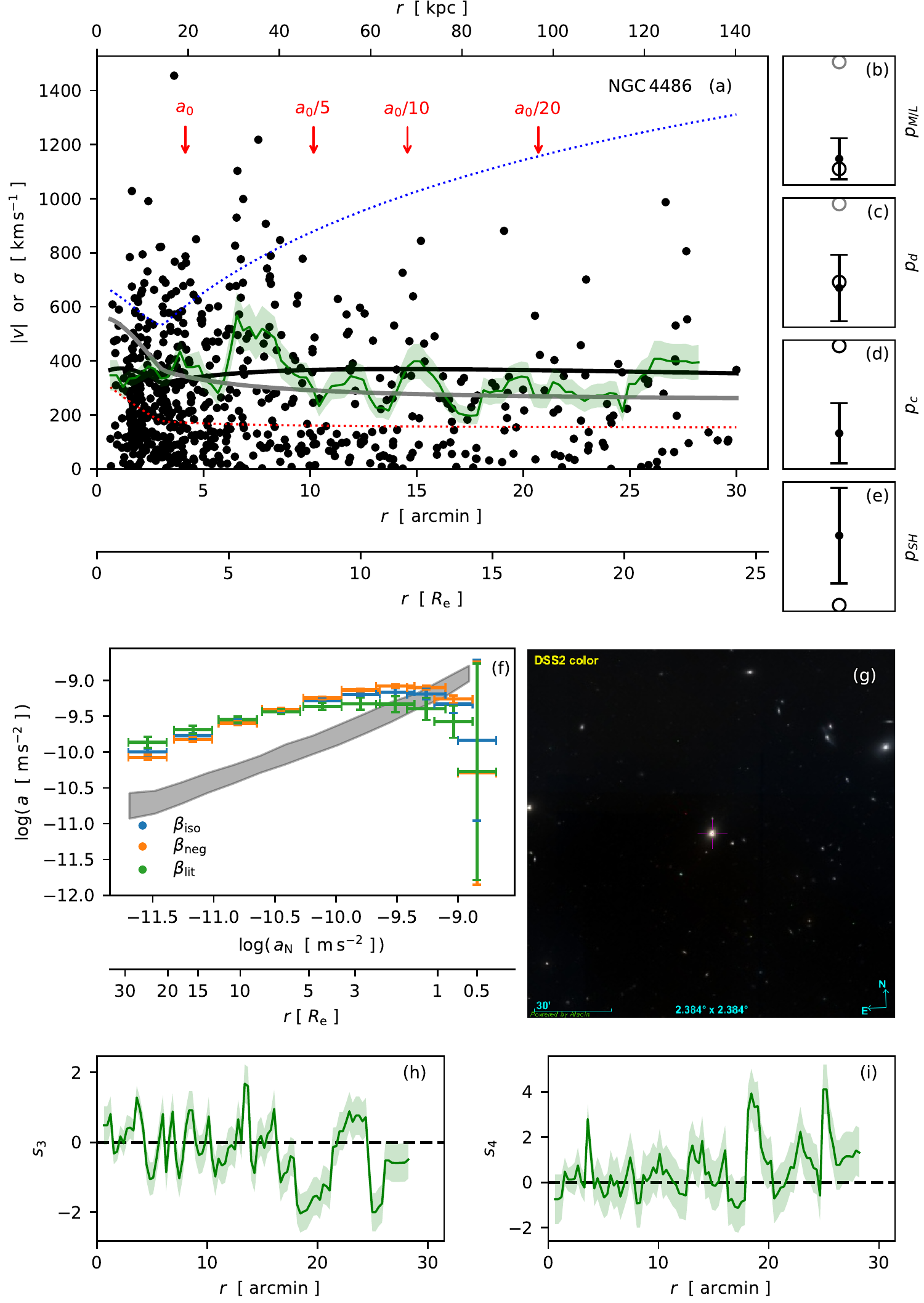}
 \caption{\protect\input{"./tables/latex/galaxycapt11.tex"}}
 \label{fig:g11}
  \end{figure*}
\begin{figure*}
 \centering
 \includegraphics[width=17cm]{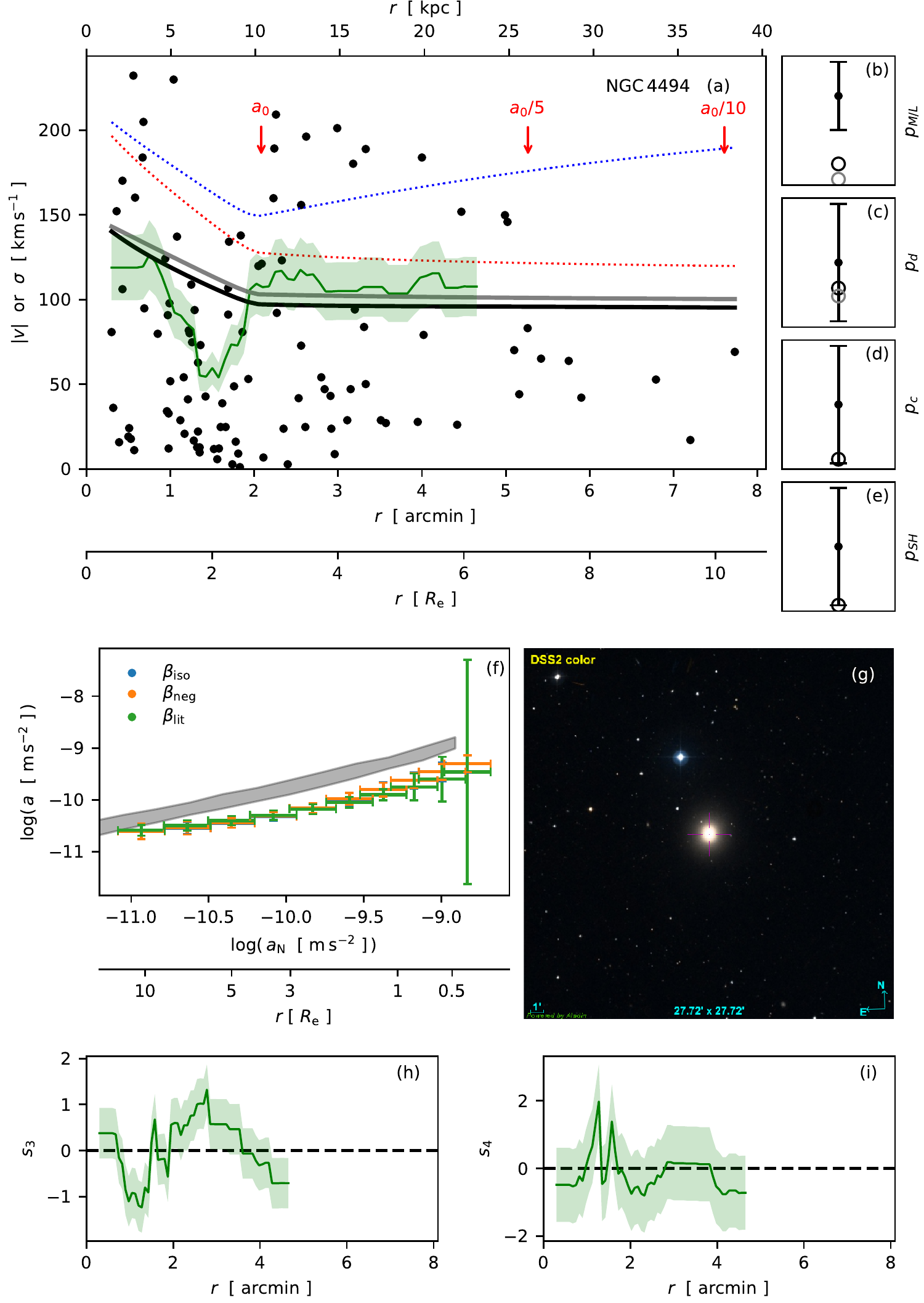}
 \caption{\protect\input{"./tables/latex/galaxycapt12.tex"}}
 \label{fig:g12}
  \end{figure*}
\begin{figure*}
 \centering
 \includegraphics[width=17cm]{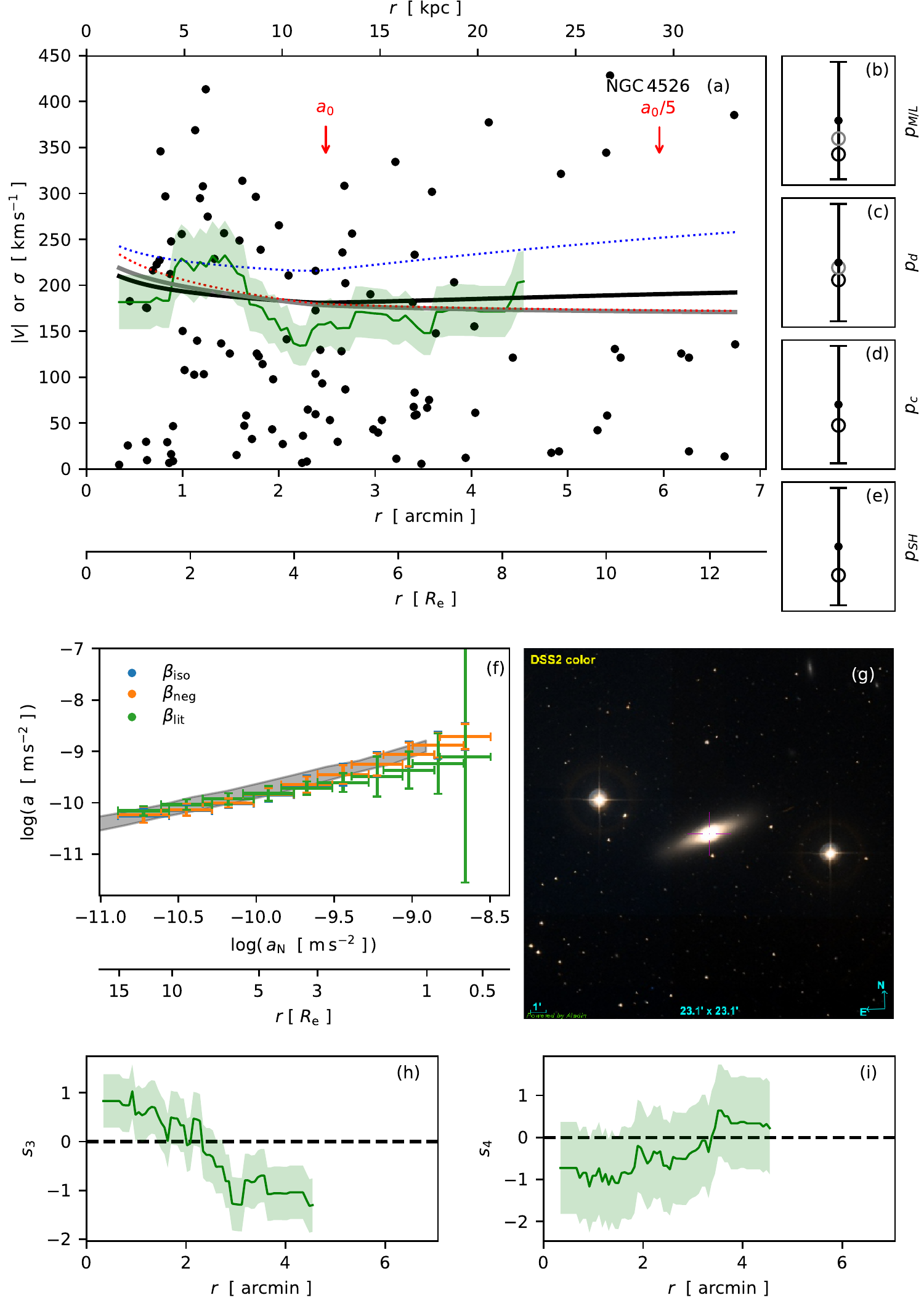}
 \caption{\protect\input{"./tables/latex/galaxycapt13.tex"}}
 \label{fig:g13}
  \end{figure*}
\begin{figure*}
 \centering
 \includegraphics[width=17cm]{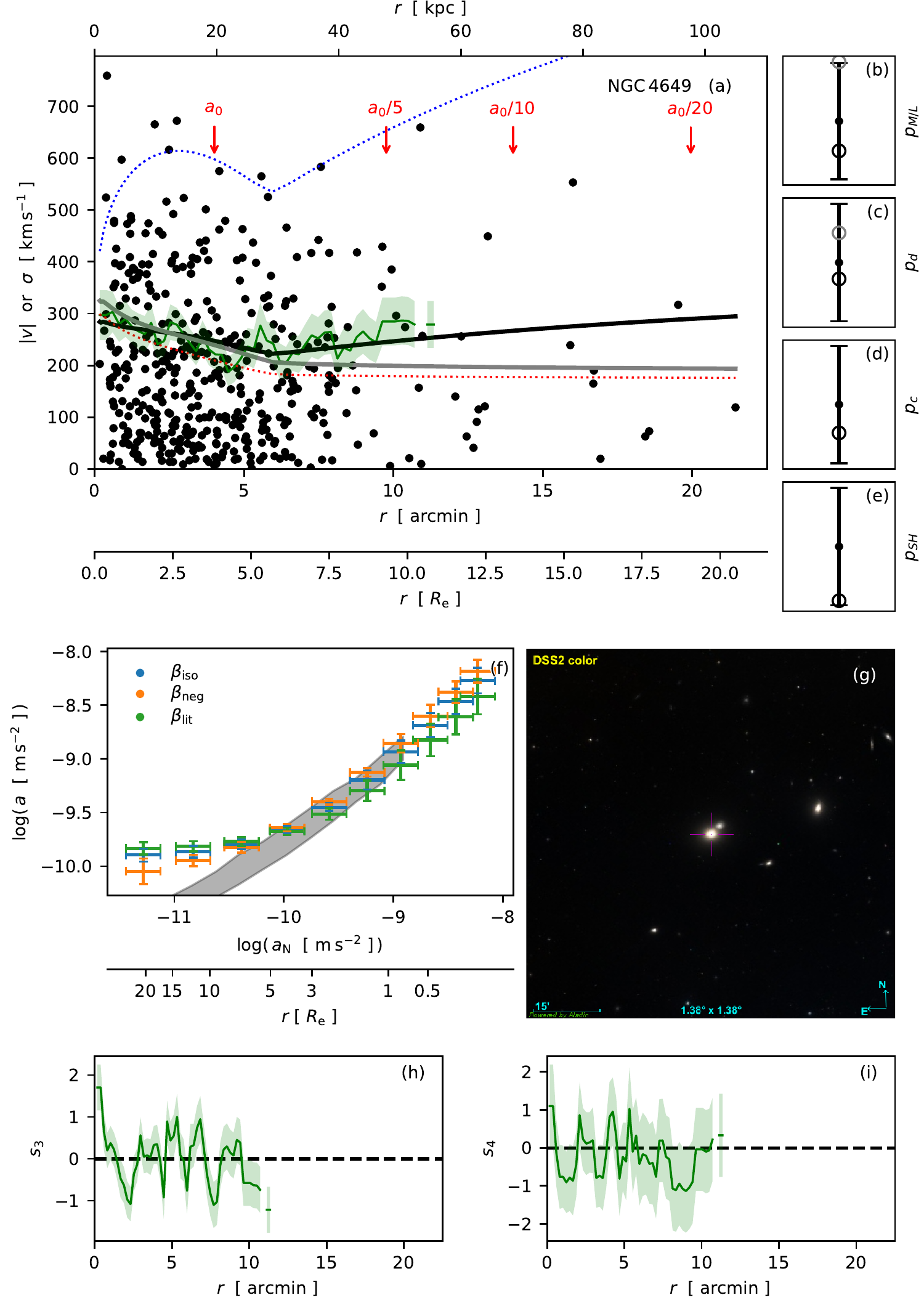}
 \caption{\protect\input{"./tables/latex/galaxycapt14.tex"}}
 \label{fig:g14}
  \end{figure*}
\begin{figure*}
 \centering
 \includegraphics[width=17cm]{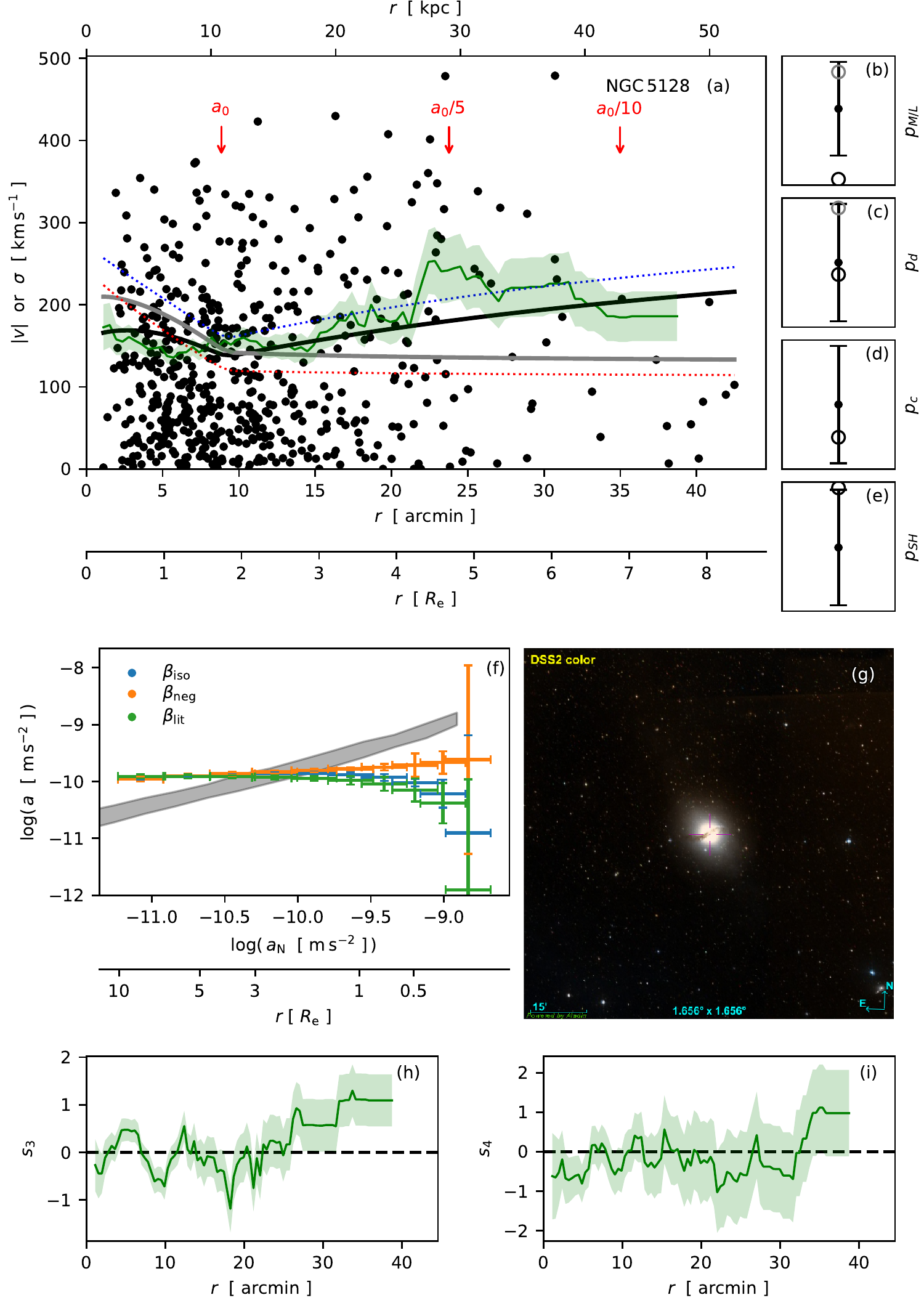}
 \caption{\protect\input{"./tables/latex/galaxycapt15.tex"}}
 \label{fig:g15}
  \end{figure*}
\begin{figure*}
 \centering
 \includegraphics[width=17cm]{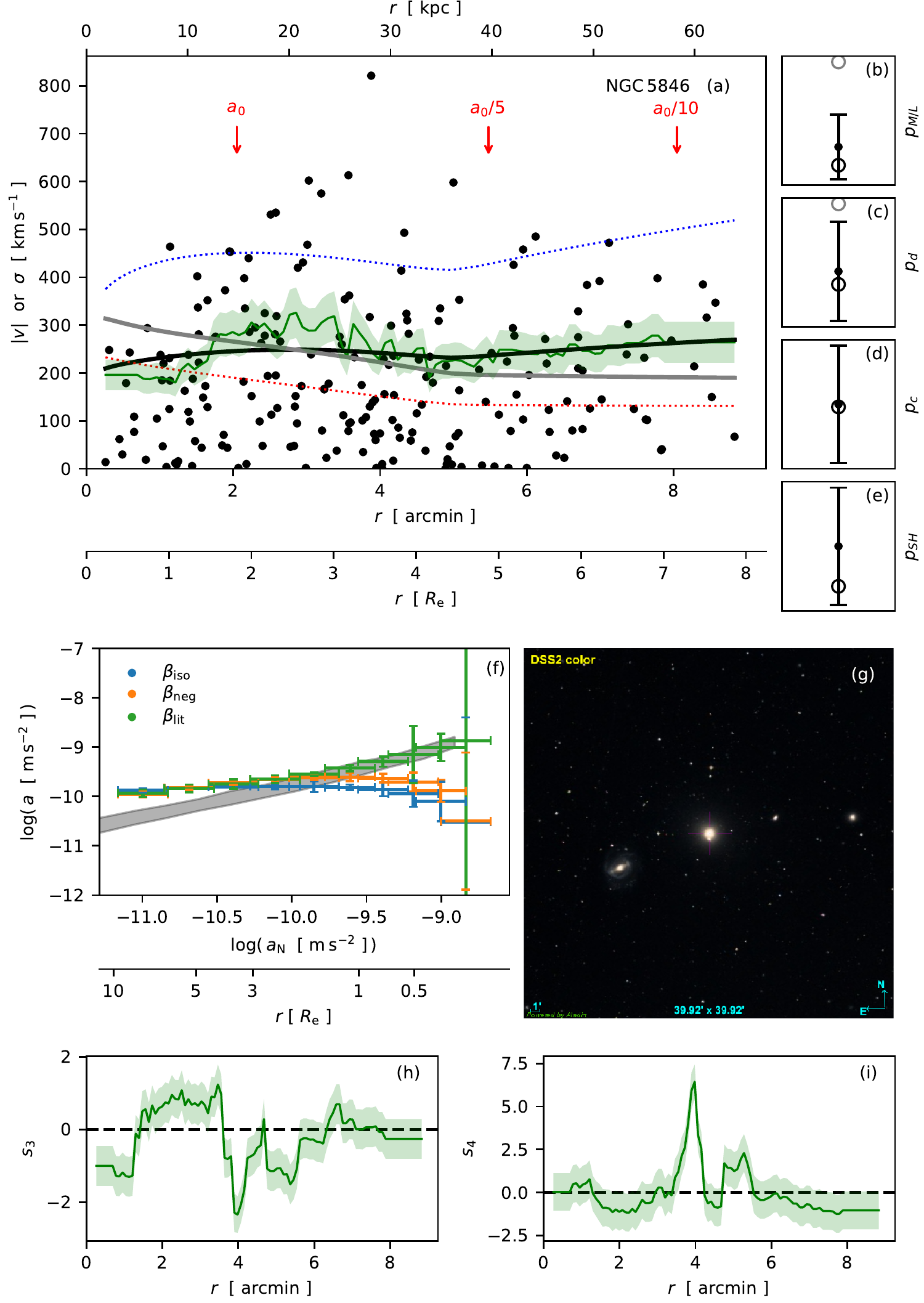}
 \caption{\protect\input{"./tables/latex/galaxycapt16.tex"}}
 \label{fig:g16}
  \end{figure*}
\begin{figure*}
 \centering
 \includegraphics[width=17cm]{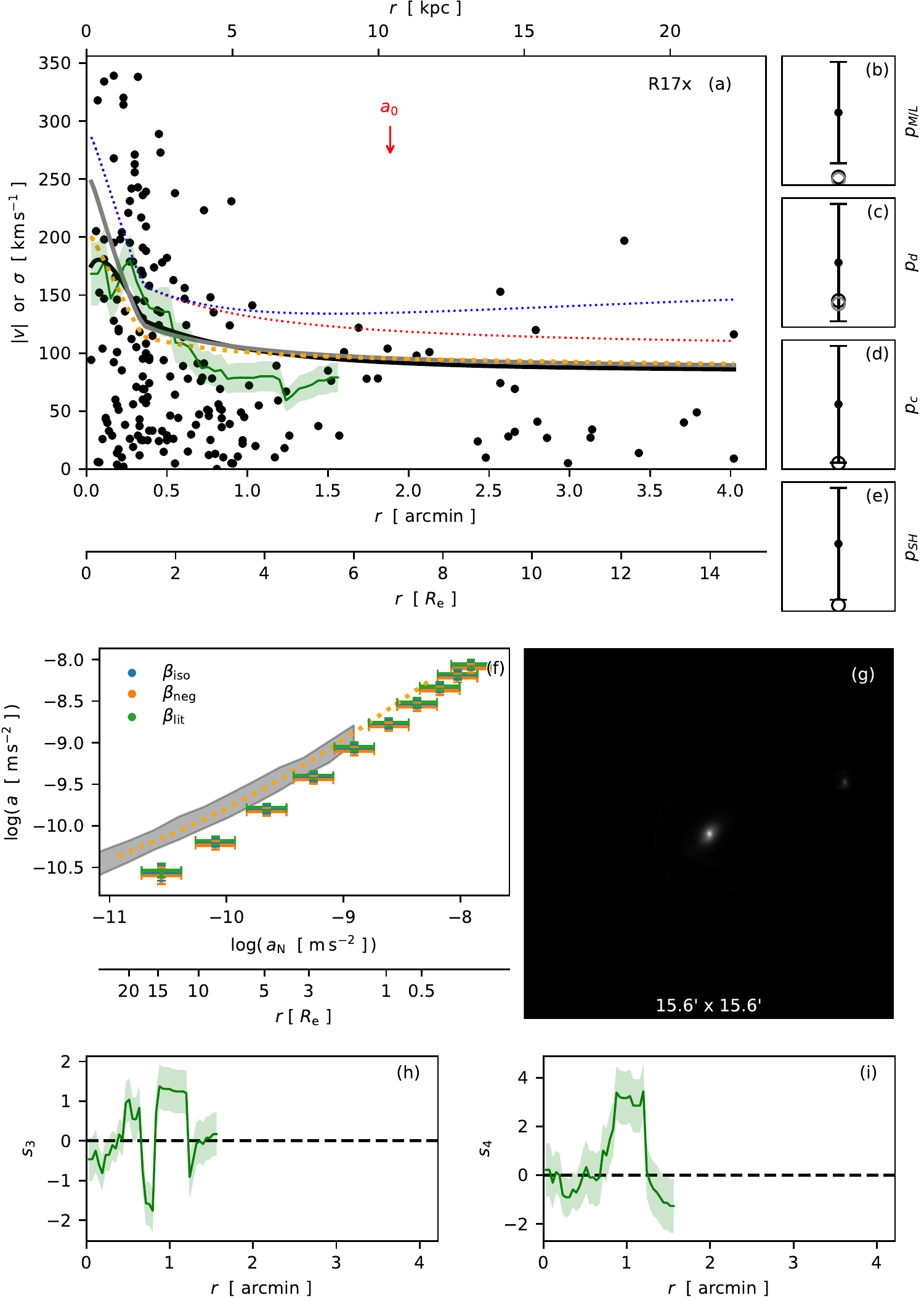}
 \caption{\protect\input{"./tables/latex/galaxycapt17.tex"}}
 \label{fig:g17}
  \end{figure*}
\begin{figure*}
 \centering
 \includegraphics[width=17cm]{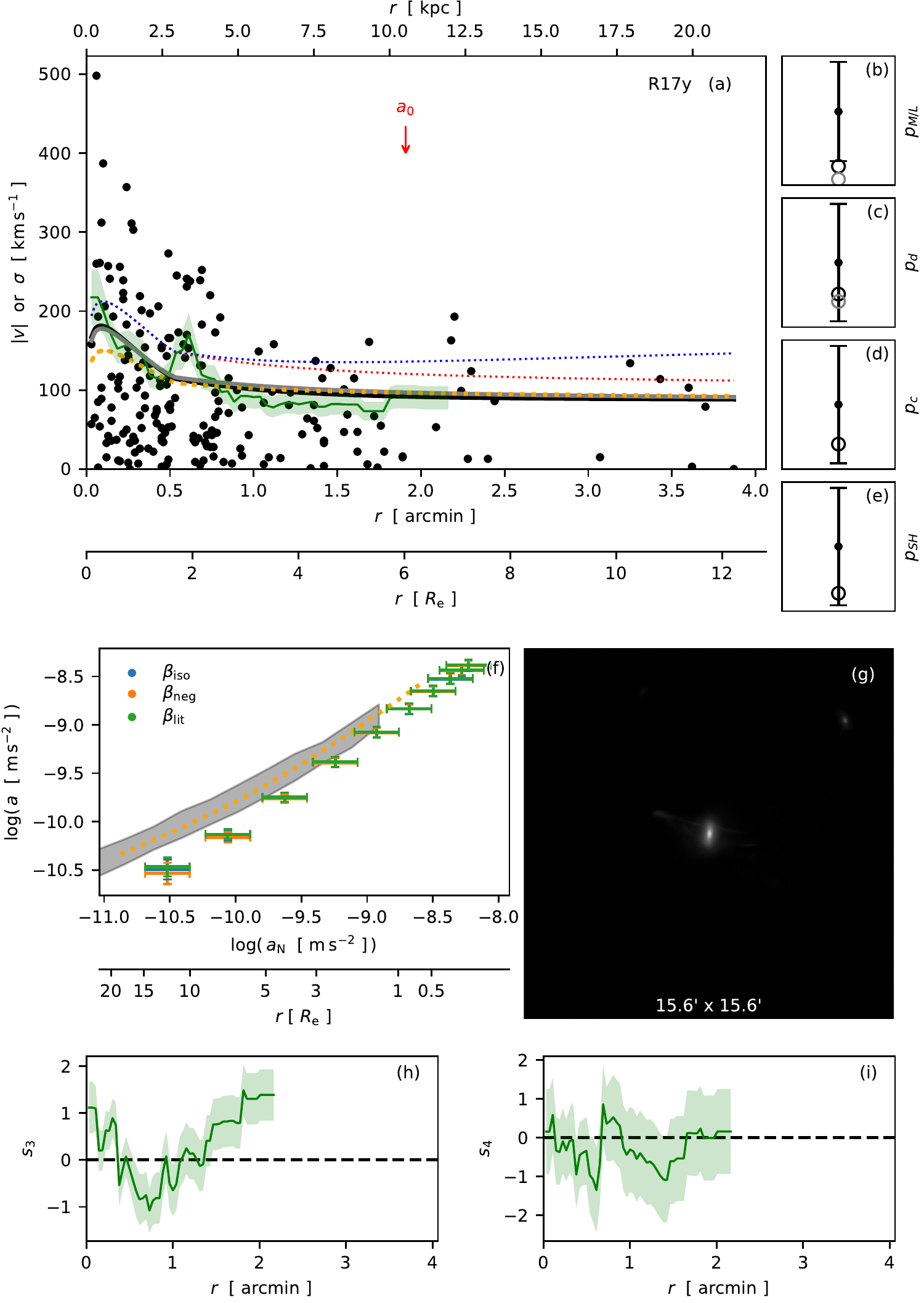}
 \caption{\protect\input{"./tables/latex/galaxycapt18.tex"}}
 \label{fig:g18}
  \end{figure*}
\begin{figure*}
 \centering
 \includegraphics[width=17cm]{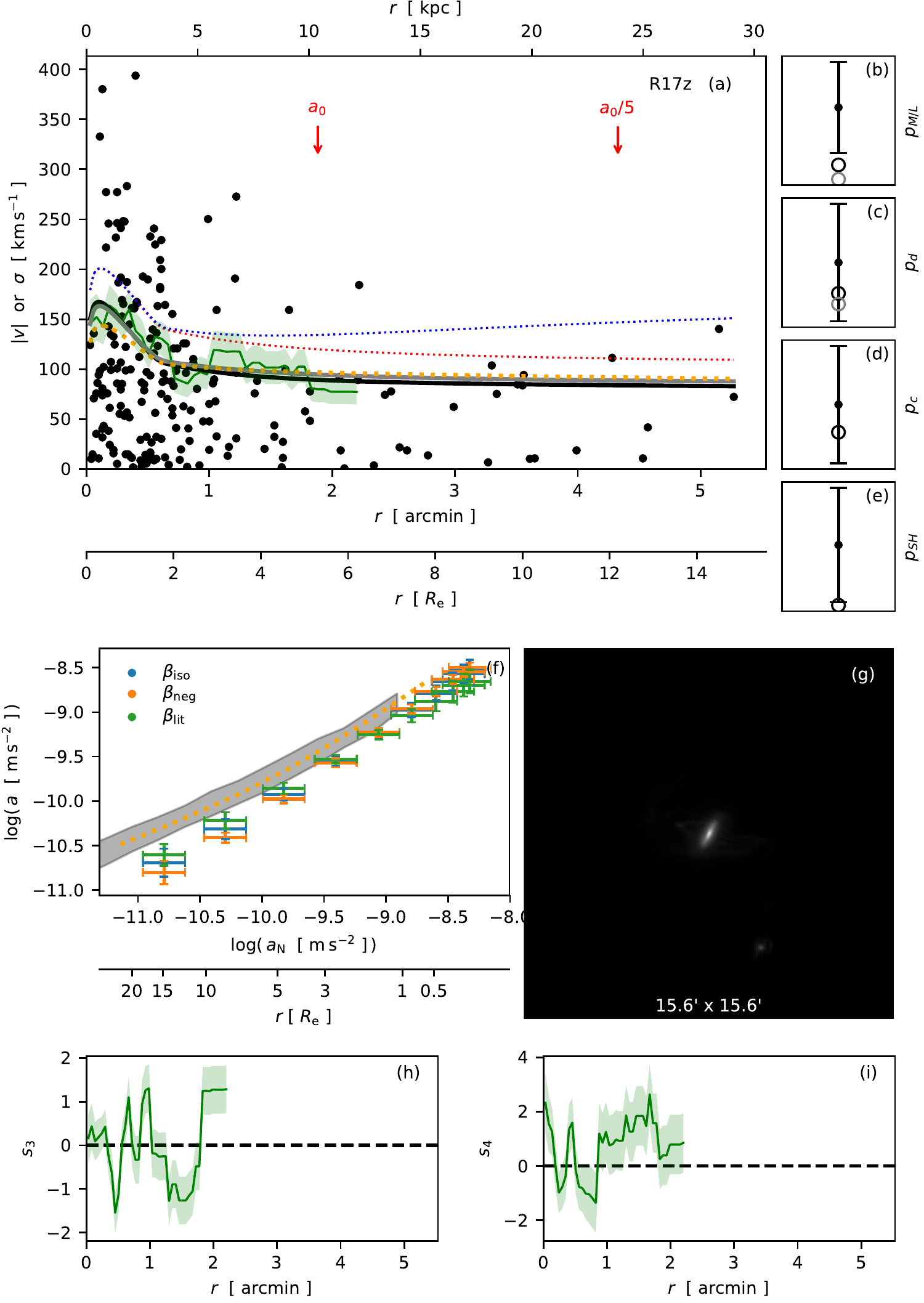}
 \caption{\protect\input{"./tables/latex/galaxycapt19.tex"}}
 \label{fig:g19}
  \end{figure*}

\end{appendix}

\end{document}